\def\frac#1#2{{{{#1}}\over{{#2}}}}
\documentclass[12pt]{article}
\usepackage{amsmath}
\usepackage{wasysym}    
\usepackage{textcomp}
\usepackage{pict2e}
 \usepackage[absolute]{textpos} 
\usepackage{cite}
\usepackage{slashbox}
\usepackage[font=small,labelfont=bf]{caption} 
\usepackage{graphics}
\usepackage{tikz-feynman}
\usepackage{float}

\usepackage{xcolor, cancel, ulem}
\usepackage{graphicx}
\usepackage{hyperref}
\usepackage{pdfpages}
\usepackage{cite}

\newsavebox{\ns}
\newsavebox{\dbrane}
\newsavebox{\dbshort}

\usepackage{epsfig}
\usepackage{amssymb}

\def\appendix{{\newpage\section*{Appendix}}\let\appendix\section%
        {\setcounter{section}{0}
        \gdef\thesection{\Alph{section}}}\section}

\newcommand\ba{\begin{eqnarray}}
\newcommand\ea{\end{eqnarray}}
\newcommand{\bbb}{\begin{eqnarray}\begin{array}{c}}
\newcommand{\bbl}[1]{\ba\label{#1}\begin{array}{c}{c}}
\newcommand{\eee}{\end{array}\end{eqnarray}}
\def\xxx{\eee\bbb}

\def\een#1{\label{#1} \eee}
\def\xxn#1{\een{#1}  \bbb}

\def\eenn{\nonumber\eee}

\definecolor{DarkGreen}{rgb}{0,.64,0}
\definecolor{gunmetal}{rgb}{0.171875, 0.207031, 0.222656}
\definecolor{chartreuse}{rgb}{.49,.98,0}
\definecolor{amethyst}{rgb}{0.59375,0.398438,0.792969}
\definecolor{brownrust}{rgb}{0.6875, 0.316406, 0.242188}
\definecolor{Violet}{rgb}{0.5,0,1}
\definecolor{BurntOrange}{rgb}{0.792969,0.332031,0}
\definecolor{FreshEggplant}{rgb}{0.59375, 0., 0.414063}
\definecolor{salmon}{rgb}{0.996094,0.507813,0.410156}
 \definecolor{FrenchRose}{rgb}{0.96875, 0.292969, 0.5625}
\definecolor{Cabaret}{rgb}{0.808594, 0.242188, 0.46875}
\definecolor{Shamrock}{rgb}{0.242188, 0.808594, 0.582031}
\definecolor{RobinsEggBlue}{rgb}{0., 0.792969, 0.792969}
\definecolor{GuardsmanRed}{rgb}{0.792969, 0., 0.}
\definecolor{Sapphire}{rgb}{0.183594, 0.328125, 0.621094}
\definecolor{Sorbus}{rgb}{0.996094, 0.429688, 0.0273438}
\definecolor{Red}{rgb}{1,0,0}
\definecolor{Blue}{rgb}{0,0,1}
\definecolor{Black}{rgb}{0,0,0}
\definecolor{Green}{rgb}{0,1,0}
\definecolor{thistle3}{rgb}{0.800781, 0.707031, 0.800781}
\definecolor{thistle4}{rgb}{0.542969, 0.480469, 0.542969}
\definecolor{DarkTurquoise}{RGB}{0,206,209}
\definecolor{turquoise4}{RGB}{0,134,139}

\definecolor{Purple}{rgb}{0.808594, 0.242188, 0.46875}
\newcommand{\purple}[1]{{\color{Purple} {#1}}}

\def\emm{\rwa}

\newcommand{\sshg}[1]{}

\newcommand{\sisg}[1]{}

\def\muchgreaterthan{{> \hskip-.07in >}}
\def\muchlessthan{< \hskip-.07in <}

\def\Dslash{\,\,{\raise.15ex\hbox{/}\mkern-12mu D}}
\def\Dbarslash{\,\,{\raise.15ex\hbox{/}\mkern-12mu {\bar D}}}
\def\delslash{\,\,{\raise.15ex\hbox{/}\mkern-9mu \partial}}
\def\delbarslash{\,\,{\raise.15ex\hbox{/}\mkern-9mu {\bar\partial}}}
\def\pslash{\,\,{\raise.15ex\hbox{/}\mkern-9mu p}}
\def\calDslash{\,\,{\raise.15ex\hbox{/}\mkern-12mu {\cal D}}}

\newcommand{\hh}{{1\over 2}}

\renewcommand{\ll}{_}
\newcommand{\uu}{^}
\newcommand{\pp}{\partial}

\renewcommand{\exp}[1]{{\rm exp}\{#1\}}

\renewcommand{\dag}{{}^\dagger{}}

\newcommand{\s}{\sigma}
\renewcommand{\t}{\tau}
\newcommand{\G}{\Gamma}
\newcommand{\g}{\gamma}
\renewcommand{\a}{\alpha}

\newcommand{\sqd}{^2}

\newcommand{\pb}{{\bar{\partial}}}
\renewcommand{\hh}{{1\over 2}}

\renewcommand{\th}{\theta}
\renewcommand{\t}{\tau}

\def\D{\Delta}

\newcommand{\llsk}{\hskip .5in}

\newcommand{\pr}{^\prime {}}

\newcommand{\IZ}{\relax\ifmmode\mathchoice
{\hbox{\cmss Z\kern-.4em Z}}{\hbox{\cmss Z\kern-.4em Z}}
{\lower.9pt\hbox{\cmsss Z\kern-.4em Z}} {\lower1.2pt\hbox{\cmsss
Z\kern-.4em Z}}\else{\cmss Z\kern-.4em Z}\fi} \font\cmss=cmss10
\font\cmsss=cmss10 at 7pt
\newcommand{\inbar}{\,\vrule height1.5ex width.4pt depth0pt}
\newcommand{\IC}{{\relax\hbox{$\inbar\kern-.3em{\rm C}$}}}
\newcommand{\IQ}{{\relax\hbox{$\inbar\kern-.3em{\rm Q}$}}}
\newcommand{\IP}{\relax{\rm I\kern-.18em P}}

\def\co{{\cal O}}

\renewcommand{\l}{\lambda}

\renewcommand{\pr}{^\prime}

\newcommand{\IR}{\relax{\rm I\kern-.18em R}}
\def\blfootnote{\xdef\@thefnmark{}\@footnotetext}
\newcommand{\bm}{\begin{matrix}}
\renewcommand{\em}{\end{matrix}}
\newcommand{\ee}[1]{\ba {#1} \ea}

\newcommand{\upp }[1]{^{({#1})}{}}

\newcommand{\rr}[1]{(\ref{{#1}})}

\newcommand{\prpr}{^{\prime\prime}}
\newcommand{\prprpr}{^{\prime\prime\prime}}
\newcommand{\prprprpr}{^{\prime\prime\prime\prime}}
\newcommand{\prprprprpr}{^{\prime\prime\prime\prime\prime}}

\newcommand{\heading}[1]{\begin{center}\it \blue{#1} \rm \end{center}}

\newcommand{\us}[2]{^{({#1}{#2})}}

\def\BeginItemize{\begin{itemize}}
\def\ei{\end{itemize}}

\def\ed{\end{document}}

\def\cc{{\cal I}}

\renewcommand{\rr}[1]{(\ref{#1})}

\def\cc{\,}

\def\cc{\,}

\newcommand{\lp}[1]{_{({#1})}}

\newcommand{\lrm}[1]{_{{\rm {#1}}}}
\newcommand{\urm}[1]{^{{\rm {#1}}}}
\newcommand{\uprm}[1]{^{({\rm {#1}})}}

\newcommand{\ls}[1]{_{[{#1}]}}

\newcommand{\redd}[1]{{\color{Red} {#1}}}

\definecolor{Purple}{rgb}{0.808594, 0.242188, 0.46875}

\renewcommand{\bm}{\begin{matrix}}
\renewcommand{\em}{\end{matrix}}

\def\be{\begin{eqnarray}}
\def\ee{\end{eqnarray}}

\newcommand{\blue}[1]{{\color{Blue}{#1}}}

\def\redlowdash{{\color{Red}{\rule[-0.5ex]{2pt}{0.4pt}}}}
\def\redmiddash{{\color{Red}{\rule[+0.5ex]{2pt}{0.4pt}}}}

\def\cute{{\lower3.5pt\hbox{\sixly
  \kern-.21pt \char58 \kern-.21pt }}}

\def\midcute{{\lower-1.0pt\hbox{\sixly
  \kern-.21pt \char58 \kern-.21pt }}}

  \def\lowcute{{\lower3.5pt\hbox{\sixly
  \kern-.21pt \char58 \kern-.21pt }}}
  
  \def\redmidcute{{\color{Red} \midcute}}
  
  \def\redlowcute{{\color{Red} \lowcute}}
   
    \def\bluelowcute{{\color{Blue} \lowcute}}

   \def\swave{\bgroup \markoverwith \midcute \ULon} 
  
  \def\redswave{\bgroup \markoverwith \redmidcute \ULon} 
  
  \def\reduline{\bgroup \markoverwith \redlowdash \ULon}
  
   \def\blueuline{\bgroup \markoverwith \bluelowdash \ULon}
  
   \def\reduwave{\bgroup \markoverwith \redlowcute \ULon}
   
   \def\blueuwave{\bgroup \markoverwith \bluelowcute \ULon}
  
  \def\redsout{\bgroup \markoverwith \redmiddash \ULon}
  
   \def\bluesout{\bgroup \markoverwith \bluemiddash \ULon}
   
   \def\Irrel{\bgroup \markoverwith {{\color{Red} {\bf X}}} \ULon}

\newcommand{\eqirrel}[1]{\rdots}
  
\let\oldcancel\cancel
\renewcommand\cancel[1][black]{%
  \def\CancelColor{\color{#1}}%
  \oldcancel}

   \let\oldbcancel\bcancel
\renewcommand\bcancel[1][black]{%
  \def\CancelColor{\color{#1}}%
  \oldbcancel}

\def\rdots{\redd{\circ\circ\circ}}

\def\dag{^\dagger}

\renewcommand{\upp }[1]{^{({#1})}}

\def\silent#1{}
\newcommand{\defout}[1]{}

\newcommand{\OLDcentpicWidthHeightAngleFile}[4]{ \begin{figure}[htb]
\begin{center}
\includegraphics[width={#1},height={#2},angle={#3}]{#4}\end{center}
\end{figure} }

\newcommand{\centpicWidthHeightAngleFileCaption}[5]{ \begin{figure}[!htb]
\begin{center}
\includegraphics[width={#1},height={#2},angle={#3}]{#4}\end{center}
\captionof{figure}{{#5}}
\end{figure} }

\newcommand{\centpicWidthHeightAngleFileCaptionFiglab}[6]{ \begin{figure}[!htb]
\begin{center}
\includegraphics[width={#1},height={#2},angle={#3}]{#4}\end{center}
\captionof{figure}{{#5}}
\label{#6}
\end{figure} }

\def\us#1^{[{#1}]}

{\color{Purple}}%
{\color{Black}}
{\color{Blue}}%
{\color{Black}}
{\color{Red}}%
{\color{Black}}

\def\blue#1{{\color{Blue}{#1}}}

{\color{Red}}%
{\color{Black}}

\def\bbsk{\hskip-.5in}

\usepackage{mathtools}

\def\tb{{\bar{\tau}}}

\def\JJM{\purple{{\cal J}}}

\makeatletter
\newcommand*{\Relbarfill@}{\arrowfill@\Relbar\Relbar\Relbar}
\newcommand*{\xeq}[2][]{\ext@arrow 0055\Relbarfill@{#1}{#2}}
\makeatother

\makeatletter
\newcommand*{\rxeq}[2][]{{\color{Red} \ext@arrow 0055\Relbarfill@{#1}{#2}}}
\makeatother

\def\QuestionableEqualityLevelOne{ \xeq[\purple{\uparrow}]{ \redd{?}}}

\def\qeqOne{\QuestionableEqualityLevelOne}

\def\qeqA{\qeqOne}

\def\sb{{\bar{\sigma}}}

\def\xxnn{\nonumber\xxx}

\def\bii{\begin{itemize}}

\def\G{\Gamma}

\newcommand{\dol}[1]{$#1$}
\def\bbd{\dol}

\usepackage{setspace}

\def\ordinary#1{#1}
\def\ddsc{\purple{\lambda}}
\def\DDSL{{\rm lim}_{ {{n\to\infty}\atop{\ddsc{\rm ~fixed}}}}}

\def\AlienSection#1{\blue{#1}}
\def\alsec#1{\AlienSection{#1}}

\def\EVENSOMEWHATSHORTERMACRO#1#2#3#4#5#6#7#8{\blue{{#1}} & \blue{{#2}} &  \blue{{#3}} & \blue{{#4}} & \blue{{#5}} & \blue{{#6}} & \blue{{#7}} & \blue{{#8}} \cr \hline}

\def\NEWDSCMACRO#1#2#3#4#5#6#7{\blue{{#1}} & \blue{{#2}} &  \blue{{#3}} & \blue{{#4}} & \blue{{#5}} & \blue{{#6}} & \blue{{#7}} \cr \hline}
\def\NEWERDSCMACRO#1#2#3#4#5#6#7#8{{#1} & {#2} &  {#3} & {#4} & {#5} & {#6} & {#7} & \blue{{#8}} \cr \hline}

\def\ReUpdatedNEWESTMACRO#1#2#3#4#5#6#7#8#9{{#1}  &  {#3} & {#4} & {#5} & {#6} & {#7} & {#8} & \blue{{#9}} \cr \hline}

\def\smgkt{\hbox{\tiny{ref. \cite{Grassi:2019txd}}}}

\def\smovie{\hbox{\scriptsize{ref. \cite{Bourget:2018obm}}}}

\def\MacroFLPLower#1{F\ls{#1}}

\def\sotp{(s/2\pi)}
\def\shgOFMW#1{}
\def\WFunxLocation{eqs. \rr{W1Formula}-\rr{W5Formula}}

\def\emm#1{{\it #1}}

\begin{document}

\begin{titlepage}
\vspace{-8 mm}
\begin{center}

  {\large \bf  \vspace{2mm}   On the exponentially small corrections \\ \vspace{4mm}  to \bbd{{\cal N} = 2} superconformal  correlators at large R-charge }
\end{center}
\vspace{2 mm}
\begin{center}
{Simeon Hellerman$^{1,2}$}\\
\vspace{6mm}
{\it $^1$Kavli Institute for the Physics and Mathematics of the Universe \textsc{(wpi)}\\
The University of Tokyo \\
 Kashiwa, Chiba  277-8582, Japan\\}
 \vspace{6mm}
{\it $^2$Department of Physics, Faculty of Science,\\
University of Tokyo, Bunkyo-ku, Tokyo 133-0022, Japan\\}
\vspace{6mm}

\end{center} 
\vspace{-4 mm}
\begin{center}
{\large Abstract}
\end{center}
\noindent
In this note we consider Coulomb-branch chiral primary correlation functions in \bbd{{\cal N} = 2} superconformal QCD with gauge group \bbd{SU(2)},
in the limit of large R-charge \bbd{\JJM = 2n} for the chiral primary operators \bbd{[\co(x)]\uu n} with the inverse gauge coupling \bbd{\t} held fixed.
In previous work \cite{Hellerman:2017sur, Hellerman:2018xpi, Hellerman:2020sqj, Hellerman:2021yqz}   
  these correlation functions were determined to all orders in \bbd{n}, up to unknown exponentially
small corrections.  In this paper we determine the first several orders of the asymptotic expansion of
the exponentially small correction itself.  To do this we use: the physical interpretation of the exponentially small correction as 
the virtual propagation of a massive BPS particle, to fix the leading term in the expansion; the supersymmetric recursion relations of
\cite{Baggio:2014ioa, Baggio:2014sna, Baggio:2015vxa, Gerchkovitz:2016gxx}
to derive differential
equations for the
coupling-dependence of the subleading terms; and the double-scaling limit of \cite{Bourget:2018obm, Grassi:2019txd}, to fix undetermined coefficients in the solution of the differential equation.
We calculate the expansion of the exponentially small term up to and including relative order \bbd{n\uu{-{5\over 2}}}.
We also use the recursion relations to calculate the subleading large-\bbd{\JJM} corrections to the exponentially small correction in the double-scaling limit, up to
and including relative order \bbd{n\uu{-5}} at fixed double-scaled coupling \bbd{\ddsc}.  We compare the expansion
to exact results from supersymmetric localization \cite{LocData} at the coupling \bbd{\t = {{25}\over \pi}\cc i}, up to \bbd{n=150}.   At values \bbd{n\sim 100-150}, we find
the fixed-coupling and double-scaled large-R-charge expansions are accurate to within one part in \bbd{10\uu 6} and \bbd{10\uu 8}, respectively, of the size of the exponentially small correction itself.  Relative to the full correlator including the dominant \textsc{eft} contribution, these estimates give results accurate to one part in \bbd{10\uu{15}} and
\bbd{10\uu{17},} respectively.

\vspace{0cm}
\begin{flushleft}
\today
\end{flushleft}
\end{titlepage}

\newpage
\section*{Dedication}

This paper is dedicated to the memory of Fred Hellerman, and to his granddaughter Dymond Henry on the occasion of her thirteenth birthday.

\OLDcentpicWidthHeightAngleFile{160mm}{90mm}{0}{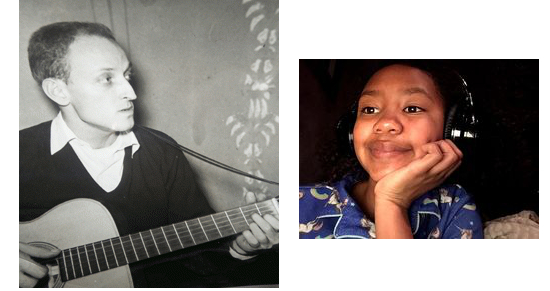}

\newpage

\tableofcontents
\newpage
\numberwithin{equation}{section}

\section{Introduction}

Recently there has been development of the use of large quantum number as a limit in which observables can be computed in systems with global symmetries with
control over quantum effects beyond the usual weak-coupling regime.  In systems with no weak-coupling parameter in the Hamiltonian, inverse large quantum number can play
the role of a small coupling constant \cite{Hellerman:2015nra, Monin:2016jmo, Cuomo:2017vzg,delaFuente:2018qwv, Loukas:2016ckj, Loukas:2017lof, Loukas:2017hic, Nakayama:2015hga, Nakayama:2020dle, Nakayama:2019jvm, Loukas:2018zjh, Sharon:2020mjs, Gaume:2020bmp, Cuomo:2020rgt, Orlando:2019hte, Cuomo:2019ejv, Orlando:2020yii, Alvarez-Gaume:2016vff, Hellerman:2017veg, Jafferis:2017zna, Alvarez-Gaume:2019biu, Kumar:2018nkf, Watanabe:2019adh,
Arias-Tamargo:2019xld, Arias-Tamargo:2019kfr, Arias-Tamargo:2020fow, Badel:2019oxl, Badel:2019khk, Giombi:2020enj, Antipin:2021akb, Cuomo:2021qws, Komargodski:2021zzy, Cuomo:2021ygt, Kravec:2018qnu, Favrod:2018xov, Kravec:2019djc, Orlando:2020idm, Hellerman:2020eff, Jack:2021ypd, Dupuis:2021yej, Dondi:2021buw, Hellerman:2017sur, Bourget:2018obm, Hellerman:2018xpi, Beccaria:2018xxl, Beccaria:2018owt, Beccaria:2020azj, Grassi:2019txd,  Hellerman:2020sqj, Hellerman:2021yqz} In a wide range of systems one finds that the limit of large quantum number dramatically simplifies the computation of many observables in the theory, while preserving 
interesting dynamical behaviors of the theory.  In all examples, the physical picture is rather straightforward: The large-charge state is well-described by a classical ground state of a set of Nambu-Goldstone
degrees of freedom\footnote{The formal description of the Abelian large-charge \textsc{eft} in terms of the NG degrees of freedom {\it via} the CCWZ formalism \cite{Coleman:1969sm, Callan:1969sn}was explored in \cite{Monin:2016jmo}.  Large-global-charge \textsc{eft}s are themselves special cases of a more general subject of large quantum-number
effective theories including large spin~\cite{Berenstein:2003gb, Alday:2007qf, Alday:2007mf, Alday:2013cwa, Hellerman:2013kba, Sonnenschein:2014jwa, Sonnenschein:2014bia, Sonnenschein:2015zaa, Dubovsky:2016cog, Sonnenschein:2018aqf, Sonnenschein:2019bca, Sonnenschein:2020jbe, Conkey:2019blu, Alday:2015eya, Fitzpatrick:2012yx, Komargodski:2012ek, Caron-Huot:2016icg} and more general high-energy~\cite{Son:1995wz, Srednicki, Deutsch, Cardy:1986ie, Hartman:2014oaa, Delacretaz:2020nit, Mukhametzhanov:2020swe} and
high-particle-number~\cite{Bachas:1991fd, Libanov:1994ug, Son:1995wz, Jaeckel:2018ipq, Monin:2018cbi, Khoze:2018mey, Dine:2020ybn, Panagopoulos:2020sxp} limits.}
~\cite{Hellerman:2015nra, Monin:2016jmo, Cuomo:2017vzg, Sharon:2020mjs, Gaume:2020bmp, Cuomo:2020rgt, Orlando:2019hte, Cuomo:2019ejv, Orlando:2020yii, Alvarez-Gaume:2016vff, Hellerman:2017veg, Hellerman:2017efx, Hellerman:2018sjf}.
In such theories, quantum
fluctuations about the dominant classical solution are suppressed by negative powers of the total charge \bbd{\JJM}.  In principle this picture gives an all-orders asymptotic expansion in inverse powers of the charge for operator 
dimensions, OPE coefficients, and other large-charge observables.  The only unknown ingredients are the Wilson coefficients in the action of the large-charge \textsc{eft}.  Beyond an all-orders asymptotic expansion,
the \textsc{eft} picture is modified only by exponentially small corrections associated with virtual propagation of massive non-Goldstone degrees of freedom over distances of order the infrared scale.

 One particularly nice testing ground for these ideas has been superconformal field theory in \bbd{D=4} with \bbd{{\cal N} = 2} extended supersymmetry. \cite{Hellerman:2017sur, Bourget:2018obm, Hellerman:2018xpi, Beccaria:2018xxl, Beccaria:2018owt, Beccaria:2020azj, Grassi:2019txd, Hellerman:2020sqj, Hellerman:2021yqz}, particularly
 those with a one-complex-dimensional Coulomb branch.       
For such theories there are tools available associated with exact supersymmetry that allow the computation of supersymmetrically-protected observables at a nonperturbative level, with the computation
of the \bbd{S\uu 4} partition function as a starting point.  These methods \cite{Gerchkovitz:2016gxx},
associated with supersymmetric localization  \cite{Nekrasov:2002qd, Pestun:2007rz, Alday:2009aq} make it possible to check the predictions of the large-quantum-number expansion against an independent calculation. 

For \bbd{D=4,~{\cal N} = 2} superconformal theories, the simplest nontrivial large quantum number calculation is the two-point function of a BPS chiral primary operator carrying large \bbd{U(1)\ll R} R-charge.
For the special case of rank-one gauge group \bbd{G=SU(2)}, the chiral ring of the Coulomb branch is generated by a single operator \bbd{\co\ll\D
} of dimension \bbd{\D} and \bbd{U(1)\lrm R} R-charge\footnote{In this paper as in \cite{Hellerman:2017sur, Hellerman:2018xpi, Hellerman:2020sqj, Hellerman:2021yqz} we use a slightly nonstandard normalization convention in which the supercharges have \bbd{U(1)\lrm R} R-charge \bbd{\pm \hh} and the lowest component of a free chiral superfield has R-charge \bbd{+1}.  This differs 
by a factor of \bbd{2} from the usual convention for \bbd{{\cal N} = 2} theories in \bbd{D=4}.} \bbd{J=\D}.   In previous work \cite{Hellerman:2017sur, Hellerman:2018xpi, Hellerman:2020sqj, Hellerman:2021yqz}  the correlation functions
of the \bbd{n\uu{\rm{\underline{th}}}} powers of these operators have been computed in a large-\bbd{n} expansion.  In the expansion at large \bbd{n} with the coupling held fixed, there is a simple
asymptotic formula for the logarithm of the correlation function that holds to all orders in \bbd{{1\over n}}, and gives universal, theory-independent predictions for all power-law terms \bbd{\JJM\uu{-k},~k\geq 1} down to corrections exponentially small in \bbd{\sqrt{\JJM}}.  These exponentially
small corrections are the leading \emm{non-}universal scheme-independent
contributions to the log of the correlation function, and represent the contribution of virtual macroscopic propagation of BPS particles which are massive on the Coulomb
branch.  Their masses scale as \bbd{\sqrt{\JJM}} because the theory is conformal, so their masses come purely from the expectation value of the vector multiplet
scalar \bbd{a} which is in turn set by the R-charge density in the classical solution describing the charged ground state.  Concrete formulae for those relationships
were given in \cite{Hellerman:2021yqz} and will be reviewed in sec. \ref{WorldlineInstantonActionsInTheTwoLimits}.

These massive macroscopic propagation (\textsc{mmp}) corrections are interesting, and we would like to study them in order to probe the physics associated with these
tiny but distinctly visible corrections, and also to increase the numerical precision of the large-R-charge expansion.  In the case of non-Lagrangian theories such as Argyres-Douglas theories \cite{Argyres:1995jj, Argyres:1995xn, Argyres:2016xua, Argyres:2015ffa, Argyres:2015gha, Argyres:2016xmc, Xie:2012hs}
we do not yet have the tools to compute these corrections.  The two interacting Lagrangian superconformal theories in \bbd{D=4} with \bbd{{\cal N} = 2} supersymmetry, are both \bbd{G=SU(2)} gauge theories:
\bbd{{\cal N} = 4} super-Yang-mills, and superconformal QCD with \bbd{4} hypermultiplets in the fundamental representation.  In the case of \bbd{{\cal N} = 4} super-Yang-Mills,
the MMP corrections vanish.  In this paper we will give asymptotic formulae for the leading large-\bbd{n} asymptotics of the massive macroscopic propagation term, in the case of \bbd{{\cal N} = 2}
superconformal QCD (\textsc{sqcd}) with \bbd{G=SU(2)} and \bbd{N\lrm F = 2N\lrm c = 4}.

The plan of the paper is as follows.  In sec. \ref{ReviewSection} we will review relevant material on chiral primary correlation functions for \bbd{{\cal N} = 2} superconformal theories in \bbd{D=4} with one-dimensional Coulomb branch  In sec. \ref{AsymptoticsOfQMMPAtFixedTau} we will apply the recursion relations of \cite{Baggio:2014ioa, Baggio:2014sna, Baggio:2015vxa, Gerchkovitz:2016gxx} to the formula for the exponentially small correction \bbd{q\ll n\uprm{\textsc{mmp}}}, which will give an order by order differential equation for the 
coupling-dependence of each order of the large-\bbd{n} expansion of \bbd{q\ll n\uprm{\textsc{mmp}}}.  We will find it is algorithmic to solve this equation in closed form up to a single undetermined term \bbd{c\ll p ({\rm Im}[\s] / n)\uu{{p\over 2}}} in the logarithm of \bbd{q\ll n\uprm{\textsc{mmp}}} at each order \bbd{p}.  By taking
the double-scaling limit, we will see that we can determine the coefficient \bbd{c\ll p} precisely by matching it with a term in the large-\bbd{\ddsc} expansion of the logarithm of the function \bbd{F\urm{inst}[\ddsc]} appearing
in \cite{Grassi:2019txd}.  We carry out this algorithm to give the fixed-coupling, large-\bbd{n} corrections to the exponentially small correction up to \bbd{p=5,} the term of order \bbd{n\uu{-{5\over 2}}} in the exponent.  In sec \ref{FixedCouplingAccuracy} we will test our subleading large-\bbd{n} corrections by comparison with high-precision
numerical data\footnote{None of the data was generated by ourselves; we thank Domenico Orlando for sharing the results of an exact computation of the correlators \cite{LocData} by the algorithm of \cite{Gerchkovitz:2016gxx}, with the
matrix elements of the large matrix of derivatives numerically to a precision of 200 digits.} representing the output of the algorithm of \cite{Gerchkovitz:2016gxx} at one particular value of the
complexified gauge coupling (\bbd{\t = {{25}\over\pi}\cc i}) and large values of \bbd{n}.  At this value of \bbd{\t} and
large \bbd{n,} we find highly precise agreement between the exact results from localization on the one hand, and the large-R-charge expansion of the exponentially small correction on the other hand.  Even at 
small values of the R-charge \bbd{n=1,2,3\cdots,} we find startlingly precise agreement between the exact result and the large-R-charge asymptotic estimate.  

In sec. \ref{AsymptoticsOfQMMPDoubleScaledLargeCharge} we will apply the recursion relations directly to find the subleading large-\bbd{n} corrections in the double-scaling limit, and solve for the subleading large-\bbd{n} corrections
to the double-scaling limit up to and including the six-loop contribution to the MMP term, which is order \bbd{n\uu{-5}} at fixed \bbd{\ddsc}.  We will show that the large-\bbd{\ddsc} limit of each order in the expansion has an increasing power-law growth at large \bbd{\ddsc} on top of the exponential suppresson, which explains why there is a distinction between the large-\bbd{\ddsc} expansion of the double-scaling limit on the one hand,
and the large-\bbd{n} expansion at fixed \bbd{\t} on the other hand.  We will show that the large-\bbd{\ddsc} growth of the subleading large-\bbd{n} corrections to the double scaling limit, 
are all directly connected with the \bbd{{\cal N} = 2} one-loop threshold correction to the effective Abelian gauge coupling \cite{Dorey:1996bn, Grimm:2007tm, Alday:2009aq} that gives the
entire perturbative difference between the UV coupling \bbd{\t} and the effective IR coupling \bbd{\s \simeq 2\t + {{4i}\over{\pi}}\cc {\tt log}[2]}.
Then in sec \ref{DoubleScaledEstimatesVsExact} we go on to compare the data with our subleading large-\bbd{n}
corrections to the double-scaled amplitude.  Finally we comment on some outstanding issues in
the conclusions (sec. \ref{ConclusionsSection}), including the possible large-order behavior of the asymptotic series in \bbd{1\over{\sqrt{n}}} and its physical interpretation \cite{Dondi:2021buw}

\section{Review of chiral primary correlation functions for \bbd{{\cal N} = 2} superconformal theories in \bbd{D=4} with one-dimensional Coulomb branch}\label{ReviewSection}

\subsection{Large R-charge expansion at fixed coupling}
In the special case of Lagrangian superconformal \bbd{SU(2)} gauge theories with marginal coupling constant, the chiral ring generator has dimension and R-charge \bbd{\D = J = 2} and its Lagrangian representation is \bbd{\co\propto {\rm Tr}[\hat{\phi}\sqd]} where \bbd{\hat{\phi}} is the lowest component of the adjoint-valued \bbd{SU(2)} gauge vector multiplet superfield.       In Lagrangian and non-Lagrangian theories alike, if the gauge symmetry is rank one, the Coulomb branch has complex dimension \bbd{1} and its chiral ring is spanned by the operators \bbd{\co\ll\D\uu n} for
\bbd{n =0,1,2,\cdots}.  The correlation functions
\bbb
G\ll{2n} \equiv |x-y|\uu{+2\D}\cc \langle \co\uu n\ll\D(x) \cc \overline{\co}\uu n\ll\D(y) \rangle
\eee
are special observables of the theory that are particularly robust, because they are unaffected by any possible \bbd{D-}term deformations of the theory; they are related to special supersymmetrically
protected properties of the theory, and can be calculated \cite{Baggio:2014ioa, Baggio:2014sna, Baggio:2015vxa, Gerchkovitz:2016gxx} {\it via} supersymmetric localization \cite{Nekrasov:2002qd, Pestun:2007rz, Alday:2009aq}

At large \bbd{n} the R-charge \bbd{\JJM\equiv n\D} carried by these operators, is large, and one can use the large-charge \textsc{eft} to calculate the asymptotic expansion of the correlation function \bbd{G\ll{2n}} at large 
\bbd{n}.  For Lagrangian and non-Lagrangian theories alike, one finds
\bbb
G\ll{2n} = 2\uu{{+2n\D}}\cc [Z\ll{S\uu 4}]\uu{-1}\cc \exp{q\ll n}\ , 
\xxn{DefOfG}
q\ll n \equiv q\ll n\uprm{\textsc{eft}}  + ({\rm exponentially~small~at~large~}n)\ ,
\xxn{ExpansionOfLittleQ}
q\ll n\uprm{\textsc{eft}} \equiv A n + B + {\rm log}\big [\Gamma[n\D + \a + 1]\big ] 
\een{LittleQEFTFormula}
where \bbd{Z\ll{S\uu 4}} is the four-sphere partition function,
\bbd{A} and \bbd{B} may depend on the theory and on the marginal
parameters if any, and \bbd{\a} is a coefficient proportional to the Weyl \bbd{a-}anomaly mismatch between
the interacting CFT and the effective theory of the Coulomb branch.  The \bbd{\a-}coefficient can be expressed convention-independently in terms of the \bbd{a-}anomaly coefficient \bbd{a\lrm{favm}} of a free \bbd{U(1)} vector multiplet:
\bbb
\a = {5\over{12}}\cc  {{a\lrm{CFT} - a\lrm{\textsc{eft}}}\over{a\lrm{favm}}}\ .
\eee
A table of \bbd{\a} coefficients of all known rank-one \bbd{{\cal N} = 2} superconformal theories was given in \cite{Hellerman:2017sur}.\footnote{The table was copied wholesale from \cite{Argyres:2016xmc} with the sole addition of a column entry for the \bbd{\a-}coefficients.}  For \bbd{{\cal N} = 4} super-Yang-Mills theory the value of \bbd{\a} is \bbd{+1}; for the case we will principally study in this paper, \bbd{{\cal N} = 2} superconformal QCD with 
\bbd{N\lrm F = 4,} the value of \bbd{\a} is \bbd{+{3\over 2}}. 

\subsection{Scheme-dependence and marginal couplings}

The \bbd{A} and \bbd{B} coefficients also depend on the normalization of the chiral ring generator and on the marginal coupling if any, and also, relatedly, on scheme choices.  When there is a marginal coupling \bbd{\t,}
the chiral ring generator lies in the same supermultiplet as the marginal operator, whose normalization and phase transform as holomorphic cotangent vectors under reparametrizations of the holomorphic coupling \bbd{\t}; 
so we define \bbd{\co \equiv \co\ls\t} with a (sometimes implicit) lower index \bbd{\t} to indicate its transformation under reparametrizations of the holomorphic coupling constant,
\bbb
\co\ls\t = {{d\t\pr}\over{d\t}}\cc \co\ls{\t\pr}\ .
\een{HolomorphicReparametrizationTransformationOfOperator}
It follows from the definition of \bbd{A} that \bbd{A} also transforms under holomorphic reparametrizations, as the logarithm of the norm-squared of a holomorphic cotangent vector:
\bbb
\exp{A\ls\t} = |{{d\t\pr}\over{d\t}}|\sqd \cc \exp{A\ls{\t\pr}}\ , \llsk\llsk A\ls\t = A\ls{\t\pr} +  {\rm log}|{{d\t\pr}\over{d\t}}|\sqd
\een{TransformationLawACoeff}
The coefficient \bbd{B} is also scheme-dependent under the choice of Euler density counterterm \cite{Gerchkovitz:2014gta, Gomis:2014woa}, which can depend on the marginal coupling as a holomorphic plus antiholomorphic function of the complex coupling \bbd{\t}:
\bbb
Z\to \exp{f(\t) + \bar{f}(\tb)}\ , \llsk\llsk
B\to B + f(\t) + \bar{f}(\tb)\ .
\een{KahlerTrans}
This follows automatically from the definitions \rr{DefOfG}, \rr{ExpansionOfLittleQ}, the fact that \bbd{Z\ll{S\uu 4}} has the same scheme-dependence, and from the fact that the correlators \bbd{G\ll{2n}} are independent of this scheme choice.  The combination \bbd{\exp{\tilde{B}} = \exp{B} / Z\ll{S\uu 4}} is scheme-independent.

For theories with marginal couplings, it is useful to know the functional form of the \bbd{A} and \bbd{B} coefficients as functions of the couplings.  

In the case of \bbd{{\cal N} = 4} super-Yang-Mills with \bbd{G=SU(2)} the \bbd{A} and \bbd{B} coefficients are essentially trivial; their values \cite{Pestun:2007rz, Gerchkovitz:2016gxx} are
\bbb
G\ll{2n}\upp{{\cal N} = 4} = {{\G[2n+2]}\over{({\rm Im}[\t])\uu{2n}}}
\eee
In superconformal \bbd{{\cal N} = 2} super-QCD with \bbd{G = SU(2)} and \bbd{N\ll F = 4} hypermultiplets in the fundamental representation, the dependence is considerably more complex. 

\subsection{The case of \bbd{{\cal N}=2} superconformal QCD with \bbd{G=SU(2)} and \bbd{N\lrm F = 4}}

In \cite{Hellerman:2020sqj, Hellerman:2021yqz} the authors studied the \bbd{A} and \bbd{B} coefficients in the case of super-QCD.
Refs. \cite{Hellerman:2020sqj, Hellerman:2021yqz} solved for the full functional form of the \bbd{A} and \bbd{B} coefficients as functions of the gauge coupling,
taking the scheme-dependences into account and using the S-duality invariance \cite{Seiberg:1994aj, Gaiotto:2009we, Alday:2009aq} of the theory.  The result for the \bbd{A-}coefficient as a function of the coupling is
\bbb
\exp{A\ls\t} = {{ |{{d\s}\over{d\t}}|\sqd}\over{16\cc {\rm Im}[\s]\sqd}}\ , 
\eee
where \bbd{\s} is the S-duality-covariant "infrared coupling" that is related to the Lagrangian "ultraviolet coupling" \cite{Gaiotto:2009we, Dorey:1996bn, Grimm:2007tm, Alday:2009aq} by a map
\bbb
\l[\s] = e\uu{2\pi i \t}\ ,
\een{IRToUVMap}
involving the modular lambda function \bbd{\l[\s]}.  At weak coupling
the map takes the form
\bbb
\bbsk
\t = {1\over{2\pi i }}\cc {\rm Log}[\l[\s]] = {\s\over 2} - {{2 \cc {\rm Log}[2]}\over{\pi}}\cc i + O(e\uu{\pi i \s})\ , \llsk\llsk
\s = 2\t + {{4 \cc {\rm log}[2]}\over\pi}\cc i + O(e\uu{2\pi i \t})\ .
\llsk
\eee
The solution for the \bbd{B-}coefficient is expressed scheme-independently in terms of \bbd{\tilde{B}}:
\bbb
\exp{\tilde{B}} = \exp{B\lrm {{ {Pestun-}\atop{Nekrasov}}}} / Z\lrm {{ {Pestun-}\atop{Nekrasov}}}\ ,
\xxnn
\exp{B\lrm {{ {Pestun-}\atop{Nekrasov}}}} \equiv
 \g\lrm G\uu{+12}\cc e\uu{-1}\cc 2\uu{-{9\over 2}}\pi\uu{-{3\over 2}}\cc
{{ |\l[\s]|\uu{+{2\over 3}}\cc |1-\l[\s]|\uu{+ {8\over 3}}}\over{|\eta(\s)|\uu 8\cc [{\rm Im}(\s)]\sqd}}\ ,
\een{ExponentiatedBCoefficientInPestunNekrasovScheme}
where \bbd{\l[\s]} is the modular Lambda function, \bbd{\eta(\s)} is the Dedekind eta function, \bbd{\g\lrm G} is the Glaisher constant \bbd{\g\lrm G = 1.282427129} and \bbd{Z\lrm {{ {Pestun-}\atop{Nekrasov}}}} is
the \bbd{S\uu 4} partition function computed by supersymmetric localization with the one-loop determinant of the localization integrand as given in \cite{Pestun:2007rz}  and the instanton factor of the integrand
given by the Nekrasov \bbd{U(2)} (as opposed to \bbd{SU(2)} as in \cite{Alday:2009aq}) partition function \cite{Nekrasov:2002qd} without the \bbd{U(1)} factor being removed (as it is
in \cite{Alday:2009aq}).  The perturbative expansion, including the overall normalization,
is given explicitly in \cite{Gerchkovitz:2016gxx} and reviewed in \cite{Hellerman:2021yqz}.

Both the \bbd{A[\t]} and \bbd{B[\t]} functions were solved almost completely using the constraints of the \textsc{eft} itself, including S-duality invariance imposed as a symmetry.  Almost, because for
each function, one overall \bbd{\t-}independent coefficient was left undetermined by \textsc{eft} and duality considerations; in each case those coefficients were determined by matching wtih
double-scaled perturbation theory introduced in \cite{Bourget:2018obm}

\subsection{Large R-charge in the double-scaling limit}

Having completely solved for all power-law terms in the inverse R-charge, it is natural to turn one's attention to the exponentially small corrections in the expansion \rr{ExpansionOfLittleQ}.  Since
these terms are exponentially small, it is natural to infer they should lie outside the \textsc{eft} description altogether, and describe virtual propagation of \emm{massive} degrees of freedom
on distances of the infrared scale.  The numerical analysis of correlation functions compared to the universal \textsc{eft} result in \cite{Hellerman:2018xpi} supports this inference: Defining
the difference 
\bbb
q\ll n\uprm{\textsc{mmp}} \equiv q\ll n - q\ll n\uprm{eft}
\eee
ref \cite{Hellerman:2018xpi} computed \bbd{q\ll n\uprm{\textsc{mmp}}} numerically based on the algorithm of \cite{Gerchkovitz:2016gxx} and showed that
\bbb
q\ll n\uprm{\textsc{mmp}} \sim \exp{- ({\rm const.})\times \ddsc\uu\hh}
\eee
to good approximation, where 
\bbb
\ddsc \equiv {n\over{4\pi\cc {\rm Im}[\t]}}\ ,
\een{GKTNormalizationOfDoubleScaledLambda}
a quantity introduced\footnote{up to a power of \bbd{4\pi}, a change in convention introduced in \cite{Grassi:2019txd}.  The actual normalization of \bbd{\ddsc} introduced in \cite{Bourget:2018obm}
is \bbd{\ddsc\ll{\smovie} \equiv {{4\pi n}\over{{\rm Im}[\t]}}}.
The normalization \rr{GKTNormalizationOfDoubleScaledLambda} is one of \emm{two} novel normalizations introduced for the parameter \bbd{\ddsc} in \cite{Grassi:2019txd}.  In the first half of \cite{Grassi:2019txd}, the parameter \bbd{\ddsc} is defined as
\bbd{\ddsc\lrm{{{first~half}\atop{of~\smgkt}}} \equiv{n\over{ {\rm Im}[\t]}} = (4\pi)\uu{-1}\cc\ddsc\ll{\smovie}}.  In the second half of \cite{Grassi:2019txd}, the authors introduce \emm{another} normalization convention
\bbd{\ddsc\lrm{ {{second~half}\atop{of~\smgkt}}} =
{n\over{4\pi\cc {\rm Im}[\t]}} = (4\pi)\uu{-2}\cc \ddsc\ll{\smovie}.}  In this paper we will always use the convention \bbd{\ddsc\lrm{here} = \ddsc\lrm{ {{second~half}\atop{of~\smgkt}}} =
{n\over{4\pi\cc {\rm Im}[\t]}} = (4\pi)\uu{-2}\cc \ddsc\ll{\smovie}.} } in \cite{Bourget:2018obm}).
The observation of the exponential behavior was only numerical, but was also theoretically motivated by the idea that the size of the leading breakdown of the \textsc{eft} should is proportional to \bbd{\exp{- S\lrm{WLI}}} where the
exponent \bbd{S\lrm{WLI}} is the scale of the geometry times the mass of the lowest massive excitation in the background Coulomb branch vev created by the large-charge operator insertions;
with the identification of the lightest massive excitation as an electrically charged BPS state, the worldline instanton action \bbd{S\lrm{WLI}} scales as \bbd{\ddsc\uu\hh} at large \bbd{n}.

With the identification of the lightest state as a fundamental hypermultiplet, and the trajectory of the particle as an equator of the \bbd{S\uu 3} spatial slice in the conformal frame of the cylinder in which
the R-charge density is constant, the worldline instanton action is 
\bbb
S\lrm{WLI} = \sqrt{{8\pi n}\over{{\rm Im}[\s]}}\ ,
\eee
where again \bbd{\s} is the IR coupling whose relationship to \bbd{\t} is given above in \rr{IRToUVMap}.

In the limit \bbd{n\to\infty} with \bbd{\ddsc \equiv {n\over{4\pi\cc {\rm Im}[\t]}}} held fixed, the worldline instanton action becomes
\bbb
S\lrm{WLI} = 4\pi \ddsc\uu\hh\ ,
\een{DoubleScaledWLIActionValue}
agreeing with the qualitative theoretical prediction of \cite{Hellerman:2018xpi} and numerical evidence supporting it, as well as giving a specific theoretical prediction 
for the exponent.

In two beautiful and important papers \cite{Bourget:2018obm, Grassi:2019txd} the limit \bbd{n\to\infty} and \bbd{\ddsc} held fixed, was studied more generally.  In this limit
the perurbation theory can be reorganized into a new double-scaled perturbation theory with 
\bbd{\ddsc \equiv {n\over{4\pi\cc {\rm Im}[\t]}}} as an adjustable classical parameter and \bbd{n} as the quantum loop-counting parameter.  This reorganized perturbation theory treats large and small \bbd{\ddsc}
on equal footing, with quantum effects uniformly suppressed when the total R-charge \bbd{\JJM } of the insertions is large.  The correlation functions \bbd{G\ll{2n}} were completely solved up to order \bbd{n\uu 0} 
in the logarithm \bbd{{\rm Log}[G\ll{2n}]} at fixed \bbd{\ddsc}.  Expressing the result in terms of our own function \bbd{q\ll n\uprm{\textsc{mmp}}}, ref. \cite{Grassi:2019txd} found
\bbb
\DDSL q\ll n\uprm{\textsc{mmp}} = F\urm{inst}\lp{\smgkt}[\ddsc] \ ,
  \een{DoubleScalingRelationMMPFunctionGKTWLIFunction}
 with \bbd{F\urm{inst}\lp{\smgkt}[\ddsc] } given as an explicit function in closed form:
\bbb
F\urm{inst}\lp{\smgkt}[\ddsc]  \equiv \sum\ll{w \in \{{ \rm positive~odd~integers}\}}\cc I\ll w[\ddsc]\ ,
\xxnn
I\ll w[\ddsc] \equiv {8\over{\pi\sqd w\sqd}}\cc \bigg [\cc
K\ll 0[4\pi w\sqrt{\ddsc}] + (4\pi w\sqrt{\ddsc})\cc K\ll 1[4\pi w\sqrt{\ddsc}]
\cc \bigg ]\ ,
\eee
where \bbd{K\ll 0} and \bbd{K\ll 1} are modified Bessel functions of the second kind.

At large \bbd{\ddsc}, the function \bbd{F\urm{inst}\lp{\smgkt}[\ddsc] } behaves as
\bbb
F\urm{inst}\lp{\smgkt}[\ddsc]  \sim e\uu{-4\pi \ddsc\uu\hh}\ ,
\eee
confirming the expected physical behavior of the MMP function, including the exponent of the leading exponential.

\subsection{Large-charge, fixed \bbd{\t} limit, versus the large-\bbd{\ddsc}
limit of the double-scaling limit}\label{WorldlineInstantonActionsInTheTwoLimits}

Ultimately we would like to solve for the fixed-\bbd{\t} large-\bbd{n} limit, to gain insight into theories that have no adjustable perturbative parameter, and also to explore issues such as \bbd{S-}duality
at large charge that cannot be understood directly in the double-scaling limit.  The double-scaling picture is clearly physically relevant, in the sense that its behavior at strong coupling is physically
similar to the large-\bbd{n} fixed coupling behavior.  However the large-\bbd{n}, fixed-\bbd{\t} limit is not quite the same as the large-\bbd{\ddsc} limit of the double-scaling limit. 

To illustrate the distinction, consider the macroscopic massive
propagation function \bbd{q\ll n\uprm{\textsc{mmp}} \equiv q\ll n
- q\ll n\uprm{eft}}, in each of these two limits.  In both
limits, the MMP function is exponentially small, and the exponent
is given by the action of a "worldline instanton" describing a
BPS particle circumnavigating the equator of an \bbd{S\uu 3}
spatial slice of the cylinder, in the conformal frame in which
the vector multiplet scalar has constant magnitude.

The worldline instanton action was worked out, in both
limits, in sec. \alsec{6} of \cite{Hellerman:2021yqz}. The mass of the hyper is
\bbd{M\lrm{hyper} =  |a|,} where the vector multiplet scalar \bbd{a} is normalized in the conventions of
\cite{Pestun:2007rz} and the mass is defined in the cylinder frame \bbd{S\uu 3 \times {\rm time}} where the cylinder has radius \bbd{R}.
The numerical value of \bbd{|a|} in the large-R-charge classical solution depends on
the kinetic term and differs between the
double-scaling limit and the fixed-coupling large-charge limit.  In the
double-scaling limit, the inverse coupling \bbd{{\rm \t}} is going to infinity and
the relationship is given by the tree-level kinetic term for
the vacuum modulus in the nonabelian theory,
\bbb
\hat{\Phi} = i a \s\uu 3\ , \llsk\llsk |a| =  \sqrt{- \hh\cc {\rm Tr}\lrm F(\hat{\Phi}\dag\hat{\Phi})} \ ,
\xxnn
M\lrm{hyper}\uprm{double-scaled} = |a| \lrm{tree~level} 
=   {1\over R}\cc \sqrt{{\JJM}\over {2\pi\cc \cc {\tt Im}[\t] }} =  {1\over R}\cc \sqrt{{n}\over {\pi\cc \cc {\tt Im}[\t] }} = {{2\ddsc\uu\hh}\over R}\ ,
\eee
in the normalization convention of \cite{Pestun:2007rz}.
In the fixed-coupling large-R-charge limit, the nonabelian degrees of freedom are becoming
infinitely heavy and the formula for \bbd{|a|} is given in terms of the infrared effective coupling \bbd{\s} which controls the kinetic term for \bbd{a} in the effective Abelian gauge theory:
\bbb
 M\lrm{hyper} =  |a| = {1\over R}\cc \sqrt{{\JJM}\over {\pi\cc \cc {\tt Im}[\s] }} =  {1\over R}\cc \sqrt{{2n}\over {\pi\cc \cc {\tt Im}[\s] }}\ .
\eee

The worldline instanton action \bbd{S\lrm{WLI} = 2\pi R M\lrm{hyper}} corresponding to a particle with the tree-level hypermultiplet mass, is \bbd{S\lrm{WLI}\uprm{double-scaled} = 4\pi\ddsc\uu\hh} while the fixed-coupling
large-\bbd{n} expression retains the full, exact loop- and instanton-corrected BPS mass, and the formula in that limit is 
\bbb
S\lrm{WLI} = \sqrt{{8\pi\cc n}\over{{\rm Im}[\s]}}
\een{ExactBPSWLI}  If we take the double-scaling limit first and then take \bbd{\ddsc} to infinity, the corrections to the worldline-instanton action, including the gauge-instanton corrections and their \bbd{\th-}angle dependence, is lost.  This illustrates physically why the large-\bbd{\ddsc} limit of the double-scaling limit is not the same thing as the fixed-coupling large-charge limit.

At the same time, the double-scaling limit \cite{Bourget:2018obm, Grassi:2019txd} has already proven useful for practical calculations at fixed coupling and large charge,  by fixing numerical coefficients in the \textsc{eft} results not determined by \bbd{S-}duality \cite{Hellerman:2021yqz}.  We would like to use the same kind of strategy to study the exponentially small correction
\bbd{q\ll n\uprm{\textsc{mmp}}} in the fixed-coupling, large-\bbd{n} limit while making use of the results of \cite{Grassi:2019txd} to fix undetermined coefficients.

 \section{Large-R-charge asymptotic expansion for the massive macroscopic propagation function \bbd{q\ll n\uprm{\textsc{mmp}}} at fixed \bbd{\t}}\label{AsymptoticsOfQMMPAtFixedTau}

In this section we will use supersymmetric
recursion relations and boundary conditions set by
\textsc{eft} behavior and the double-scaling limit, to find an asymptotic
expansion for \bbd{q\ll n\uprm{\textsc{mmp}}[\t]} to the first several orders
in \bbd{1\over{\sqrt{n}}} at fixed \bbd{\t} correcting the leading exponentially
small term given by \bbd{e\uu{-S\lrm{WLI}}}.

 \subsection{Recursion relations for subleading large-charge corrections at fixed \bbd{\t}}

Now we want to derive recursion relations for the loop corrections to the macroscopic massive propagation amplitude \bbd{q\ll n\uprm{\textsc{mmp}}} at fixed gauge coupling \bbd{\t}  We start with the ansatz
\bbb
Z\ll n = \exp{q\ll n}\ ,
\xxnn
q\ll n = q\ll n\uprm{eft} + q\ll n\uprm{\textsc{mmp}}\ .
\eee
The recursion relations give a second-order "equation of variation" (\textsc{eov}) for the coupling-dependence
the \textsc{mmp} term.
The \textsc{eov} of \bbd{q\ll n\uprm{\textsc{mmp}}} is
\bbb
e\uu{-A}\cc \pp\bar{\pp} q\ll n\uprm{\textsc{mmp}}= (2n+ {7\over 2})(2n + {5\over 2}) \cc \biggl [ \cc
 {{Z\ll{n+1}\uprm{\textsc{mmp}} }\over{   Z\ll n\uprm{\textsc{mmp}}}}  - 1
 \cc \biggl ]
-    (2n+ {3\over 2})(2n + {1\over 2}) \cc  \biggl [ \cc
 {{Z\ll n\uprm{\textsc{mmp}} }\over{   Z\ll {n-1}\uprm{\textsc{mmp}}}}  - 1
 \cc \biggl ]\ ,\llsk
\een{BasicMMPRecursionRelation}
where we have defined \bbd{Z\ll n\uprm{\textsc{mmp}} \equiv e\uu{q\ll n\uprm{\textsc{mmp}}}}.

Given the transformation law \rr{TransformationLawACoeff}, eq.  \rr{BasicMMPRecursionRelation} is covariant under holomorphic reparametrizations of the coupling.  It will be more convenient to analyze it in terms
of the infrared coupling \bbd{\s}. The expression for \bbd{\exp{-A}} is much simpler in terms of \bbd{\s} than in terms of \bbd{\t}, as is the
leading-order behavior of \bbd{q\ll n\uprm{\textsc{mmp}}}.  We have
\bbb
\exp{-A\ls\s} = 16\cc {\rm Im}[\s]\sqd\ ,
\eee
so the LHS of \rr{BasicMMPRecursionRelation} is
\bbb
e\uu{-A}\cc \pp\bar{\pp} q\ll n\uprm{\textsc{mmp}} = e\uu{-A\ls\s}\cc \pp\ll\s\bar{\pp}\ll{\bar{\s}} q\ll n\uprm{\textsc{mmp}}=16\cc {\rm Im}[\s]\sqd\cc \pp\ll\s\bar{\pp}\ll{\bar{\s}} q\ll n\uprm{\textsc{mmp}} 
\eee
So eq. \rr{BasicMMPRecursionRelation} becomes 
\bbb
16\cc {\rm Im}[\s]\sqd\cc \pp\ll\s\bar{\pp}\ll{\bar{\s}} q\ll n\uprm{\textsc{mmp}}  = (2n+ {7\over 2})(2n + {5\over 2}) \cc \biggl [ \cc
 {{Z\ll{n+1}\uprm{\textsc{mmp}} }\over{   Z\ll n\uprm{\textsc{mmp}}}}  - 1
 \cc \biggl ]
-    (2n+ {3\over 2})(2n + {1\over 2}) \cc  \biggl [ \cc
 {{Z\ll n\uprm{\textsc{mmp}} }\over{   Z\ll {n-1}\uprm{\textsc{mmp}}}}  - 1
 \cc \biggl ]\llsk
\een{BasicMMPRecursionRelationInTermsOfSigma}

Now expand the MMP function \bbd{q\ll n\uprm{\textsc{mmp}}} as an asymptotic series at large \bbd{n} and fixed coupling.  On physical grounds, by virtue of the BPS formula we know the exponential representing the large-\bbd{n} behavior exactly 
but not the prefactor.  Because of this, it is easiest for us to expand the logarithm
of \bbd{q\ll n\uprm{\textsc{mmp}}} in powers of \bbd{n} rather than \bbd{q\ll n\uprm{\textsc{mmp}}} itself, with the BPS formula as the leading term of order \bbd{n\uu{+\hh}} and the subleading term representing the
fluctuation determinant.  We have
\bbb
q\ll n\uprm{\textsc{mmp}} = e\uu{- \mathfrak{W}}
\een{DefOfScriptW}
with
\bbb
\mathfrak{W} = m\cc\sqrt{n\over{{\rm Im}[\s]}} + \g[\s]\cc {\rm log}[n] + \sum\ll{p \geq 0}\cc w\ll p[\s]\cc n\uu{-{p\over 2}}\ .
\een{ParametrizationOfScriptW}
 We know from the BPS formula that \bbd{m = \sqrt{8\pi}} but we leave it as a constant for now, so that we can understand clearly the \bbd{m-}dependence of various terms in the expansion.

Next we expand the recursion relation \rr{BasicMMPRecursionRelationInTermsOfSigma} at large \bbd{n} and fixed \bbd{\s}, which gives PDEs for the coefficient functions \bbd{w\ll p[\s]} in terms of the lower \bbd{w\ll {p\pr < p}[\s]}.
Then write the recursion relation \rr{BasicMMPRecursionRelationInTermsOfSigma} and expand each side at large \bbd{n},
\bbb
[{\rm LHS}] = [{\rm RHS}]
\eee
with
\bbb
[{\rm LHS}] \equiv 16\cc {\rm Im}[\s]\sqd\cc \pp\ll\s\bar{\pp}\ll{\bar{\s}} q\ll n\uprm{\textsc{mmp}}  =  \sum\ll{k \geq 0}\cc [{\rm LHS}]\ll{1 - {k\over 2}}\cc n\uu{1 - {k\over 2}}\ ,
\xxnn
\bbsk
[{\rm RHS}] \equiv  (2n+ {7\over 2})(2n + {5\over 2}) \cc \biggl [ \cc
 {{Z\ll{n+1}\uprm{\textsc{mmp}} }\over{   Z\ll n\uprm{\textsc{mmp}}}}  - 1
 \cc \biggl ]
-    (2n+ {3\over 2})(2n + {1\over 2}) \cc  \biggl [ \cc
 {{Z\ll n\uprm{\textsc{mmp}} }\over{   Z\ll {n-1}\uprm{\textsc{mmp}}}}  - 1
 \cc \biggl ] =  \sum\ll{k \geq 0}\cc [{\rm RHS}]\ll{1 - {k\over 2}}\cc n\uu{1 - {k\over 2}}\ ,
 \llsk
\eee
and set each order in \bbd{n} on the LHS equal to the corresponding order on the RHS.

At order \bbd{n\uu{+1}} we have
\bbb
 [{\rm LHS}]\ll 1 =   [{\rm RHS}]\ll 1 =  {{m\sqd}\over {{\rm Im}[\s]}}
\eee
So the \textsc{eov} at order \bbd{n\uu{+1}} is identically satisfied for our choice of the leading term in the exponent \bbd{\mathfrak{W}}, for
any value of \bbd{m}.
At order \bbd{n\uu{+\hh}} we have
\bbb
 [{\rm LHS}]\ll \hh = 8 i m\cc \sqrt{{\rm Im}[\s]}\cc   {\rm Log}[n]  \cc (\pp\ll\s - \pb\ll\sb)\g - 4im\cc \sqrt{{\rm Im}[\s]}\cc \bigg (\cc
  (\pp\ll\s - \pb\ll\sb)\cc w\ll 0 - {{3i}\over {4\cc {\rm Im}[\s]}}
 \cc \bigg )
 \xxnn
  [{\rm RHS}]\ll \hh = 
  - 4im\cc \sqrt{{\rm Im}[\s]}\cc \bigg (\cc
+ {{i \g}\over{{\rm Im}[\s]}} - {{3i}\over{4\cc {\rm Im}[\s]}}
 \cc \bigg )
\eee
The cancellation of the logarithmic term gives the PDE
\bbb
(\pp\ll\s - \pb\ll\sb)\g = 0
\een{LogTermEOM}
and the cancellation of the nonlogarithmic term gives the PDE
\bbb
(\pp\ll\s - \pb\ll\sb)\cc w\ll 0 = + {{i \g}\over{{\rm Im}[\s]}}
\een{NonLogTermEOM}
Parametrize \bbd{\s \equiv {{\th\lrm{IR}}\over{2\pi}}+ is} where \bbd{\th\lrm{IR}} is the infrared \bbd{\th-}angle.  Then using \bbd{\pp\ll\s - \pp\ll\sb = - i \pp\ll s} we have
\bbb
\pp\ll s \g = 0\ , \llsk\llsk \pp\ll s w\ll 0 = -  {\g\over s}\ .
\eee
The general solution to the first equation is \bbd{\g = \g[\th\lrm{IR}]}, independent of \bbd{s}.  In the weak-coupling limit
\bbd{s\to\infty} we know the coefficient functions must all be independent of \bbd{\th\lrm{IR} \equiv 2\pi \cc {\rm Re}[\s]} up
to exponentially small corrections due to gauge instantons.  But \bbd{\g} is independent of \bbd{s}, so its \bbd{\th\lrm{IR}} dependence would be
of order \bbd{s\uu 0} unless it is exactly independent of \bbd{\th\lrm{IR}} as
well.  So we have
\bbb
\g[\s] = \g = {\rm constant}.
\een{GammaGeneral}
Then the general solution to the nonlogarithmic \textsc{eov} \rr{NonLogTermEOM} is  \bbd{w\ll 0 = - \g\cc {\rm log}[s] + c\ll 0[\th\lrm{IR}]}\ .
Again, the constraint that \bbd{w\ll 0} must be independent of \bbd{{\rm Re}[\s]} up to terms exponentially small in \bbd{s,} forces \bbd{c\ll 0[\th\lrm{IR}]} to be a constant, so we have
\bbb
w\ll 0 = - \g\cc {\rm log}[s] + c\ll 0\ , \llsk\llsk c\ll 0 {\rm ~constant}.
\eee

At higher order, the \textsc{eov} simplify if we absorb a power of \bbd{s} into the definition of the coefficient functions.  Define
\bbb
\hat{w}\ll p[\s] \equiv s\uu{-{p\over 2}}\cc w\ll p[\s]\ , \llsk\llsk w\ll p[\s] = s\uu{+{p\over 2}}\cc \hat{w}\ll p[\s]\ .
\een{w0General}
In terms of these rescaled coefficient functions, the \textsc{eov} simplify somewhat.

Now plug the solutions \rr{GammaGeneral}, \rr{w0General} for \bbd{\g[\s]} and \bbd{w\ll 0[\s]} into the large \bbd{n} expansion of \bbd{\mathfrak{W}} and expand the recursion relation to order \bbd{n\uu 0}.  This gives:
\bbb
[{\rm LHS}]\ll 0 = -4 m s\cc \big [\cc \pp\ll s \hat{w}\ll 1 + {{\hat{w}\ll 1}\over{2s}} + {{\g - \g\sqd}\over{ms}} \cc \big ] 
\xxnn
 [{\rm RHS}]\ll 0 = -4 m s\cc \big [\cc + {{\hat{w}\ll 1}\over{2s}} + {{\g - \g\sqd}\over{ms}} - {m\over{2s\sqd}} - {{m\uu 3}\over{192\cc s\uu 3}} \cc \big ] 
\eee
so the \textsc{eov} at order \bbd{n\uu 0} is
\bbb
\pp\ll s \hat{w}\ll 1 = - {m\over{2s\sqd}} - {{m\uu 3}\over{192\cc s\uu 3}} \ , 
\eee
to which the general solution is \bbd{\hat{w}\ll 1 = +{m\over{2s}} + {{m\uu 3}\over{384\cc s\uu 3}} + c\ll 1[\th\lrm{IR}]}.  Again the constraint that \bbd{w\ll 1} be independent of \bbd{\th\lrm{IR}} up to exponentially
small corrections forces the unfixed function \bbd{c\ll 1[\th\lrm{IR}]} to be a constant, so
the general solution satisfying the asymptotic condition in the weak coupling region is
\bbb
\hat{w}\ll 1 = c\ll 1 + {m\over {2s}} + {{m\uu 3}\over{384\cc s\sqd}}
\xxnn
w\ll 1 = c\ll 1\cc s\uu{+\hh} + {m\over 2}\cc s\uu{-\hh} + {{m\uu 3}\over{384}}\cc s\uu{-{3\over 2}}\ ,
\eee
where \bbd{c\ll 1} is a constant.  Note that several terms cancelled between the LHS and RHS of the recursion relation at order \bbd{n\uu 0}; the difference between the LHS and RHS is a simpler expression
than either side individually.  So at order \bbd{n\uu{-\hh}} we will just give the difference between the two terms; 
the \textsc{eov} at order \bbd{n\uu{-\hh}} is
\bbb
0 = [{\rm LHS} - {\rm RHS}]\ll{-\hh }= - 4ms\uu{{3\over 2}} \cc \bigg ( \cc
\pp\ll s\cc \hat{w}\ll 2 + {\g\over{s\sqd}} + {{m\uu 2\cc\g }\over{32\cc s\uu 3}}\ ,
\cc \bigg )
\eee
to which the general solution is \bbd{\hat{w}\ll 2 = c\ll 2 [\th\lrm{IR}]+\frac{m^2 \gamma }{64 s^2}+\frac{\gamma }{s}}.  Again the condition that the coefficient functions must be independent of \bbd{\th\lrm{IR}} up to exponentially small corrections at large \bbd{s,}
forces \bbd{c\ll 2} to be a constant, so the general solution consistent with the asymptotic condition is
\bbb
\hat{w}\ll 2 = c\ll 2 +\frac{m^2 \gamma }{64 s^2}+\frac{\gamma }{s}
\eee
with \bbd{c\ll 2} independent of the coupling.

The same pattern continues at each higher order: At order \bbd{n\uu{{{1-p}\over 2}}}, the recursion relation gives first order linear inhomogeneous differential equation for \bbd{\hat{w}\ll p} of the form
\bbd{\pp\ll s \hat{w}\ll p = \cdots,} where the RHS of the PDE is some fixed function of \bbd{s} determined by the solutions for the \bbd{w\ll {p\pr}} with \bbd{p\pr < p}; the RHS is a finite series in powers of 
\bbd{s\uu{-1}} with leading term \bbd{s\uu{-2}}, and coefficients depending on \bbd{m,\g} and the coefficients \bbd{c\ll {p\pr}} with \bbd{p\pr < p}.  The general solution for \bbd{\hat{w}\ll p} is then given
by the antiderivative of the RHS plus an undetermined function \bbd{c\ll p[\th\lrm{IR}]}.  The condition that \bbd{\hat{w}\ll p} must be independent of \bbd{\th\lrm{IR}} up to exponentially small
corrections, then forces \bbd{c\ll p} to be a constant, independent of \bbd{\s} altogether.  The form of \bbd{\hat{w}\ll p} is then \bbd{c\ll p + ({\rm constant})\times s\uu{-1} + \cdots + ({\rm constant})\times s\uu{-\hat{k}\ll p}} where
\bbd{\hat{k}\ll p = p} for \bbd{p} even and \bbd{\hat{k}\ll p = p+1} for \bbd{p} odd.  Multiplying
by \bbd{s\uu{+{p\over 2}}} to obtain then unhatted \bbd{w\ll p[s],} we then obtain
\bbb
w\ll p[\s] = w\ll p[s] = c\ll p\cc s\uu{+{p\over 2}} + ({\rm constant})\times s\uu{{p\over 2} - 1} + \cdots +  ({\rm constant})\times s\uu{-k\ll p}
\eee
where  \bbd{k\ll p = \hat{k}\ll p - {p\over 2}} is given by \bbd{k\ll p= {p\over 2}} for \bbd{p} even and \bbd{k\ll p= 1 + {p\over 2}} for \bbd{p} odd.

Carrying out this algorithm up to \bbd{p=5} we find the first five hatted coefficient functions \bbd{\hat{w}\ll p} are:
  \bbb
\hat{w}\ll   1 =  \bigg ( c\ll 1 +\frac{m^3}{384 \cc s^2}+\frac{m}{2 s}\bigg )
\eee

 \bbb
\hat{w}\ll   2 =  \bigg (c\ll 2 +\frac{m^2 \gamma }{64 \cc s^2}+\frac{\gamma }{s}\bigg )
\eee

 \bbb
\hat{w}\ll   3 =  \bigg (c\ll 3 -\frac{m^5}{163840 \cc s^4}-\frac{m^3}{768 \cc s^3}-\frac{m}{8
\cc s^2}-\frac{c\ll 1  m^2}{128 \cc s^2}-\frac{c\ll 1 }{2 s}+\frac{m \gamma }{64 \cc s^2}+\frac{m \gamma ^2}{32 \cc s^2}\bigg )
\een{WHat3Solution}

 \bbb
\hat{w}\ll   4 =  \bigg (c\ll 4 -\frac{c\ll 1  m}{64 \cc s^2}-\frac{c\ll 2  m^2}{64
\cc s^2}-\frac{c\ll 2 }{s}-\frac{m^4 \gamma }{8192 \cc s^4}-\frac{m^2 \gamma }{64 \cc s^3}-\frac{47 \gamma }{96 \cc s^2}-\frac{c\ll 1  m \gamma }{32 \cc s^2}+\frac{\gamma
^2}{32 \cc s^2}+\frac{\gamma ^3}{48 \cc s^2}\bigg )
\eee

 \bbb
\hat{w}\ll   5 =  \bigg (c\ll 5 +\frac{m^7}{29,360,128  \cc s^6}+\frac{3 m^5}{32,7680
 \cc s^5}+\frac{25 m^3}{24,576  \cc s^4}+\frac{3 c\ll 1  m^4}{32,768  \cc s^4}+\frac{m}{16  \cc s^3}+\frac{3 c\ll 1  m^2}{256  \cc s^3}+\frac{23 c\ll 1 }{64  \cc s^2}+\frac{c\ll 1 ^2
m}{128  \cc s^2}
\xxnn
-\frac{3 c\ll 2  m}{64  \cc s^2}-\frac{3 c\ll 3  m^2}{128  \cc s^2}-\frac{3 c\ll 3 }{2 s}-\frac{3 m^3 \gamma }{8192  \cc s^4}-\frac{3 m \gamma
}{128  \cc s^3}-\frac{3 c\ll 1  \gamma }{64  \cc s^2}-\frac{c\ll 2  m \gamma }{16  \cc s^2}-\frac{3 m^3 \gamma ^2}{4096  \cc s^4}-\frac{3 m \gamma ^2}{64  \cc s^3}-\frac{c\ll 1 
\gamma ^2}{32  \cc s^2}\bigg )
\eee

and the unhatted \bbd{w\ll p} are given by \bbd{w\ll p = s\uu{+{p\over 2}}\cc \hat{w}\ll p}.

\subsection{Fixing the integration constants using the strong coupling expansion of the double-scaling limit}

Now we want to fix the integration constants \bbd{c\ll{1,2,3,4,5,\cdots}}.  We can do this by taking the double-scaling limit.  In the double-scaling limit we have \bbd{s\to\infty} and 
\bbb
{n\over {{\rm im}[s]}}\to {n\over{2\cc {\rm Im}[\t]}} + O(n\uu{-1}) = 2\pi\ddsc + O(n\uu{-1})
\eee
So then in the double-scaling limit we have
\bbb
-\DDSL {\rm log}[q\ll n] = \DDSL \mathfrak{W} 
\xxnn
= \DDSL\cc \biggl \{\cc m (n/s)\uu\hh  + \g\cc {\rm log}[n / s] + c\ll 0 + \sum\ll{p \geq 1}\cc \hat{w}\ll p[s]\cc (n/s)\uu{-{p\over 2}}\cc \biggl \}
\eee
In the double-scaling limit, we have \bbd{{n\over s}\to 2\pi\ddsc} and \bbd{\hat{w}\ll p[s] = c\ll p + O(s\uu{-\hh})} goes to \bbd{c\ll p}.  So
\bbb
-\DDSL {\rm log}[q\ll n] = \DDSL \mathfrak{W} 
\xxnn
=  m (2\pi\ddsc )\uu\hh  + \g\cc {\rm log}[2\pi\ddsc ] + c\ll 0 + \sum\ll{p \geq 1}\cc c\ll p\cc (2\pi\ddsc )\uu{-{p\over 2}} 
\eee

Also by comparing double-scaling limits, it was shown in \cite{Hellerman:2021yqz} that the double scaling limit of \bbd{q\ll n\uprm{\textsc{mmp}}} at any value of \bbd{\ddsc,} is exactly equal to 
the quantity \bbd{F\lp{\smgkt}\urm{inst}}, as defined in \cite{Grassi:2019txd} as the \bbd{n\uu 0} term in their \bbd{\Delta C\ll 1} with the linear, logarithmic, and constant terms of its large-\bbd{\ddsc} expansion
removed.  The authors computed this quantity exactly and its large-\bbd{\ddsc} expansion can be computed easily.  In \cite{Grassi:2019txd} the authors gave the formula
\bbb
F\lp{\smgkt}\urm{inst} = \sum\ll{ {{{\rm positive~odd}}\atop{{\rm integers~}w}}}\cc I\ll w[\ddsc]\ ,
\xxnn
I\ll w[\ddsc] \equiv {8\over{\pi\sqd \cc w\sqd}}\cc \bigg [ \cc K\ll 0(4\pi w \ddsc\uu\hh ) + (4\pi w\ddsc\uu\hh )\cc K\ll 1(4\pi w \ddsc\uu\hh) \cc \bigg ] .
\een{FormulaForF0}
At large \bbd{\ddsc} the term \bbd{I\ll w} goes as \bbd{e\uu{-4\pi w \ddsc\uu\hh}} so the \bbd{I\ll 3} and higher terms are exponentially smaller than the leading \bbd{I\ll 1} term.  So we have
\bbb
\bbsk
m (2\pi\ddsc )\uu\hh  + \g\cc {\rm log}[2\pi\ddsc ] + c\ll 0 + \sum\ll{p \geq 1}\cc c\ll p\cc (2\pi\ddsc )\uu{-{p\over 2}} = 
-\DDSL {\rm log}[q\ll n]  = - {\rm log}[F\lp{\smgkt}\urm{inst}[\ddsc]] \llsk
\xxnn
=  - {\rm log}[I\ll 1[\ddsc]] + ({\rm exponentially~small~in~}\ddsc\uu\hh)\ . 
\eee
So the integration constants \bbd{c\ll p} can simply be read off from the large-\bbd{\ddsc} asymptotic expansion of the negative of the logarithm of the function \bbd{I\ll 1[\ddsc]}.  We have
\bbb
\bbsk
I\ll 1[\ddsc] = 2\uu{{7\over 2}}\cc \pi\uu{-1}\cc e\uu{-4\pi\ddsc\uu\hh}\cc \ddsc\uu{+{1\over 4}} \times \biggl \{ \cc 1
+\frac{11}{2\uu 5 \pi  \sqrt{\ddsc }}-\frac{31}{2\uu{11} \pi ^2 \ddsc } +\frac{177}{2\uu{16} \pi ^3 \ddsc ^{3/2}} -\frac{7125}{2\uu{23} \pi ^4 \ddsc ^2} 
+ {{102,165 }\over{2\uu{28}\cc \pi\uu 5\cc \ddsc\uu{5\over 2} }}
+ O(\ddsc\uu{-3})
  \cc \biggl \} \llsk
\eee

So, taking the negative of the logarithm, we have
\bbb
\bbsk\bbsk
- {\rm log}[I\ll 1[\ddsc]] = 4\pi\cc \ddsc\uu{+\hh} - {1\over 4}\cc {\rm log}[\ddsc] - {\rm log}[2\uu{{7\over 2}}\cc \pi\uu{-1} ] - {\rm log}\biggl [\cc 
1
+\frac{11}{2\uu 5 \pi  \sqrt{\ddsc }}-\frac{31}{2\uu{11} \pi ^2 \ddsc } +\frac{177}{2\uu{16}\cc \pi ^3 \ddsc ^{3/2}} -\frac{7125}{2\uu{23} \pi ^4 \ddsc ^2}  + {{102,165 }\over{2\uu{28}\cc \pi\uu 5\cc \ddsc\uu{5\over 2} }}
+ O(\ddsc\uu{-3})
\cc   \cc \biggl ]\ .
\llsk
\eenn
The strong-coupling expansion of (the negative of) the log of \bbd{I\ll 1[\ddsc]} is
\bbb
- {\rm log}[I\ll 1[\ddsc]] = 4\pi\cc \ddsc\uu{+\hh} - {1\over 4}\cc {\rm log}[\ddsc] - {7\over 2}\cc {\rm log}[2] + {\rm log}[\pi]  + \sum\ll{p \geq 1}\cc b\ll p\cc \ddsc\uu{-{p\over 2}}\ ,
\eee
with

\bbb
  b\ll 1\equiv -\frac{11}{2\uu 5 \cc\pi }  
\xxnn
  b\ll 2\equiv \frac{19}{2\uu 8 \cc\pi\sqd}  
\xxnn
  b\ll 3\equiv -\frac{527}{2\uu{13}\times 3 \cc\pi\uu 3}  
\xxnn
  b\ll 4\equiv \frac{235}{2\uu{15} \cc\pi\uu 4}  
 \xxnn
  b\ll 5\equiv -\frac{14,083}{2\uu{20}\times 5 \cc\pi\uu 5}  
 \eee

Matching coefficients of \bbd{\ddsc\uu{+\hh}} we get
\bbb
m = \sqrt{8\pi}\ ,
\eee
as anticipated in \cite{Hellerman:2021yqz} from the \textsc{eft} derivation of the worldline instanton action via the BPS formula.  Matching coefficients of \bbd{{\rm log}[\ddsc]} we get
\bbb
\g = -{1\over 4}\ .
\eee
Then the equality between constant coefficients gives
\bbb
\bbsk\bbsk
c\ll 0 = - {\rm log}[2\uu{{7\over 2}}\cc \pi\uu{-1}] - \g\cc {\rm log}[2\pi] = - {7\over 2}\cc {\rm Log}[2] + {\rm Log}[\pi]  +{1\over 4} \cc {\rm log}[2\pi] = -{{13}\over 4}\cc {\rm Log}[2] + {5\over 4}\cc {\rm Log}[\pi]
\llsk
\eee 
Matching the coefficient of \bbd{\ddsc\uu{-{p\over 2}}} gives
\bbb
c\ll p = (2\pi)\uu{+{p\over 2}}\cc b\ll p\ .
\eee
So then we have
\bbb
m = \sqrt{8\pi} \ ,
\xxnn
\g = - {1\over 4}\ ,
\xxnn
c\ll 0 = -{{13}\over 4}\cc {\rm Log}[2] + {5\over 4}\cc {\rm Log}[\pi]
\xxnn
c\ll 1 = (2\pi)\uu{+\hh}\cc b\ll 1 = -\frac{11}{2\uu {{9\over 2}} \cc\pi\uu\hh }  
\xxnn
c\ll 2 = (2\pi)\cc b\ll 2 =   \frac{19}{2\uu 7 \cc\pi}  
\xxnn
c\ll 3 =  (2\pi)\uu{+{3\over 2}} \cc b\ll 3 = -\frac{527}{2\uu{ {{23}\over 2} }\times 3 \cc\pi\uu {3\over 2}}  
\xxnn
c\ll 4 = (2\pi)\sqd \cc b\ll 4 = \frac{235}{2\uu{13} \cc\pi\uu 2}  
\xxnn
c\ll 5 =  (2\pi)\uu{+{5\over 2}} \cc b\ll 5 =  -\frac{14,083}{2\uu{ {{35}\over 2}}\times 5 \cc\pi\uu {5\over 2}}  
\een{CCoeffList}

So then the (unhatted) \bbd{w\ll p-}functions \bbd{w\ll p = s\uu{+{p\over 2}}\cc \hat{w}\ll p} for \bbd{p = 1,\cdots, 5} are:
\bbb

w\ll{1} =

\frac{1}{48 \sotp^{3/2}}+\frac{1}{\sqrt{\sotp}}-\frac{11 \sqrt{\sotp}}{16}

\xxn{W1Formula}

w\ll{2} =

-\frac{1}{4}-\frac{1}{64 \sotp}+\frac{19 \sotp}{64}

\xxn{W2Formula}

w\ll{3} =

-\frac{1}{5120 \sotp^{5/2}}-\frac{1}{96 \sotp^{3/2}}-\frac{119}{512 \sqrt{\sotp}}+\frac{11 \sqrt{\sotp}}{32}-\frac{527
\sotp^{3/2}}{3072}

\xxn{W3Formula}

w\ll{4} =

\frac{119}{1024}+\frac{1}{2048 \sotp^2}+\frac{1}{64 \sotp}-\frac{19 \sotp}{64}+\frac{235 \sotp^2}{2048}

\xxn{W4Formula}

\bbsk\bbsk
w\ll{5} = 
\frac{1}{229,376 \sotp^{7/2}}+\frac{3}{10,240 \sotp^{5/2}}+\frac{737}{98,304 \sotp^{3/2}}+\frac{101}{1024 \sqrt{\sotp}}-\frac{8,155
\sqrt{\sotp}}{32,768}+\frac{527 \sotp^{3/2}}{2048}-\frac{14,083 \sotp^{5/2}}{163,840}
\llsk

\een{W5Formula}

\section{Accuracy of the asymptotic estimates of the MMP function at large \bbd{n} and fixed \bbd{\t}}\label{FixedCouplingAccuracy}

Coulomb branch correlation functions can be computed exactly via supersymmetric localization, by an algorithm given in \cite{Gerchkovitz:2016gxx}.
We now compare our asymptotic expansion with the results of the localization calculation. 

\subsection{Limitations on the accuracy of agreement between the fixed-\bbd{\t} large-charge asymptotic series and the localization calculation}\label{FixedCouplingSourcesOfErrorSec}
 Before doing this,
we consider several possible sources of error, beyond the intrinsic imprecision of carrying an asymptotic series
to finite order.

\heading{Numerical errors}

To compute the two-point function of \bbd{[\co(x)]\uu n,} the algorithm involves taking the determinant of an \bbd{n\times n} matrix, where
each matrix element is a multi-partial derivative of the \bbd{S\uu 4} partition
function with respect to \bbd{\t} and \bbd{\bar{\t}}, each of which
is calculated as an integral over a real section of the Coulomb branch.  
The major difficulty in the computation of correlation functions by this method to any given precision, comes from the 
fact that the matrix elements of the \bbd{(n+1)\times (n+1)} matrix have wildly different orders of magnitude,
and the alternating signs in the sum defining the determinant cause cancellation of the largest terms.  As a result,
the computation of the determinant to any desired precision, involves calculating the individual matrix elements to a
far \emm{higher} precision as \bbd{n} becomes large.  This difficulty is the main limitation on the practical calculation
of exact correlation functions by the method of  \cite{Gerchkovitz:2016gxx}.  The result is that the precision 
of the localization calculation relative to the final result is generally far lower than the working precision of the numerical
evaluation of the individual matrix elements.  This difficulty has limited the guaranteed accuracy of the
numerical evaluation of no better than one part in \bbd{10\uu{10}} of the size of the overall MMP function.  At \bbd{n=150,} attaining this accuracy requires
evaluating individual matrix elements to an accuracy of several hundred digits.

\heading{Omission of gauge instantons from the localization calculation}
The inclusion of gauge instanton effects in the localization integrand, while well-understood in principle \cite{Nekrasov:2002qd}, in practice
significantly increases the computational cost of numerical evaluation of the integrals; so far all
practical computations at large \bbd{n} have omitted those contributions, which demands that any
comparison with \textsc{eft} large charge predictions be done at a value of \bbd{\t} small enough
that gauge instanton effects are negligible compared to the desired precision in the MMP function.
For that reason we are considering correlators evaluated at a ratier small value of the coupling,
\bbd{\t = {{25}\over \pi}\cc i,} where the instanton factor \bbd{e\uu{2\pi i}} is \bbd{e\uu{-50} \simeq  2\times 10\uu{-22}}.  For the
range of \bbd{n} we consider, \bbd{n \leq 150,} the value of \bbd{q\ll n} is roughly \bbd{3.6\times 10\uu 2}, so the absolute error from the
omission of gauge instantons should cause an error no larger than \bbd{ e\uu{- 2\pi \cc {\rm Im}[\t]}\cc q\ll n \lesssim O(10\uu{-20}-10\uu{-19})}. 

\heading{Omission of two-worldline-instanton terms as limitation on the accuracy of the fixed-coupling estimates}

The fixed-coupling estimates are based on the idea that the leading contribution at large \bbd{n} and fixed \bbd{\t} to the MMP function is a single
worldline instanton of a virtual BPS particle whose mass is controlled by the BPS formula and proportional to \bbd{\sqrt{{n\over{{\rm Im}[\s]}}}}.
That is, the behavior of \bbd{{\rm Log}[q\ll n] = - \mathfrak{W} = - \mathfrak{W}[\t,n]} is assumed to be \bbd{- \mathfrak{W} = - \sqrt{{{8\pi n}\over{{\rm Im}[\s]}}} + O({\rm Log}[n])}
with the \bbd{{\rm Log}[n]} and \bbd{n\uu{-{p\over 2}}} terms determined by the recursion relations and asymptotic behaviors in various limits.
As a series in \bbd{n\uu{-{p\over 2}}} it is extremely unlikely that \bbd{\mathfrak{W}} has a finite radius of convergence and even understanding
its properties as an asymptotic series at fixed coupling is challenging.  We expect that there will be contributions to the amplitude from two massive worldline
instantons connected by a massless propagator.  As discussed in sec. \ref{TwoLoopPhysicalInterpretation}, these effects contribute at order \bbd{n\uu{-1}}
in the double-scaling limit and would have a parametric scaling \bbd{n\uu{-\hh}\cc \exp{- 2\sqrt{{{8\pi n}\over{{\rm Im}[\s]}}}}}.  And indeed the numerical
data show a maximum accuracy of the fixed-coupling large-\bbd{n} expansion of approximately this amount. 
For \bbd{n = 150} and \bbd{\t = {{25}\over \pi}\cc i} the accuracy
of the estimates stops improving for \bbd{p \geq 6.}   On the other hand, there is no clear \emm{internal} sign of a breakdown of the asymptotic series when the size of the corrections reaches the level comparable to
the two-worldline-instanton effect.

This outcome is somehwat in tension with the
\emm{generic} predictions of resurgence theory, in which the point at which the size of the perurbative terms stop decreasing agrees with the optimally accurate truncation of perturbation theory, 
at which the error is comparable to the contribution of the lowest-action omitted saddle point contributing to the path integral.  In many supersymmetric examples,
however (see {\it e.g. \rm} \cite{Behtash:2015zha, Honda:2016mvg}) these three criteria for an optimal truncation may not agree with one another.
It would be interesting to see whether the accuracy of the fixed-coupling asymptotic series can be improved by the inclusion of higher-action worldline-instanton effects.
There is no corresponding puzzle for the double-scaled estimates,
which already include all multi-winding contributions at a given order in \bbd{n\uu{-1}}.

\heading{Macroscopic virtual propagation of monopoles and dyons}
The omitted instanton effects we have mentioned so far, appear
only though the relationship between the ultraviolet coupling
\bbd{\t} and the infrared coupling \bbd{\s}.  These effects
depend on the ultraviolet \bbd{\th-}angle but are independent
of the infrared \bbd{\th}-angle \bbd{\th\lrm{IR} \equiv 2\pi {\rm Re}[\s]}.
However S-duality predicts there must be additional exponentially small terms that corresponding
to macroscopic virtual propagation of BPS monopole and dyon particles carrying magnetic charge \bbd{Q\lrm{magnetic}}, whose masses depend on the infrared 
theta-angle \bbd{\th\lrm{IR}} {\it via\rm} the Witten effect \cite{Witten:1979ey}
with the BPS mass obeying the formula
\bbb
\bbsk
M\lrm{BPS} = |a|\cc |Q\lrm {electric} + \s\cc Q\lrm {magnetic}| = 
\sqrt{{2n}\over {\pi}}\times
 {{|Q\lrm {electric} + \s\cc Q\lrm {magnetic}|}\over {R\sqrt{{\tt Im}[\s]}}}
\eee
For nonzero \bbd{Q\lrm{magnetic}} the virtual effects of these particles are parametrically smaller than all terms in the double-scaled
large-\bbd{n} expansion and even smaller than gauge 
instanton effects, so we have neglected them.  And
indeed at \bbd{\t = {{25}\over \pi} i} these effects are far smaller numerically
than the ten-digit window of relative accuracy we have maintained
for comparison of our asymptotic expansion with exact results. 
However even at moderately
small values of \bbd{{\rm Im}[\t]} these effects may contribute significantly to
correlation functions.  For instance, for \bbd{\s = i s} between
\bbd{s=1} and \bbd{s = 2}, a macroscopic
virtual monopole contribution is a larger effect than the leading
doubly-exponentially suppressed contribution \bbd{\sim n\uu{-\hh}\cc \exp{- 2\sqrt{{8\pi n}\over{{\rm Im}[\s]}}}}  with two macroscopic massive 
electric hypermultiplet propagators.
  It would be
interesting to understand how to incorporate these effects
systematically as contributions to
some kind of hyperasymptotic large-R-charge series, along with
the multiple-macroscopic massive electric hypermultiplet effects
discussed above.  It is possible that the resugence-theoretic \cite{Ecalle} ideas explored in the nonsupersymmetric Wilson-Fisher models \cite{Dondi:2021buw}
may play a role, although the distinct resurgence-theoretic issues associated with nonperturbative effects in supersymmetric theories ({\it e.g.\rm} \cite{Behtash:2015zha, Honda:2016mvg})
caution against too simplistic an extrapolation of those results to SQCD.

\subsection{Summary and highlights of the results}

We evaluate the asymptotic expansion up to and including \bbd{n=150} and
compare with exact results from localization \cite{LocData}, as computed by the
method of \cite{Gerchkovitz:2016gxx}.

Before presenting the comparison of the estimate with data systematically, we first give some highlights  of the comparison at \bbd{\t = {{25}\over \pi}
\cc i} .

\heading{n=1}

At \bbd{n=1} the exact value of the connected MMP term is
\bbb
q\ll 1\uprm{\textsc{mmp}} = 0.42631\ .
\eee
The best estimate obtained from the fixed-coupling large-charge
asymptotic expansion is the expansion with the \bbd{-n\uu{-{3\over 2}}\cc w\ll 3[\s]} term included in the exponent of the MMP estimate;
after that, the series begins to diverge at \bbd{n=1}.

Even at the lowest possible nonzero value of the R-charge, the 
best estimate is reasonably accurate,
\bbb
q\ll 1\uprm{\textsc{mmp}} \cc \bigg |\lrm{ {{fixed-coupling}\atop{estimate~incl.~{\it w}\ll 3[\s]} }} = 0.44737\ ,
\eee
a relative error of only 4.94\%.

The relative accuracy of the
estimates of the MMP function gives a rather understated picture
of the accuracy of the estimate of the correlator as a whole, since
the correlator is parametrically dominated by the \textsc{eft} term,
which is exponentially larger than the exact MMP term.  At
\bbd{n=1}, the \textsc{eft} term in the \bbd{\t-}frame as computed in the
Pestun-Nekrasov scheme, is 
\bbb
q\ll 1\uprm{eft} =-11.22042\ ,
\eee
some 26 times larger than the MMP factor.  So, compared
to the full correlator as computed
in this scheme, the relative error of the Pestun-
Nekrasov scheme is about two tenths of a percent.

\heading{n=2}

For \bbd{n=1} however there is no natural scheme-independent
measure of the accuracy of the full correlator so instead we
can consider \bbd{n=2}.  Here we have
\bbb
q\ll 2\uprm{\textsc{mmp}} = 0.292443\ .
\eee
Here, even the NLO estimate of the MMP factor, the exponentiated
WLI action with determinantal prefactor, is already accurate
to less than 20 percent of the MMP term itself:
\bbb
q\ll 2\uprm{\textsc{mmp}} \cc \bigg |\lrm{ {{estimated by~WLI}\atop{with~prefactor} }} =
0.236916\ .
\eee
For larger \bbd{n} the best fixed-coupling estimate is almost
always the estimate with terms up to and including \bbd{-n\uu{-{5\over 2}}\cc
w\ll 5[\s]} included in the exponent of the estimate, with the
few exceptions being a set of values with \bbd{n\lesssim 60}
for which the estimates up to and including \bbd{n\uu{-{3\over 2}}\cc w\ll 3[\s]} are best due to an "accidental" agreement, as the sign of the error
changes sign before heading into the asymptotic regime.  The estimate
with terms up to and including \bbd{-n\uu{-{5\over 2}}\cc w\ll 5[\s]} in
the exponent, is
\bbb
q\ll 2\uprm{\textsc{mmp}} \cc \bigg |\lrm{ {{fixed-coupling}\atop{estimate~incl.~{\it w}\ll 5[\s]~term} }} = 0.292332\ ,
\eee
a relative error of less than 0.04\% of the size of the MMP correction
itself.  

\heading{Relative errors}

That is, defining
\bbb
\big [ \cc {{{\rm error~of}\atop{{\rm estimate~for~}q\uprm{\textsc{mmp}}\ll n}}}\cc 
\big ] \equiv (q\ll n\uprm{\textsc{mmp}})\lrm{estimate}
- q\ll n\uprm{\textsc{mmp}}\ ,
\eee
and the relative error of the asymptotic estimates,
\bbb
\big [ \cc {{{\rm relative~error~of}\atop{{\rm estimate~for~}q\uprm{\textsc{mmp}}\ll n}}}\cc 
\big ] \equiv \biggl |\cc {{(q\ll n\uprm{\textsc{mmp}})\lrm{estimate}
- q\ll n\uprm{\textsc{mmp}}\ }\over{ q\ll n\uprm{\textsc{mmp}}}} \cc \biggl |\ ,
\eee
we find
\bbb
\big [ \cc {{{\rm relative~error~of}\atop{{\rm N\uu 4 LO~estimate~for~}q\uprm{\textsc{mmp}}\ll 1}}}\cc 
\big ] = 4.94\%
\eee
for the fixed-coupling estimate including the \bbd{ w\ll 3[\s]} term in the exponent,
and
\bbb
\big [ \cc {{{\rm relative~error~of}\atop{{\rm N\uu 6 LO~estimate~for~}q\uprm{\textsc{mmp}}\ll 2}}}\cc 
\big ] = 3.8\times 10\uu{-4}
\eee
for the fixed-coupling estimate including the \bbd{ w\ll 5[\s]} term in the exponent.  (Here, we are considering \bbd{e\uu{- S\lrm{WLI}}} without
the prefactor to be the leading-order estimate of the MMP function 
and the exponential with the prefactor included to be the NLO estimate,
so that including the term \bbd{- n\uu{-{p\over 2}}\cc w\ll p[\s]} in
the exponent gives the \bbd{{\rm N\uu{p+1}LO}} estimate.)

\heading{Some larger values of \bbd{n}}

The estimates get quickly better from there.  At \bbd{n=40} we
have
\bbb
\big [ \cc {{{\rm relative~error~of}\atop{{\rm N\uu 6 LO~estimate~for~}q\uprm{\textsc{mmp}}\ll {40}}}}\cc 
\big ] = 1.1\times 10\uu{-4}\ ,
\eee
at \bbd{n=100} we have
\bbb
\big [ \cc {{{\rm relative~error~of}\atop{{\rm N\uu 6 LO~estimate~for~}q\uprm{\textsc{mmp}}\ll {100}}}}\cc 
\big ] = 1.2\times 10\uu{-6}\ ,
\eee
and at \bbd{n=150} we have
\bbb
\big [ \cc {{{\rm relative~error~of}\atop{{\rm N\uu 6 LO~estimate~for~}q\uprm{\textsc{mmp}}\ll {150}}}}\cc 
\big ] = 3.8\times 10\uu{-9}\ ,
\eee
less than four parts in a billion.

\heading{Errors relative to the (log of the) full (scheme-independent version of the) correlation function}

To give an idea of the accuracy of the large-R-charge expansion
overall, we can compute the relative error of the logarithm
of the \emm{full} correlator with the \textsc{eft} term included,
as normalized by powers of the lowest correlator \bbd{G\ll 2} in
order to cancel the dependence on the holomorphic coordinate 
in which the gauge coupling is expressed.  That is,
we can define
\bbb
\tilde{G}\ll{2n} \equiv {{G\ll{2n}}\over{(G\ll 2)\uu n}}\ ,
\eee
which is independent of the holomorphic coordinate frame:
\bbb
\tilde{G}\ll{2n}  = {{G\ll{2n[\t]}}\over{(G\ll 2[\t])\uu n}} = {{G\ll{2n[\t\pr]}}\over{(G\ll 2[\t\pr])\uu n}}
\eee
for any holomorphic coordinate \bbd{\t\pr}.  This combination is
also independent of the Euler-density counterterm choice \rr{KahlerTrans}, and so it is
fully scheme-independent.  In terms of the \bbd{q\ll n}, it is given by
\bbb
\tilde{G}\ll{2n} = e\uu{q\ll n - n q\ll 1 + (n-1)\cc q\ll 0\ .
}\eee
So defining the scheme-independent combination
\bbb
\tilde{q}\ll n \equiv {\rm Log}[\tilde{G}\ll{2n}] = q\ll n- n q\ll 1 + (n-1)\cc q\ll 0\ ,
\eee
we can compute the asymptotic estimates,
\bbb
\bbsk
(\tilde{q}\ll n)\lrm{estimate} \equiv (q\ll n)\lrm{estimate}- n q\ll 1+ (n-1)\cc q\ll 0 = q\ll n\uprm{eft}+ (q\ll n\uprm{\textsc{mmp}})\lrm{estimate}- n q\ll 1+ (n-1)\cc q\ll 0\ ,
\llsk
\eee
the error of the asymptotic estimates,
\bbb
\big [ \cc {{{\rm error~of}\atop{{\rm estimate~for~}\tilde{q}\ll n}}}\cc 
\big ] \equiv(\tilde{q}\ll n)\lrm{estimate} - \tilde{q\ll n} = (q\ll n\uprm{\textsc{mmp}})\lrm{estimate}
- q\ll n\uprm{\textsc{mmp}}\ ,
\eee
and the relative error of the asymptotic estimates,
\bbb
\big [ \cc {{{\rm relative~error~of}\atop{{\rm estimate~for~}\tilde{q}\ll n}}}\cc 
\big ] \equiv
\bigg |\cc {{(\tilde{q}\ll n)\lrm{estimate} - \tilde{q}\ll n}\over{\tilde{q}\ll n}}
\cc \bigg | = \bigg |\cc {{(q\ll n\uprm{\textsc{mmp}})\lrm{estimate} - q\ll n\uprm{\textsc{mmp}}}\over{q\ll n - n q\ll 1 + (n-1) q\ll 0}} \cc \bigg |\ .
\eee

At \bbd{n=2} we have
\bbb
\tilde{q}\ll 2 = q\ll 2 - 2 q\ll 1 + {\rm Log}{[Z\ll{S\uu 4}]} = 
1.19311
\xxnn
(\tilde{q}\ll 2)\lrm{estimate} = (q\ll 2\uprm{\textsc{mmp}})\lrm{estimate} + q\ll 2\uprm{eft} - 2 q\ll 1 + {\rm Log}[Z\ll{S\uu 4}] =1.19300
\xxnn
\big [ \cc {{{\rm relative~error~of}\atop{{\rm estimate~for~}\tilde{q}\ll 2}}}\cc 
\big ] = 9.32\times 10\uu{-5}
\eee
already less than one part in ten thousand at \bbd{n=2}.

\heading{Some larger values of \bbd{n}}

At \bbd{n=10} we have
\bbb
\tilde{q}\ll {10} = 27.15040
\xxnn
(\tilde{q}\ll {10})\lrm{estimate} = 27.15023
\xxnn
\big [ \cc {{{\rm relative~error~of}\atop{{\rm estimate~for~}\tilde{q}\ll {10}}}}\cc 
\big ] = 6.2\times 10\uu{-6}
\eee

At \bbd{n=40} we have
\bbb
\tilde{q}\ll {40} = 203.998642600
\xxnn
(\tilde{q}\ll {40})\lrm{estimate} = 203.9986424593
\xxnn
\big [ \cc {{{\rm relative~error~of}\atop{{\rm estimate~for~}\tilde{q}\ll {40}}}}\cc 
\big ] = 6.9\times 10\uu{-10}
\eee

At \bbd{n=100} we have
\bbb
\cc\cc\cc\cc\cc\cc\cc\cc\cc\cc\cc\cc\cc\cc\cc\cc\cc\cc\cc\cc\tilde{q}\ll {100} = 682.005901930669
\xxnn
(\tilde{q}\ll {100})\lrm{estimate} = 682.005901930647
\xxnn
\big [ \cc {{{\rm relative~error~of}\atop{{\rm estimate~for~}\tilde{q}\ll {100}}}}\cc 
\big ] = 3.2\times 10\uu{-14}\ ,
\eee
and at \bbd{n=150} we have
\bbb
\cc\cc\cc\cc\cc\cc\cc\cc\cc\cc\cc\cc\cc\cc\cc\cc\cc\cc\cc\cc\tilde{q}\ll {150} = 1140.19247443414906186
\xxnn
(\tilde{q}\ll {150})\lrm{estimate} = 1140.19247443414906669
\xxnn
\big [ \cc {{{\rm relative~error~of}\atop{{\rm estimate~for~}\tilde{q}\ll {150}}}}\cc 
\big ] = 4.2\times 10\uu{-18}\ .
\eee

\subsection{Plots}

 \subsubsection{On the visual display of quantitative information}

 When communicating results on the accuracy of the large-quantum-number
expansion in general, one acute challenge has been to present
the results graphically in a way that makes the errors visible 
to the naked eye.\footnote{See the plots of large-charge predictions
versus exact results in \cite{Giombi:2020enj, Banerjee:2017fcx, Banerjee:2019jpw} for examples of this difficulty.}  "Errors too small to see on a graph"
is certainly a target one should aspire to reach, and it has been
reached consistently by large quantum number expansions, even 
when extrapolated to surprisingly low quantum numbers; and yet actually
achieving it is a mixed blessing, since invisible
deviations may seem to lose their meaning, but we wish to keep our
scientific focus on the meaning of them.

With that in mind, we will plot
our results in a way adapted to meet this peculiar difficulty as best we can.
Instead of plotting data points indicating exact results and theory curves
representing the approximations, we will plot the negative of the logarithm,
base ten, of the absolute value of the difference between the estimate
and the exact result for the log of the correlator, divided by the log
of the correlator.  That is, for each estimate \bbd{(q\uprm{\textsc{mmp}})\lrm{estimate}}
we will plot the \bbd{y}-coordinate \bbd{y\ll n} as
\bbb
y(n) \equiv - {1\over{{\rm Log}[10]}}\cc {\rm Log} \bigg | \cc {{q\ll n\uprm{\textsc{mmp}} - (q\uprm{\textsc{mmp}})\lrm{estimate}}\over{q\ll n\uprm{\textsc{mmp}}}}\cc \bigg |
\eee 
This means the \bbd{y-}coordinate on figure \ref{FixedCouplingAccuracyPlotA}, plots the effective number of
accurate significant digits of the estimate of the MMP function \bbd{q\ll n\uprm{\textsc{mmp}}}.  In other words, the absolute accuracy of the estimate is
\bbb
\big |\cc q\ll n\uprm{\textsc{mmp}}- (q\uprm{\textsc{mmp}})\lrm{estimate} \big |= 
10\uu{- y(n)}\cc  \big | q\ll n\uprm{\textsc{mmp}} \big |\ .
\eee
For the best estimate we have computed, which expands
the exponent \bbd{\mathfrak{W}[n,\s]} up to order \bbd{n\uu{-{5\over 2}}},
and values of \bbd{n} above \bbd{130} or so, we have
\bbd{y(n) \gtrsim 7} digits of accuracy for our estimates of
\bbd{q\ll n\uprm{\textsc{mmp}}}.

Denominating the errors relative to the MMP function somewhat
understates the accuracy of the large-R-charge approximation,
since the MMP term is itself exponentially small in \bbd{\sqrt{n}} relative
to the full logarithm of the correlator, with the exact closed-form
expression for the \textsc{eft} contribution dominating the full correlator.  So to give perspective, we have
also plotted, on a separate graph, the effective number 
of accurate significant digits of the log of the scheme-independently
normalized full correlator \bbd{\widetilde{G\ll{2n}} \equiv G\ll{2n} / (G\ll 2)\uu n
= e\uu{q\ll n - n q\ll 1 + (n-1) q\ll 0}}.  The denominator is
chosen to cancel the dependence on the choice
of holomorphic coordinatization of the gauge coupling; that is,
this normalized correlator is invariant under a holomorphic reparametrization \rr{HolomorphicReparametrizationTransformationOfOperator} of the complexified inverse gauge
coupling, and also under a change of Euler-counterterm choice, the K\"ahler transformation \rr{KahlerTrans}.
 The \bbd{y-}axis for the second plot, figure \ref{FixedCouplingAccuracyPlotBRelativeToFull}, is
\bbb
 y(n) \equiv
- {1\over{{\rm Log}[10]}}\cc {\rm Log}\left [ \cc {{{\rm Log}[\tilde{G}\ll{2n}\urm{exact}] - {\rm Log}[\tilde{G}\ll{2n}\urm{estimate}]}\over{{\rm Log}[\tilde{G}\ll{2n}\urm{exact}]}} \cc \right ] = - {1\over{{\rm Log}[10]}}\cc {\rm Log} \left [ \cc {{| q_n - (q_n)\lrm{estimate} | }\over{ |q_n- n\cc q\ll 1 + (n-1) q\ll 0 |}} \cc  \right ]
\eee
For the best estimate we have computed, which expands
the exponent \bbd{\mathfrak{W}[n,\s]} up to order \bbd{n\uu{-{5\over 2}}},
and values of \bbd{n} above \bbd{130} or so, we have
\bbd{y(n) \gtrsim 15} digits of accuracy for our estimates of the full 
\bbd{{\rm log}[\widetilde{G\ll{2n}} ] = q\ll n - n q\ll 1 + (n-1) q\ll 0 }.

\newpage

  \subsubsection{Plotting the log of relative error of the fixed-\bbd{\t} estimates through order \bbd{n\uu{-{5\over 2}}} in the exponent, evaluated at \bbd{\t = {{25}\over \pi}, 1 \leq n \leq 150.}}

\centpicWidthHeightAngleFileCaptionFiglab{120mm}{80mm}{0}
{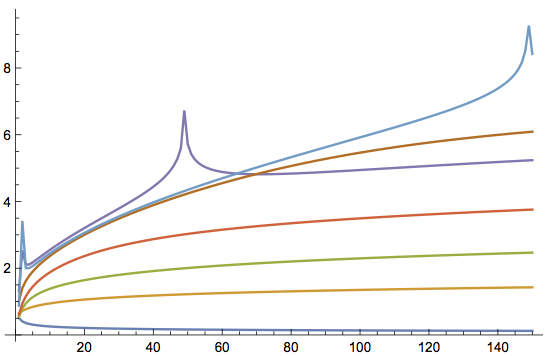}
{Plot giving the accuracy of the double-scaled estimates of the MMP 
function through N${}\uu 6$LO, plotted as the number of
digits of accuracy of each of the estimates, as a function
of \bbd{n}.  The quantity
being plotted is \bbd{- {1\over{{\rm Log}[10]}}} the \emm{logarithm} of the relative error in the estimate of the MMP function.  The 
horizontal axis is \bbd{n}, and the vertical axis is 
\bbd{- {1\over{{\rm Log}[10]}}\cc {\rm Log} \left | \cc {{ q_n\uprm{\textsc{mmp}} - (q_n\uprm{\textsc{mmp}})\lrm{estimate}  }\over{ q_n\uprm{\textsc{mmp}}}} \cc  \right |}.  The LO, NLO, N${}\sqd$LO, N${}\uu 3$LO, N${}\uu 4$LO, N${}\uu 5$LO and N${}\uu 6$LO estimates
are given by the blue, yellow, green, red, and purple, brown, and light blue curves respectively, which are in ascending order on the chart for
\bbd{n\gtrsim 65}.   The
upward spikes on the accuracy plots of the N${}\uu 4$LO and
N${}\uu 6$LO estimates are "accidental accuracies" in which the estimate transitions between 
slightly overestimating and slightly underestimating the exact result as \bbd{n} is varied, generating an atypically precise estimate at N${}\uu 4$LO and N${}\uu 6$LO in certain limited ranges of \bbd{n}.   At \bbd{n=150} the N${}\uu 5$LO fixed-\bbd{\t} large-\bbd{n} estimate
of the MMP function is accurate to one part in \bbd{10\uu 6} and the N${}\uu 6$LO fixed-\bbd{\t} large-\bbd{n} estimate
is accurate to one part in \bbd{10\uu 8}.} 
{FixedCouplingAccuracyPlotA}
\newpage

\newpage

 \subsubsection{Plotting the log of relative error of the fixed-\bbd{\t} estimates through order \bbd{n\uu{-{5\over 2}}} in the exponent, evaluated at \bbd{\t = {{25}\over \pi}, 1 \leq n \leq 150.}, relative to the full log of the scheme-independent correlator \\  \bbd{\tilde{G}\ll{2n} \equiv {{G\ll{2n} }\over{ (G\ll 2)\uu n}} = e\uu{ q\ll n - n\cc q\ll 1 + (n-1) q\ll 0}}}

\centpicWidthHeightAngleFileCaptionFiglab{120mm}{80mm}{0}
{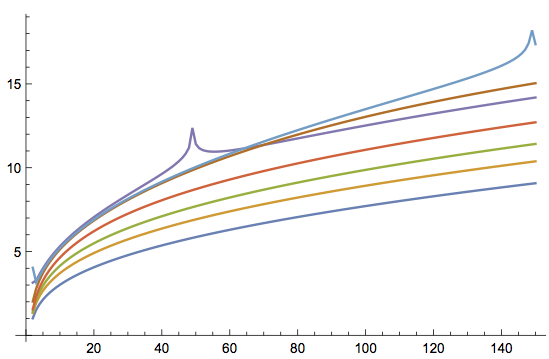}
{Plot giving the accuracy of the fixed-coupling large-R-charge estimates of the full log of the scheme-independent correlator,  \bbd{\tilde{q}\ll n\equiv {\rm Log}[\tilde{G}\ll{2n}] \equiv  q\ll n - n\cc q\ll 1 + (n-1) q\ll 0}.
The estimate for \bbd{\tilde{q}\ll n} is defined to be
\bbd{(\tilde{q}\ll n)\lrm{estimate} \equiv(q\ll n\uprm{\textsc{mmp}})\lrm{estimate} + q\ll n\uprm{eft} - n
q\ll 1 + (n-1)q\ll 0}, with the MMP contribution as estimated by eqs. \rr{DefOfScriptW},\rr{ParametrizationOfScriptW},\rr{W1Formula}-\rr{W5Formula}   through N${}\uu 6$LO.  The quantity
being plotted is \bbd{- {1\over{{\rm Log}[10]}}} times the logarithm of the relative error in the estimate of the correlator.  The 
horizontal axis is \bbd{n}, and the vertical axis is 
\bbd{ y(n) \equiv
- {1\over{{\rm Log}[10]}}\cc {\rm Log}\left | \cc {{{\rm Log}[\tilde{G}\ll{2n}\urm{exact}] - {\rm Log}[\tilde{G}\ll{2n}\urm{estimate}]}\over{{\rm Log}[\tilde{G}\ll{2n}\urm{exact}]}} \cc \right | = - {1\over{{\rm Log}[10]}}\cc {\rm Log} \left | \cc {{ q_n\uprm{\textsc{mmp}} - (q_n\uprm{\textsc{mmp}})\lrm{estimate}  }\over{ q_n- n\cc q\ll 1 + (n-1) q\ll 0 }} \cc  \right |}. The leading-order, NLO, N${}\sqd$LO, N${}\uu 3$LO, N${}\uu 4$LO, N${}\uu 5$LO and N${}\uu 6$LO estimates
are given by the dark blue, yellow, green, red, and purple, brown, and light blue curves respectively.   The
upward spikes on the accuracy of the N${}\uu 4$LO and N${}\uu 6$LO estimates are "accidental accuracies" in which an estimate transitions between 
slightly overestimating and slightly underestimating the exact result as \bbd{n} is varied, generating an atypically precise estimate in some limited range of \bbd{n}.  At \bbd{n=150} the N${}\uu 5$LO fixed-\bbd{\t} large-\bbd{n} estimate
of the scheme-independent correlator is accurate to one part in \bbd{10\uu{15}} and the N${}\uu 6$LO estimate is accurate to one part in \bbd{10\uu{17}}. } 
{FixedCouplingAccuracyPlotBRelativeToFull}

\newpage

\subsection{Table of values of the fixed-\bbd{\t} large-charge asymptotic estimates at \bbd{\t = {{25}\over \pi}\cc i}, up to \bbd{n=150}}

 \begin{table}[H]
\begin{center}
\resizebox{\columnwidth}{!}{%
\begin{tabular}{ |c||c|c|c|c|c|c|c| } 
 \hline\hline
 ~ & ${{e\uu{-S\lrm{WLI}}{\rm ~w}/}\atop{\rm prefactor}}$ 
 & ${{\rm estimate~w/}\atop{O(n\uu{-\hh})}}$ &${{\rm estimate~w/}\atop{O(n\uu{-1})}}$  &${{\rm estimate~w/}\atop{O(n\uu{-{3\over 2}})}}$ &${{\rm estimate~w/}\atop{O(n\uu{-2})}}$ & ${{\rm estimate~w/}\atop{O(n\uu{-{5\over 2}})}} $ & ${{\rm exact}\atop{q\ll n\uprm{\textsc{mmp}}}}$ \\ \hline\hline

\ReUpdatedNEWESTMACRO {1} {0.2942925813} {0.3306547971} {0.5494056108} {0.3208548662} {0.4473688062} {0.3855992091} {0.4826779760} {0.4263073863} 
\ReUpdatedNEWESTMACRO{2} {0.1773129689} {0.2369155213} {0.3392528251} {0.2592574487} {0.2915876823} {0.2809550113} {0.2923319506} {0.2924432054} 
\ReUpdatedNEWESTMACRO{3} {0.1201984264} {0.1777356756} {0.2382818085} {0.1991725121} {0.2123297908} {0.2088531261} {0.2118834217} {0.2140116919} 
\ReUpdatedNEWESTMACRO{4} {0.08660812339} {0.1376160664} {0.1773896477} {0.1550715601} {0.1616504023} {0.1601561764} {0.1612839808} {0.1629019007} 
\ReUpdatedNEWESTMACRO{5} {0.06488651297} {0.1090165541} {0.1368079122} {0.1228554207} {0.1265627818} {0.1258128005} {0.1263191935} {0.1274138410} 
\ReUpdatedNEWESTMACRO{6} {0.04997817335} {0.08788478715} {0.1081284329} {0.09885739579} {0.1011186866} {0.1007021939} {0.1009589550} {0.1016922546} 
\ReUpdatedNEWESTMACRO{7} {0.03931173162} {0.07184428016} {0.08704428588} {0.08060668448} {0.08206644241} {0.08181796572} {0.08195980455} {0.08245642599} 
\ReUpdatedNEWESTMACRO{8} {0.03143988894} {0.05940855833} {0.07109089681} {0.06646850876} {0.06745213130} {0.06729571354} {0.06737924445} {0.06772080050} 
\ReUpdatedNEWESTMACRO{9} {0.02548812819} {0.04960141702} {0.05874895841} {0.05534091587} {0.05602642358} {0.05592374401} {0.05597544585} {0.05621405675} 
\ReUpdatedNEWESTMACRO{10} {0.02089934585} {0.04175689585} {0.04903003507} {0.04646262665} {0.04695358198} {0.04688386790} {0.04691717164} {0.04708632558} 
\ReUpdatedNEWESTMACRO{11} {0.01730375289} {0.03540658081} {0.04126413514} {0.03929504552} {0.03965469749} {0.03960603245} {0.03962819994} {0.03974973039} 
\ReUpdatedNEWESTMACRO{12} {0.01444766172} {0.03021262008} {0.03498215250} {0.03344884738} {0.03371738216} {0.03368260919} {0.03369777501} {0.03378615497} 
\ReUpdatedNEWESTMACRO{13} {0.01215224674} {0.02592614410} {0.02984686301} {0.02863720688} {0.02884100969} {0.02881566382} {0.02882628483} {0.02889126648} 
\ReUpdatedNEWESTMACRO{14} {0.01028855058} {0.02236051076} {0.02561047918} {0.02464524274} {0.02480212485} {0.02478332983} {0.02479091948} {0.02483917594} 
\ReUpdatedNEWESTMACRO{15} {0.008761755521} {0.01937356121} {0.02208752376} {0.02130957226} {0.02143184640} {0.02141769798} {0.02142321768} {0.02145938050} 
\ReUpdatedNEWESTMACRO{16} {0.007500967037} {0.01685554371} {0.01913693280} {0.01850432737} {0.01860068261} {0.01858988973} {0.01859396672} {0.01862129258} 
\ReUpdatedNEWESTMACRO{17} {0.006452395834} {0.01472070962} {0.01664992496} {0.01613139844} {0.01620807852} {0.01619974753} {0.01620280064} {0.01622360696} 
\ReUpdatedNEWESTMACRO{18} {0.005574700049} {0.01290134962} {0.01454158672} {0.01411350505} {0.01417506884} {0.01416856972} {0.01417088442} {0.01418683837} 
\ReUpdatedNEWESTMACRO{19} {0.004835737458} {0.01134348831} {0.01274491628} {0.01238919441} {0.01243901837} {0.01243389962} {0.01243567409} {0.01244798710} 
\ReUpdatedNEWESTMACRO{20} {0.004210259565} {0.01000372820} {0.01120653279} {0.01090917838} {0.01094979540} {0.01094572872} {0.01094710280} {0.01095666332} 
\ReUpdatedNEWESTMACRO{21} {0.003678247909} {0.008846904394} {0.009883536354} {0.009633615187} {0.009666947415} {0.009663690908} {0.009664764735} {0.009672229925} 
\ReUpdatedNEWESTMACRO{22} {0.003223696321} {0.007844320098} {0.008741177896} {0.008530067806} {0.008557589292} {0.008554962571} {0.008555808820} {0.008561668609} 
\ReUpdatedNEWESTMACRO{23} {0.002833707955} {0.006972404298} {0.007751108180} {0.007571953743} {0.007594805756} {0.007592672836} {0.007593344899} {0.007597967228} 
\ReUpdatedNEWESTMACRO{24} {0.002497817812} {0.006211680108} {0.006890046993} {0.006737356034} {0.006756429974} {0.006754687312} {0.006755224851} {0.006758887946} 
\ReUpdatedNEWESTMACRO{25} {0.002207478951} {0.005545964521} {0.006138761199} {0.006008102884} {0.006024100625} {0.006022668651} {0.006023101434} {0.006026017027} 
\ReUpdatedNEWESTMACRO{26} {0.001955669025} {0.004961742339} {0.005481272548} {0.005369049148} {0.005382527465} {0.005381344515} {0.005381695095} {0.005384025281} 
\ReUpdatedNEWESTMACRO{27} {0.001736586205} {0.004447672504} {0.004904238255} {0.004807510976} {0.004818914601} {0.004817932510} {0.004818218124} {0.004820087697} 
\ReUpdatedNEWESTMACRO{28} {0.001545412243} {0.003994195976} {0.004396462814} {0.004312817709} {0.004322504168} {0.004321685038} {0.004321918968} {0.004323424514} 
\ReUpdatedNEWESTMACRO{29} {0.001378126419} {0.003593222175} {0.003948510385} {0.003875954302} {0.003884212735} {0.003883526547} {0.003883719104} {0.003884935751} 
\ReUpdatedNEWESTMACRO{30} {0.001231358390} {0.003237876660} {0.003552394945} {0.003489274164} {0.003496339719} {0.003495762541} {0.003495921786} {0.003496908245}

\end{tabular}
}
\captionof{table}{ The successive refined estimates for the massive macroscopic propagation function \bbd{q\ll n\uprm{\textsc{mmp}} \equiv q\ll n - q\ll n\uprm{eft}}, at the coupling \bbd{\t = {{25 i}\over{\pi}}}.  The first column, "\bbd{e\uu{- S\lrm{WLI}}} with prefactor" is the approximation
\bbd{q\ll n\uprm{\textsc{mmp}} \simeq \big[\cc {{2\uu{13}\cc n}\over{\pi\uu 5\cc {\rm Im}[\s]}}\cc \big ]\uu{+{1\over 4}}\cc \exp{- \sqrt{{8\pi n}\over{{\rm Im}[\s]}}}}, with \bbd{\s} denoting the infrared coupling related to the UV coupling \bbd{\t} by \bbd{e\uu{2\pi i \t} = \l[\s],} where \bbd{\l[\s]} is the modular lambda function.  The value of \bbd{\t} in this table is \bbd{\t = {{25}\over\pi}\cc i}.  In each successive column we add a correction of the form \bbd{-n\uu{-{p\over 2}}\cc w\ll p[\s]} to the
exponent of the exponential for \bbd{p=1,\cdots 5}.  The functions
\bbd{w\ll p[\s]} are given explictly in \WFunxLocation}.\label{TypesOfLiouvilleAmplitudeTransformationTableSpecificParameterChoiceOfInterestToUs}
\end{center}
\end{table}

\newpage
 \vskip-1in
 \begin{table}[H]
\begin{center}
\resizebox{\columnwidth}{!}{%
\begin{tabular}{ |c||c|c|c|c|c|c|c| } 
 \hline\hline
 ~ & ${{e\uu{-S\lrm{WLI}}{\rm ~w}/}\atop{\rm prefactor}}$ 
 & ${{\rm estimate~w/}\atop{O(n\uu{-\hh})}}$ &${{\rm estimate~w/}\atop{O(n\uu{-1})}}$  &${{\rm estimate~w/}\atop{O(n\uu{-{3\over 2}})}}$ &${{\rm estimate~w/}\atop{O(n\uu{-2})}}$ & ${{\rm estimate~w/}\atop{O(n\uu{-{5\over 2}})}} $ & ${{\rm exact}\atop{q\ll n\uprm{\textsc{mmp}}}}$ \\ \hline\hline

\ReUpdatedNEWESTMACRO{31} {0.001102271039} {0.002922296878} {0.003201331023} {0.003146267184} {0.003152332104} {0.003151844745} {0.003151977022} {0.003152779381} 
\ReUpdatedNEWESTMACRO{32} {0.0009884666163} {0.002641465936} {0.002889531972} {0.002841371302} {0.002846593527} {0.002846180510} {0.002846290844} {0.002846945436} 
\ReUpdatedNEWESTMACRO{33} {0.0008879111129} {0.002391076604} {0.002612045828} {0.002569818642} {0.002574328534} {0.002573977313} {0.002574069707} {0.002574605288} 
\ReUpdatedNEWESTMACRO{34} {0.0007988729862} {0.002167419511} {0.002364621035} {0.002327509260} {0.002331414888} {0.002331115242} {0.002331192901} {0.002331632319} 
\ReUpdatedNEWESTMACRO{35} {0.0007198732580} {0.001967290819} {0.002143596083} {0.002110907063} {0.002114298388} {0.002114041953} {0.002114107457} {0.002114468932} 
\ReUpdatedNEWESTMACRO{36} {0.0006496446787} {0.001787915643} {0.001945808373} {0.001916953643} {0.001919905844} {0.001919685741} {0.001919741178} {0.001920039289} 
\ReUpdatedNEWESTMACRO{37} {0.0005870981574} {0.001626884277} {0.001768518637} {0.001742996630} {0.001745572764} {0.001745383318} {0.001745430384} {0.001745676837} 
\ReUpdatedNEWESTMACRO{38} {0.0005312950456} {0.001482098879} {0.001609348000} {0.001586729899} {0.001588983046} {0.001588819551} {0.001588859632} {0.001589063854} 
\ReUpdatedNEWESTMACRO{39} {0.0004814241594} {0.001351728735} {0.001466225365} {0.001446143478} {0.001448118466} {0.001447977008} {0.001448011239} {0.001448180848} 
\ReUpdatedNEWESTMACRO{40} {0.0004367826568} {0.001234172590} {0.001337343268} {0.001319481435} {0.001321216247} {0.001321093558} {0.001321122874} {0.001321264040} 
\ReUpdatedNEWESTMACRO{41} {0.0003967600640} {0.001128026830} {0.001221120707} {0.001205206358} {0.001206733271} {0.001206626612} {0.001206651784} {0.001206769522} 
\ReUpdatedNEWESTMACRO{42} {0.0003608248851} {0.001032058507} {0.001116171746} {0.001101969286} {0.001103315811} {0.001103222881} {0.001103244551} {0.001103342946} 
\ReUpdatedNEWESTMACRO{43} {0.0003285133404} {0.0009451824252} {0.001021278896} {0.001008584201} {0.001009773851} {0.001009692709} {0.001009711409} {0.001009793798} 
\ReUpdatedNEWESTMACRO{44} {0.0002994198641} {0.0008664415919} {0.0009353705035} {0.0009240063010} {0.0009250592249} {0.0009249882309} {0.0009250044051} {0.0009250735212} 
\ReUpdatedNEWESTMACRO{45} {0.0002731890619} {0.0007949905180} {0.0008575014645} {0.0008473134585} {0.0008482469668} {0.0008481847288} {0.0008481987498} {0.0008482568358} 
\ReUpdatedNEWESTMACRO{46} {0.0002495088849} {0.0007300809000} {0.0007868367408} {0.0007776903459} {0.0007785193470} {0.0007784646815} {0.0007784768620} {0.0007785257633} 
\ReUpdatedNEWESTMACRO{47} {0.0002281048198} {0.0006710493189} {0.0007226372331} {0.0007144148153} {0.0007151521785} {0.0007151040766} {0.0007151146799} {0.0007151559185} 
\ReUpdatedNEWESTMACRO{48} {0.0002087349292} {0.0006173066448} {0.0006642476420} {0.0006568461799} {0.0006575030396} {0.0006574606387} {0.0006574698874} {0.0006575047208} 
\ReUpdatedNEWESTMACRO{49} {0.0001911856094} {0.0005683288901} {0.0006110860113} {0.0006044151093} {0.0006050011198} {0.0006049636807} {0.0006049717634} {0.0006050012329} 
\ReUpdatedNEWESTMACRO{50} {0.0001752679504} {0.0005236492943} {0.0005626347012} {0.0005566148966} {0.0005571384461} {0.0005571053342} {0.0005571124109} {0.0005571373807} 
\ReUpdatedNEWESTMACRO{51} {0.0001608146067} {0.0004828514617} {0.0005184325769} {0.0005129938943} {0.0005134622857} {0.0005134329544} {0.0005134391614} {0.0005134603500} 
\ReUpdatedNEWESTMACRO{52} {0.0001476771008} {0.0004455633979} {0.0004780682350} {0.0004731489485} {0.0004735685520} {0.0004735425302} {0.0004735479836} {0.0004735659895} 
\ReUpdatedNEWESTMACRO{53} {0.0001357234943} {0.0004114523178} {0.0004411741180} {0.0004367196891} {0.0004370960719} {0.0004370729519} {0.0004370777512} {0.0004370930739} 
\ReUpdatedNEWESTMACRO{54} {0.0001248363716} {0.0003802201166} {0.0004074213892} {0.0004033835532} {0.0004037215894} {0.0004037010183} {0.0004037052488} {0.0004037183057} 
\ReUpdatedNEWESTMACRO{55} {0.0001149110926} {0.0003515994118} {0.0003765154626} {0.0003728514412} {0.0003731554055} {0.0003731370769} {0.0003731408118} {0.0003731519527} 
\ReUpdatedNEWESTMACRO{56} {0.0001058542731} {0.0003253500775} {0.0003481920948} {0.0003448639164} {0.0003451375640} {0.0003451212116} {0.0003451245140} {0.0003451340321} 
\ReUpdatedNEWESTMACRO{57} {0.00009758246263} {0.0003012562054} {0.0003222139647} {0.0003191878753} {0.0003194345108} {0.0003194199026} {0.0003194228267} {0.0003194309685} 
\ReUpdatedNEWESTMACRO{58} {0.00009002099070} {0.0002791234344} {0.0002983676725} {0.0002956136258} {0.0002958361615} {0.0002958230949} {0.0002958256878} {0.0002958326607} 
\ReUpdatedNEWESTMACRO{59} {0.00008310295913} {0.0002587766014} {0.0002764611036} {0.0002739523175} {0.0002741533239} {0.0002741416221} {0.0002741439244} {0.0002741499032} 
\ReUpdatedNEWESTMACRO{60} {0.00007676835981} {0.0002400576708} {0.0002563211089} {0.0002540336796} {0.0002542154292} {0.0002542049370} {0.0002542069841} {0.0002542121164}

\end{tabular}
}
\captionof{table}{The successive refined estimates for the massive macroscopic propagation function \bbd{q\ll n\uprm{\textsc{mmp}} \equiv q\ll n - q\ll n\uprm{eft}}, at the coupling \bbd{\t = {{25 i}\over{\pi}}}.  The first column, "\bbd{e\uu{- S\lrm{WLI}}} with prefactor" is the approximation
\bbd{q\ll n\uprm{\textsc{mmp}} \simeq \big[\cc {{2\uu{13}\cc n}\over{\pi\uu 5\cc {\rm Im}[\s]}}\cc \big ]\uu{+{1\over 4}}\cc \exp{- \sqrt{{8\pi n}\over{{\rm Im}[\s]}}}}, with \bbd{\s} denoting the infrared coupling related to the UV coupling \bbd{\t} by \bbd{e\uu{2\pi i \t} = \l[\s],} where \bbd{\l[\s]} is the modular lambda function.  The value of \bbd{\t} in this table is \bbd{\t = {{25}\over\pi}\cc i}.  In each successive column we add a correction of the form \bbd{-n\uu{-{p\over 2}}\cc w\ll p[\s]} to the
exponent of the exponential for \bbd{p=1,\cdots 5}.  The functions
\bbd{w\ll p[\s]} are given explictly in \WFunxLocation}.
\label{TypesOfLiouvilleAmplitudeTransformationTableSpecificParameterChoiceOfInterestToUs}
\end{center}
\end{table}

\newpage
 \vskip-1in
 \begin{table}[H]
\begin{center}
\resizebox{\columnwidth}{!}{%
\begin{tabular}{ |c||c|c|c|c|c|c|c| } 
 \hline\hline
 ~ & ${{e\uu{-S\lrm{WLI}}{\rm ~w}/}\atop{\rm prefactor}}$ 
 & ${{\rm estimate~w/}\atop{O(n\uu{-\hh})}}$ &${{\rm estimate~w/}\atop{O(n\uu{-1})}}$  &${{\rm estimate~w/}\atop{O(n\uu{-{3\over 2}})}}$ &${{\rm estimate~w/}\atop{O(n\uu{-2})}}$ & ${{\rm estimate~w/}\atop{O(n\uu{-{5\over 2}})}} $ & ${{\rm exact}\atop{q\ll n\uprm{\textsc{mmp}}}}$ \\ \hline\hline

\ReUpdatedNEWESTMACRO{61} {0.00007096330130} {0.0002228239063} {0.0002377914602} {0.0002357040259} {0.0002358685302} {0.0002358591119} {0.0002358609343} {0.0002358653448} 
\ReUpdatedNEWESTMACRO{62} {0.00006563932971} {0.0002069462559} {0.0002207310457} {0.0002188244923} {0.0002189735347} {0.0002189650708} {0.0002189666952} {0.0002189704896} 
\ReUpdatedNEWESTMACRO{63} {0.00006075283132} {0.0001923079213} {0.0002050122741} {0.0002032694786} {0.0002034046422} {0.0002033970276} {0.0002033984774} {0.0002034017450} 
\ReUpdatedNEWESTMACRO{64} {0.00005626450649} {0.0001788030905} {0.0001905196626} {0.0001889252678} {0.0001890479594} {0.0001890411017} {0.0001890423972} {0.0001890452141} 
\ReUpdatedNEWESTMACRO{65} {0.00005213890550} {0.0001663358121} {0.0001771485847} {0.0001756888014} {0.0001758002733} {0.0001757940909} {0.0001757952498} {0.0001757976806} 
\ReUpdatedNEWESTMACRO{66} {0.00004834401852} {0.0001548189954} {0.0001648041588} {0.0001634665919} {0.0001635679601} {0.0001635623808} {0.0001635634187} {0.0001635655183} 
\ReUpdatedNEWESTMACRO{67} {0.00004485091292} {0.0001441735199} {0.0001534002598} {0.0001521737559} {0.0001522660158} {0.0001522609759} {0.0001522619064} {0.0001522637217} 
\ReUpdatedNEWESTMACRO{68} {0.00004163341200} {0.0001343274424} {0.0001428586406} {0.0001417331527} {0.0001418171936} {0.0001418126366} {0.0001418134717} {0.0001418150425} 
\ReUpdatedNEWESTMACRO{69} {0.00003866780993} {0.0001252152894} {0.0001331081479} {0.0001320746166} {0.0001321512336} {0.0001321471094} {0.0001321478597} {0.0001321492202} 
\ReUpdatedNEWESTMACRO{70} {0.00003593261876} {0.0001167774246} {0.0001240840233} {0.0001231342717} {0.0001232041768} {0.0001232004409} {0.0001232011157} {0.0001232022950} 
\ReUpdatedNEWESTMACRO{71} {0.00003340834326} {0.0001089594845} {0.0001157272792} {0.0001148539196} {0.0001149177508} {0.0001149143636} {0.0001149149711} {0.0001149159943} 
\ReUpdatedNEWESTMACRO{72} {0.00003107728063} {0.0001017118719} {0.0001079841400} {0.0001071804917} {0.0001072388213} {0.0001072357476} {0.0001072362950} {0.0001072371835} 
\ReUpdatedNEWESTMACRO{73} {0.00002892334190} {0.00009498930276} {0.0001008055423} {0.0001000655584} {0.0001001189005} {0.0001001161089} {0.0001001166027} {0.0001001173747} 
\ReUpdatedNEWESTMACRO{74} {0.00002693189252} {0.00008875039937} {0.00009414668731} {0.00009346488944} {0.00009351370610} {0.00009351116875} {0.00009351161451} {0.00009351228591} 
\ReUpdatedNEWESTMACRO{75} {0.00002508961004} {0.00008295732519} {0.00008796663846} {0.00008733805916} {0.00008738276627} {0.00008738045807} {0.00008738086086} {0.00008738144519} 
\ReUpdatedNEWESTMACRO{76} {0.00002338435676} {0.00007757545647} {0.00008222796063} {0.00008164809185} {0.00008168906396} {0.00008168696257} {0.00008168732685} {0.00008168783576} 
\ReUpdatedNEWESTMACRO{77} {0.00002180506575} {0.00007257308678} {0.00007689639567} {0.00007636114281} {0.00007639871763} {0.00007639680305} {0.00007639713279} {0.00007639757633} 
\ReUpdatedNEWESTMACRO{78} {0.00002034163869} {0.00006792116092} {0.00007194057038} {0.00007144621139} {0.00007148069366} {0.00007147894796} {0.00007147924668} {0.00007147963351} 
\ReUpdatedNEWESTMACRO{79} {0.00001898485421} {0.00006359303508} {0.00006733173349} {0.00006687488227} {0.00006690654723} {0.00006690495434} {0.00006690522518} {0.00006690556278} 
\ReUpdatedNEWESTMACRO{80} {0.00001772628561} {0.00005956426027} {0.00006304351857} {0.00006262109222} {0.00006265018869} {0.00006264873419} {0.00006264897995} {0.00006264927477} 
\ReUpdatedNEWESTMACRO{81} {0.00001655822685} {0.00005581238694} {0.00005905173003} {0.00005866091959} {0.00005868767269} {0.00005868634362} {0.00005868656680} {0.00005868682442} 
\ReUpdatedNEWESTMACRO{82} {0.00001547362605} {0.00005231678815} {0.00005533414987} {0.00005497239407} {0.00005499700765} {0.00005499579236} {0.00005499599518} {0.00005499622043} 
\ReUpdatedNEWESTMACRO{83} {0.00001446602551} {0.00004905849977} {0.00005187036297} {0.00005153532468} {0.00005155798349} {0.00005155687147} {0.00005155705594} {0.00005155725300} 
\ReUpdatedNEWESTMACRO{84} {0.00001352950768} {0.00004602007569} {0.00004864159906} {0.00004833114414} {0.00004835201573} {0.00004835099755} {0.00004835116544} {0.00004835133795} 
\ReUpdatedNEWESTMACRO{85} {0.00001265864647} {0.00004318545662} {0.00004563058954} {0.00004534276787} {0.00004536200435} {0.00004536107147} {0.00004536122438} {0.00004536137547} 
\ReUpdatedNEWESTMACRO{86} {0.00001184846318} {0.00004053985115} {0.00004282143782} {0.00004255446620} {0.00004257220572} {0.00004257135045} {0.00004257148983} {0.00004257162223} 
\ReUpdatedNEWESTMACRO{87} {0.00001109438680} {0.00003806962771} {0.00004019950164} {0.00003995174843} {0.00003996811658} {0.00003996733198} {0.00003996745910} {0.00003996757518} 
\ReUpdatedNEWESTMACRO{88} {0.00001039221797} {0.00003576221656} {0.00003775128635} {0.00003752125755} {0.00003753636860} {0.00003753564839} {0.00003753576442} {0.00003753586624} 
\ReUpdatedNEWESTMACRO{89} {9.738096494$\times 10\uu{-6}$} {0.00003360602056} {0.00003546434784} {0.00003525067466} {0.00003526463263} {0.00003526397113} {0.00003526407710} {0.00003526416645} 
\ReUpdatedNEWESTMACRO{90} {9.128471804$\times 10\uu{-6}$} {0.00003159033412} {0.00003332720441} {0.00003312863197} {0.00003314153162} {0.00003314092369} {0.00003314102053} {0.00003314109898}

\end{tabular}
}
\captionof{table}{The successive refined estimates for the massive macroscopic propagation function \bbd{q\ll n\uprm{\textsc{mmp}} \equiv q\ll n - q\ll n\uprm{eft}}, at the coupling \bbd{\t = {{25 i}\over{\pi}}}.  The first column, "\bbd{e\uu{- S\lrm{WLI}}} with prefactor" is the approximation
\bbd{q\ll n\uprm{\textsc{mmp}} \simeq \big[\cc {{2\uu{13}\cc n}\over{\pi\uu 5\cc {\rm Im}[\s]}}\cc \big ]\uu{+{1\over 4}}\cc \exp{- \sqrt{{8\pi n}\over{{\rm Im}[\s]}}}}, with \bbd{\s} denoting the infrared coupling related to the UV coupling \bbd{\t} by \bbd{e\uu{2\pi i \t} = \l[\s],} where \bbd{\l[\s]} is the modular lambda function.  The value of \bbd{\t} in this table is \bbd{\t = {{25}\over\pi}\cc i}.  In each successive column we add a correction of the form \bbd{-n\uu{-{p\over 2}}\cc w\ll p[\s]} to the
exponent of the exponential for \bbd{p=1,\cdots 5}.  The functions
\bbd{w\ll p[\s]} are given explictly in \WFunxLocation}.
\label{TypesOfLiouvilleAmplitudeTransformationTableSpecificParameterChoiceOfInterestToUs}
\end{center}
\end{table}

\newpage
 \vskip-1in
 \begin{table}[H]
\begin{center}
\resizebox{\columnwidth}{!}{%
\begin{tabular}{ |c||c|c|c|c|c|c|c| } 
 \hline\hline
 ~ & ${{e\uu{-S\lrm{WLI}}{\rm ~w}/}\atop{\rm prefactor}}$ 
 & ${{\rm estimate~w/}\atop{O(n\uu{-\hh})}}$ &${{\rm estimate~w/}\atop{O(n\uu{-1})}}$  &${{\rm estimate~w/}\atop{O(n\uu{-{3\over 2}})}}$ &${{\rm estimate~w/}\atop{O(n\uu{-2})}}$ & ${{\rm estimate~w/}\atop{O(n\uu{-{5\over 2}})}} $ & ${{\rm exact}\atop{q\ll n\uprm{\textsc{mmp}}}}$ \\ \hline\hline

\ReUpdatedNEWESTMACRO{91} {8.560076236$\times 10\uu{-6}$} {0.00002970526931} {0.00003132925656} {0.00003114463367} {0.00003115656141} {0.00003115600238} {0.00003115609094} {0.00003115615984} 
\ReUpdatedNEWESTMACRO{92} {8.029900776$\times 10\uu{-6}$} {0.00002794168866} {0.00002946071395} {0.00002928898394} {0.00002930001857} {0.00002929950422} {0.00002929958526} {0.00002929964580} 
\ReUpdatedNEWESTMACRO{93} {7.535173031$\times 10\uu{-6}$} {0.00002629114378} {0.00002771252889} {0.00002755272122} {0.00002756293471} {0.00002756246120} {0.00002756253540} {0.00002756258862} 
\ReUpdatedNEWESTMACRO{94} {7.073337222$\times 10\uu{-6}$} {0.00002474581949} {0.00002607633566} {0.00002592755843} {0.00002593701650} {0.00002593658035} {0.00002593664834} {0.00002593669514} 
\ReUpdatedNEWESTMACRO{95} {6.642035986$\times 10\uu{-6}$} {0.00002329848272} {0.00002454439522} {0.00002440582830} {0.00002441459104} {0.00002441418909} {0.00002441425141} {0.00002441429258} 
\ReUpdatedNEWESTMACRO{96} {6.239093820$\times 10\uu{-6}$} {0.00002194243589} {0.00002310954469} {0.00002298043354} {0.00002298855589} {0.00002298818526} {0.00002298824243} {0.00002298827865} 
\ReUpdatedNEWESTMACRO{97} {5.862502010$\times 10\uu{-6}$} {0.00002067147429} {0.00002176515122} {0.00002164480122} {0.00002165233348} {0.00002165199155} {0.00002165204402} {0.00002165207590} 
\ReUpdatedNEWESTMACRO{98} {5.510404894$\times 10\uu{-6}$} {0.00001947984709} {0.00002050506980} {0.00002039284119} {0.00002039982941} {0.00002039951380} {0.00002039956198} {0.00002039959005} 
\ReUpdatedNEWESTMACRO{99} {5.181087339$\times 10\uu{-6}$} {0.00001836222166} {0.00001932360471} {0.00001921890791} {0.00001922539430} {0.00001922510284} {0.00001922514711} {0.00001922517183} 
\ReUpdatedNEWESTMACRO{100} {4.872963321$\times 10\uu{-6}$} {0.00001731365088} {0.00001821547422} {0.00001811776564} {0.00001812378889} {0.00001812351960} {0.00001812356029} {0.00001812358207} 
\ReUpdatedNEWESTMACRO{101} {4.584565488$\times 10\uu{-6}$} {0.00001632954327} {0.00001717577827} {0.00001708455643} {0.00001709015203} {0.00001708990310} {0.00001708994053} {0.00001708995971} 
\ReUpdatedNEWESTMACRO{102} {4.314535640$\times 10\uu{-6}$} {0.00001540563547} {0.00001619996883} {0.00001611477088} {0.00001611997141} {0.00001611974120} {0.00001611977565} {0.00001611979255} 
\ReUpdatedNEWESTMACRO{103} {4.061616018$\times 10\uu{-6}$} {0.00001453796714} {0.00001528382277} {0.00001520422131} {0.00001520905670} {0.00001520884369} {0.00001520887541} {0.00001520889031} 
\ReUpdatedNEWESTMACRO{104} {3.824641337$\times 10\uu{-6}$} {0.00001372285780} {0.00001442341693} {0.00001434901710} {0.00001435351484} {0.00001435331766} {0.00001435334688} {0.00001435336001} 
\ReUpdatedNEWESTMACRO{105} {3.602531502$\times 10\uu{-6}$} {0.00001295688564} {0.00001361510529} {0.00001354554210} {0.00001354972747} {0.00001354954486} {0.00001354957179} {0.00001354958336} 
\ReUpdatedNEWESTMACRO{106} {3.394284932$\times 10\uu{-6}$} {0.00001223686802} {0.00001285549794} {0.00001279043386} {0.00001279433010} {0.00001279416091} {0.00001279418575} {0.00001279419595} 
\ReUpdatedNEWESTMACRO{107} {3.198972443$\times 10\uu{-6}$} {0.00001155984356} {0.00001214144179} {0.00001208056453} {0.00001208419306} {0.00001208403624} {0.00001208405915} {0.00001208406814} 
\ReUpdatedNEWESTMACRO{108} {3.015731647$\times 10\uu{-6}$} {0.00001092305564} {0.00001147000284} {0.00001141302337} {0.00001141640388} {0.00001141625845} {0.00001141627960} {0.00001141628753} 
\ReUpdatedNEWESTMACRO{109} {2.843761805$\times 10\uu{-6}$} {0.00001032393722} {0.00001083844988} {0.00001078510051} {0.00001078825118} {0.00001078811626} {0.00001078813579} {0.00001078814278} 
\ReUpdatedNEWESTMACRO{110} {2.682319115$\times 10\uu{-6}$} {9.760096897$\times 10\uu{-6}$} {0.00001024423944} {0.00001019427220} {0.00001019720974} {0.00001019708452} {0.00001019710256} {0.00001019710872} 
\ReUpdatedNEWESTMACRO{111} {2.530712374$\times 10\uu{-6}$} {9.229306029$\times 10\uu{-6}$} {9.685001983$\times 10\uu{-6}$} {9.638187027$\times 10\uu{-6}$} {9.640926874$\times 10\uu{-6}$} {9.640810611$\times 10\uu{-6}$} {9.640827288$\times 10\uu{-6}$} {9.640832718$\times 10\uu{-6}$} 
\ReUpdatedNEWESTMACRO{112} {2.388299000$\times 10\uu{-6}$} {8.729486868$\times 10\uu{-6}$} {9.158529175$\times 10\uu{-6}$} {9.114653382$\times 10\uu{-6}$} {9.117209776$\times 10\uu{-6}$} {9.117101782$\times 10\uu{-6}$} {9.117117204$\times 10\uu{-6}$} {9.117121989$\times 10\uu{-6}$} 
\ReUpdatedNEWESTMACRO{113} {2.254481374$\times 10\uu{-6}$} {8.258701627$\times 10\uu{-6}$} {8.662762128$\times 10\uu{-6}$} {8.621627798$\times 10\uu{-6}$} {8.624013880$\times 10\uu{-6}$} {8.623913529$\times 10\uu{-6}$} {8.623927795$\times 10\uu{-6}$} {8.623932012$\times 10\uu{-6}$} 
\ReUpdatedNEWESTMACRO{114} {2.128703475$\times 10\uu{-6}$} {7.815142395$\times 10\uu{-6}$} {8.195780584$\times 10\uu{-6}$} {8.157204245$\times 10\uu{-6}$} {8.159432152$\times 10\uu{-6}$} {8.159338865$\times 10\uu{-6}$} {8.159352069$\times 10\uu{-6}$} {8.159355782$\times 10\uu{-6}$} 
\ReUpdatedNEWESTMACRO{115} {2.010447781$\times 10\uu{-6}$} {7.397121821$\times 10\uu{-6}$} {7.755792923$\times 10\uu{-6}$} {7.719604237$\times 10\uu{-6}$} {7.721685182$\times 10\uu{-6}$} {7.721598428$\times 10\uu{-6}$} {7.721610654$\times 10\uu{-6}$} {7.721613924$\times 10\uu{-6}$} 
\ReUpdatedNEWESTMACRO{116} {1.899232426$\times 10\uu{-6}$} {7.003064516$\times 10\uu{-6}$} {7.341126943$\times 10\uu{-6}$} {7.307167705$\times 10\uu{-6}$} {7.309112051$\times 10\uu{-6}$} {7.309031343$\times 10\uu{-6}$} {7.309042667$\times 10\uu{-6}$} {7.309045546$\times 10\uu{-6}$} 
\ReUpdatedNEWESTMACRO{117} {1.794608567$\times 10\uu{-6}$} {6.631499111$\times 10\uu{-6}$} {6.950221352$\times 10\uu{-6}$} {6.918344563$\times 10\uu{-6}$} {6.920161895$\times 10\uu{-6}$} {6.920086781$\times 10\uu{-6}$} {6.920097276$\times 10\uu{-6}$} {6.920099809$\times 10\uu{-6}$} 
\ReUpdatedNEWESTMACRO{118} {1.696157973$\times 10\uu{-6}$} {6.281050911$\times 10\uu{-6}$} {6.581617895$\times 10\uu{-6}$} {6.551686917$\times 10\uu{-6}$} {6.553386101$\times 10\uu{-6}$} {6.553316169$\times 10\uu{-6}$} {6.553325898$\times 10\uu{-6}$} {6.553328125$\times 10\uu{-6}$} 
\ReUpdatedNEWESTMACRO{119} {1.603490790$\times 10\uu{-6}$} {5.950435111$\times 10\uu{-6}$} {6.233954090$\times 10\uu{-6}$} {6.205841867$\times 10\uu{-6}$} {6.207431107$\times 10\uu{-6}$} {6.207365976$\times 10\uu{-6}$} {6.207374999$\times 10\uu{-6}$} {6.207376957$\times 10\uu{-6}$} 
\ReUpdatedNEWESTMACRO{120} {1.516243486$\times 10\uu{-6}$} {5.638450512$\times 10\uu{-6}$} {5.905956503$\times 10\uu{-6}$} {5.879544842$\times 10\uu{-6}$} {5.881031737$\times 10\uu{-6}$} {5.880971055$\times 10\uu{-6}$} {5.880979426$\times 10\uu{-6}$} {5.880981146$\times 10\uu{-6}$}

\end{tabular}
}
\captionof{table}{The successive refined estimates for the massive macroscopic propagation function \bbd{q\ll n\uprm{\textsc{mmp}} \equiv q\ll n - q\ll n\uprm{eft}}, at the coupling \bbd{\t = {{25 i}\over{\pi}}}.  The first column, "\bbd{e\uu{- S\lrm{WLI}}} with prefactor" is the approximation
\bbd{q\ll n\uprm{\textsc{mmp}} \simeq \big[\cc {{2\uu{13}\cc n}\over{\pi\uu 5\cc {\rm Im}[\s]}}\cc \big ]\uu{+{1\over 4}}\cc \exp{- \sqrt{{8\pi n}\over{{\rm Im}[\s]}}}}, with \bbd{\s} denoting the infrared coupling related to the UV coupling \bbd{\t} by \bbd{e\uu{2\pi i \t} = \l[\s],} where \bbd{\l[\s]} is the modular lambda function.  The value of \bbd{\t} in this table is \bbd{\t = {{25}\over\pi}\cc i}.  In each successive column we add a correction of the form \bbd{-n\uu{-{p\over 2}}\cc w\ll p[\s]} to the
exponent of the exponential for \bbd{p=1,\cdots 5}.  The functions
\bbd{w\ll p[\s]} are given explictly in \WFunxLocation}.
\label{TypesOfLiouvilleAmplitudeTransformationTableSpecificParameterChoiceOfInterestToUs}
\end{center}
\end{table}

\newpage
 \vskip-1in
 \begin{table}[H]
\begin{center}
\resizebox{\columnwidth}{!}{%
\begin{tabular}{ |c||c|c|c|c|c|c|c| } 
 \hline\hline
 ~ & ${{e\uu{-S\lrm{WLI}}{\rm ~w}/}\atop{\rm prefactor}}$ 
 & ${{\rm estimate~w/}\atop{O(n\uu{-\hh})}}$ &${{\rm estimate~w/}\atop{O(n\uu{-1})}}$  &${{\rm estimate~w/}\atop{O(n\uu{-{3\over 2}})}}$ &${{\rm estimate~w/}\atop{O(n\uu{-2})}}$ & ${{\rm estimate~w/}\atop{O(n\uu{-{5\over 2}})}} $ & ${{\rm exact}\atop{q\ll n\uprm{\textsc{mmp}}}}$ \\ \hline\hline

\ReUpdatedNEWESTMACRO{121} {1.434076954$\times 10\uu{-6}$} {5.343973708$\times 10\uu{-6}$} {5.596434531$\times 10\uu{-6}$} {5.571613442$\times 10\uu{-6}$} {5.573005031$\times 10\uu{-6}$} {5.572948473$\times 10\uu{-6}$} {5.572956243$\times 10\uu{-6}$} {5.572957753$\times 10\uu{-6}$} 
\ReUpdatedNEWESTMACRO{122} {1.356674762$\times 10\uu{-6}$} {5.065953711$\times 10\uu{-6}$} {5.304274642$\times 10\uu{-6}$} {5.280941735$\times 10\uu{-6}$} {5.282244539$\times 10\uu{-6}$} {5.282191807$\times 10\uu{-6}$} {5.282199022$\times 10\uu{-6}$} {5.282200346$\times 10\uu{-6}$} 
\ReUpdatedNEWESTMACRO{123} {1.283741531$\times 10\uu{-6}$} {4.803406960$\times 10\uu{-6}$} {5.028435053$\times 10\uu{-6}$} {5.006494974$\times 10\uu{-6}$} {5.007715039$\times 10\uu{-6}$} {5.007665857$\times 10\uu{-6}$} {5.007672559$\times 10\uu{-6}$} {5.007673719$\times 10\uu{-6}$} 
\ReUpdatedNEWESTMACRO{124} {1.215001444$\times 10\uu{-6}$} {4.555412707$\times 10\uu{-6}$} {4.767940789$\times 10\uu{-6}$} {4.747304706$\times 10\uu{-6}$} {4.748447639$\times 10\uu{-6}$} {4.748401753$\times 10\uu{-6}$} {4.748407980$\times 10\uu{-6}$} {4.748408996$\times 10\uu{-6}$} 
\ReUpdatedNEWESTMACRO{125} {1.150196861$\times 10\uu{-6}$} {4.321108734$\times 10\uu{-6}$} {4.521879112$\times 10\uu{-6}$} {4.502464239$\times 10\uu{-6}$} {4.503535242$\times 10\uu{-6}$} {4.503492416$\times 10\uu{-6}$} {4.503498205$\times 10\uu{-6}$} {4.503499093$\times 10\uu{-6}$} 
\ReUpdatedNEWESTMACRO{126} {1.089087043$\times 10\uu{-6}$} {4.099687382$\times 10\uu{-6}$} {4.289395279$\times 10\uu{-6}$} {4.271124437$\times 10\uu{-6}$} {4.272128338$\times 10\uu{-6}$} {4.272088355$\times 10\uu{-6}$} {4.272093738$\times 10\uu{-6}$} {4.272094514$\times 10\uu{-6}$} 
\ReUpdatedNEWESTMACRO{127} {1.031446972$\times 10\uu{-6}$} {3.890391871$\times 10\uu{-6}$} {4.069688609$\times 10\uu{-6}$} {4.052489823$\times 10\uu{-6}$} {4.053431107$\times 10\uu{-6}$} {4.053393766$\times 10\uu{-6}$} {4.053398773$\times 10\uu{-6}$} {4.053399450$\times 10\uu{-6}$} 
\ReUpdatedNEWESTMACRO{128} {9.770662515$\times 10\uu{-7}$} {3.692512878$\times 10\uu{-6}$} {3.862008840$\times 10\uu{-6}$} {3.845814964$\times 10\uu{-6}$} {3.846697794$\times 10\uu{-6}$} {3.846662909$\times 10\uu{-6}$} {3.846667569$\times 10\uu{-6}$} {3.846668157$\times 10\uu{-6}$} 
\ReUpdatedNEWESTMACRO{129} {9.257480979$\times 10\uu{-7}$} {3.505385368$\times 10\uu{-6}$} {3.665652736$\times 10\uu{-6}$} {3.650401109$\times 10\uu{-6}$} {3.651229354$\times 10\uu{-6}$} {3.651196753$\times 10\uu{-6}$} {3.651201091$\times 10\uu{-6}$} {3.651201602$\times 10\uu{-6}$} 
\ReUpdatedNEWESTMACRO{130} {8.773084006$\times 10\uu{-7}$} {3.328385640$\times 10\uu{-6}$} {3.479960950$\times 10\uu{-6}$} {3.465593075$\times 10\uu{-6}$} {3.466370332$\times 10\uu{-6}$} {3.466339856$\times 10\uu{-6}$} {3.466343896$\times 10\uu{-6}$} {3.466344339$\times 10\uu{-6}$} 
\ReUpdatedNEWESTMACRO{131} {8.315748527$\times 10\uu{-7}$} {3.160928593$\times 10\uu{-6}$} {3.304315100$\times 10\uu{-6}$} {3.290776351$\times 10\uu{-6}$} {3.291505965$\times 10\uu{-6}$} {3.291477466$\times 10\uu{-6}$} {3.291481229$\times 10\uu{-6}$} {3.291481613$\times 10\uu{-6}$} 
\ReUpdatedNEWESTMACRO{132} {7.883861443$\times 10\uu{-7}$} {3.002465179$\times 10\uu{-6}$} {3.138135057$\times 10\uu{-6}$} {3.125374406$\times 10\uu{-6}$} {3.126059487$\times 10\uu{-6}$} {3.126032830$\times 10\uu{-6}$} {3.126036336$\times 10\uu{-6}$} {3.126036667$\times 10\uu{-6}$} 
\ReUpdatedNEWESTMACRO{133} {7.475912161$\times 10\uu{-7}$} {2.852480037$\times 10\uu{-6}$} {2.980876423$\times 10\uu{-6}$} {2.968846188$\times 10\uu{-6}$} {2.969489632$\times 10\uu{-6}$} {2.969464689$\times 10\uu{-6}$} {2.969467958$\times 10\uu{-6}$} {2.969468243$\times 10\uu{-6}$} 
\ReUpdatedNEWESTMACRO{134} {7.090485648$\times 10\uu{-7}$} {2.710489289$\times 10\uu{-6}$} {2.832028189$\times 10\uu{-6}$} {2.820683800$\times 10\uu{-6}$} {2.821288301$\times 10\uu{-6}$} {2.821264956$\times 10\uu{-6}$} {2.821268003$\times 10\uu{-6}$} {2.821268247$\times 10\uu{-6}$} 
\ReUpdatedNEWESTMACRO{135} {6.726256004$\times 10\uu{-7}$} {2.576038494$\times 10\uu{-6}$} {2.691110548$\times 10\uu{-6}$} {2.680410332$\times 10\uu{-6}$} {2.680978400$\times 10\uu{-6}$} {2.680956543$\times 10\uu{-6}$} {2.680959386$\times 10\uu{-6}$} {2.680959594$\times 10\uu{-6}$} 
\ReUpdatedNEWESTMACRO{136} {6.381980481$\times 10\uu{-7}$} {2.448700739$\times 10\uu{-6}$} {2.557672873$\times 10\uu{-6}$} {2.547577851$\times 10\uu{-6}$} {2.548111823$\times 10\uu{-6}$} {2.548091353$\times 10\uu{-6}$} {2.548094006$\times 10\uu{-6}$} {2.548094183$\times 10\uu{-6}$} 
\ReUpdatedNEWESTMACRO{137} {6.056493938$\times 10\uu{-7}$} {2.328074866$\times 10\uu{-6}$} {2.431291819$\times 10\uu{-6}$} {2.421765526$\times 10\uu{-6}$} {2.422267580$\times 10\uu{-6}$} {2.422248404$\times 10\uu{-6}$} {2.422250880$\times 10\uu{-6}$} {2.422251029$\times 10\uu{-6}$} 
\ReUpdatedNEWESTMACRO{138} {5.748703680$\times 10\uu{-7}$} {2.213783817$\times 10\uu{-6}$} {2.311569574$\times 10\uu{-6}$} {2.302577881$\times 10\uu{-6}$} {2.303050047$\times 10\uu{-6}$} {2.303032078$\times 10\uu{-6}$} {2.303034389$\times 10\uu{-6}$} {2.303034515$\times 10\uu{-6}$} 
\ReUpdatedNEWESTMACRO{139} {5.457584671$\times 10\uu{-7}$} {2.105473094$\times 10\uu{-6}$} {2.198132213$\times 10\uu{-6}$} {2.189643171$\times 10\uu{-6}$} {2.190087340$\times 10\uu{-6}$} {2.190070498$\times 10\uu{-6}$} {2.190072657$\times 10\uu{-6}$} {2.190072761$\times 10\uu{-6}$} 
\ReUpdatedNEWESTMACRO{140} {5.182175076$\times 10\uu{-7}$} {2.002809320$\times 10\uu{-6}$} {2.090628173$\times 10\uu{-6}$} {2.082611864$\times 10\uu{-6}$} {2.083029804$\times 10\uu{-6}$} {2.083014012$\times 10\uu{-6}$} {2.083016029$\times 10\uu{-6}$} {2.083016115$\times 10\uu{-6}$} 
\ReUpdatedNEWESTMACRO{141} {4.921572120$\times 10\uu{-7}$} {1.905478906$\times 10\uu{-6}$} {1.988726831$\times 10\uu{-6}$} {1.981155232$\times 10\uu{-6}$} {1.981548589$\times 10\uu{-6}$} {1.981533780$\times 10\uu{-6}$} {1.981535665$\times 10\uu{-6}$} {1.981535734$\times 10\uu{-6}$} 
\ReUpdatedNEWESTMACRO{142} {4.674928229$\times 10\uu{-7}$} {1.813186795$\times 10\uu{-6}$} {1.892117180$\times 10\uu{-6}$} {1.884964034$\times 10\uu{-6}$} {1.885334345$\times 10\uu{-6}$} {1.885320452$\times 10\uu{-6}$} {1.885322214$\times 10\uu{-6}$} {1.885322270$\times 10\uu{-6}$} 
\ReUpdatedNEWESTMACRO{143} {4.441447446$\times 10\uu{-7}$} {1.725655303$\times 10\uu{-6}$} {1.800506585$\times 10\uu{-6}$} {1.793747283$\times 10\uu{-6}$} {1.794095984$\times 10\uu{-6}$} {1.794082948$\times 10\uu{-6}$} {1.794084595$\times 10\uu{-6}$} {1.794084639$\times 10\uu{-6}$} 
\ReUpdatedNEWESTMACRO{144} {4.220382086$\times 10\uu{-7}$} {1.642623030$\times 10\uu{-6}$} {1.713619633$\times 10\uu{-6}$} {1.707231106$\times 10\uu{-6}$} {1.707559537$\times 10\uu{-6}$} {1.707547301$\times 10\uu{-6}$} {1.707548842$\times 10\uu{-6}$} {1.707548876$\times 10\uu{-6}$} 
\ReUpdatedNEWESTMACRO{145} {4.011029631$\times 10\uu{-7}$} {1.563843842$\times 10\uu{-6}$} {1.631197057$\times 10\uu{-6}$} {1.625157670$\times 10\uu{-6}$} {1.625467083$\times 10\uu{-6}$} {1.625455596$\times 10\uu{-6}$} {1.625457038$\times 10\uu{-6}$} {1.625457062$\times 10\uu{-6}$} 
\ReUpdatedNEWESTMACRO{146} {3.812729827$\times 10\uu{-7}$} {1.489085927$\times 10\uu{-6}$} {1.552994725$\times 10\uu{-6}$} {1.547284187$\times 10\uu{-6}$} {1.547575752$\times 10\uu{-6}$} {1.547564964$\times 10\uu{-6}$} {1.547566314$\times 10\uu{-6}$} {1.547566330$\times 10\uu{-6}$} 
\ReUpdatedNEWESTMACRO{147} {3.624861985$\times 10\uu{-7}$} {1.418130909$\times 10\uu{-6}$} {1.478782705$\times 10\uu{-6}$} {1.473381976$\times 10\uu{-6}$} {1.473656786$\times 10\uu{-6}$} {1.473646653$\times 10\uu{-6}$} {1.473647916$\times 10\uu{-6}$} {1.473647926$\times 10\uu{-6}$} 
\ReUpdatedNEWESTMACRO{148} {3.446842464$\times 10\uu{-7}$} {1.350773019$\times 10\uu{-6}$} {1.408344385$\times 10\uu{-6}$} {1.403235597$\times 10\uu{-6}$} {1.403494676$\times 10\uu{-6}$} {1.403485155$\times 10\uu{-6}$} {1.403486338$\times 10\uu{-6}$} {1.403486342$\times 10\uu{-6}$} 
\ReUpdatedNEWESTMACRO{149} {3.278122322$\times 10\uu{-7}$} {1.286818322$\times 10\uu{-6}$} {1.341475655$\times 10\uu{-6}$} {1.336642035$\times 10\uu{-6}$} {1.336886338$\times 10\uu{-6}$} {1.336877391$\times 10\uu{-6}$} {1.336878498$\times 10\uu{-6}$} {1.336878498$\times 10\uu{-6}$} 
\ReUpdatedNEWESTMACRO{150} {3.118185125$\times 10\uu{-7}$} {1.226083996$\times 10\uu{-6}$} {1.277984141$\times 10\uu{-6}$} {1.273409938$\times 10\uu{-6}$} {1.273640360$\times 10\uu{-6}$} {1.273631949$\times 10\uu{-6}$} {1.273632987$\times 10\uu{-6}$} {1.273632982$\times 10\uu{-6}$} 

\end{tabular}
}
\captionof{table}{The successive refined estimates for the massive macroscopic propagation function \bbd{q\ll n\uprm{\textsc{mmp}} \equiv q\ll n - q\ll n\uprm{eft}}, at the coupling \bbd{\t = {{25 i}\over{\pi}}}.  The first column, "\bbd{e\uu{- S\lrm{WLI}}} with prefactor" is the approximation
\bbd{q\ll n\uprm{\textsc{mmp}} \simeq \big[\cc {{2\uu{13}\cc n}\over{\pi\uu 5\cc {\rm Im}[\s]}}\cc \big ]\uu{+{1\over 4}}\cc \exp{- \sqrt{{8\pi n}\over{{\rm Im}[\s]}}}}, with \bbd{\s} denoting the infrared coupling related to the UV coupling \bbd{\t} by \bbd{e\uu{2\pi i \t} = \l[\s],} where \bbd{\l[\s]} is the modular lambda function.  The value of \bbd{\t} in this table is \bbd{\t = {{25}\over\pi}\cc i}.  In each successive column we add a correction of the form \bbd{-n\uu{-{p\over 2}}\cc w\ll p[\s]} to the
exponent of the exponential for \bbd{p=1,\cdots 5}.  The functions
\bbd{w\ll p[\s]} are given explictly in \WFunxLocation}.
\label{TypesOfLiouvilleAmplitudeTransformationTableSpecificParameterChoiceOfInterestToUs}
\end{center}
\end{table}
\newpage

\subsection{Table of effective digits of accuracy of the fixed-\bbd{\t} large-charge asymptotic estimates at \bbd{\t = {{25}\over \pi}\cc i}, up to \bbd{n=150}}

 \begin{table}[H]
\begin{center}
\begin{tabular}{ |c||c|c|c|c|c|c|c| } 
 \hline\hline
 ~ & ${{\rm accuracy}\atop{{\rm of~}e\uu{- S\lrm{WLI}}}}$  & ${{\rm accuracy~w/}\atop{\rm prefactor}}$ 
 & ${{\rm accuracy~w/}\atop{O(n\uu{-\hh})}}$ &${{\rm accuracy~w/}\atop{O(n\uu{-1})}}$  &${{\rm accuracy~w/}\atop{O(n\uu{-{3\over 2}})}}$ &${{\rm accuracy~w/}\atop{O(n\uu{-2})}}$ &${{\rm accuracy~w/}\atop{O(n\uu{-{5\over 2}})}}$ \\ \hline\hline

\EVENSOMEWHATSHORTERMACRO{1} {0.5091} {0.6490} {0.5395} {0.6067} {1.306} {1.020} {0.8787}
\EVENSOMEWHATSHORTERMACRO{2} {0.4049} {0.7215} {0.7957} {0.9451} {2.534} {1.406} {3.420}
\EVENSOMEWHATSHORTERMACRO{3} {0.3582} {0.7708} {0.9454} {1.159} {2.105} {1.618} {2.002}
\EVENSOMEWHATSHORTERMACRO{4} {0.3294} {0.8090} {1.051} {1.318} {2.114} {1.773} {2.003}
\EVENSOMEWHATSHORTERMACRO{5} {0.3091} {0.8405} {1.132} {1.446} {2.175} {1.901} {2.066}
\EVENSOMEWHATSHORTERMACRO{6} {0.2937} {0.8672} {1.199} {1.555} {2.249} {2.012} {2.142}
\EVENSOMEWHATSHORTERMACRO{7} {0.2813} {0.8904} {1.255} {1.649} {2.325} {2.111} {2.220}
\EVENSOMEWHATSHORTERMACRO{8} {0.2710} {0.9110} {1.303} {1.733} {2.402} {2.202} {2.297}
\EVENSOMEWHATSHORTERMACRO{9} {0.2623} {0.9295} {1.346} {1.809} {2.477} {2.287} {2.372}
\EVENSOMEWHATSHORTERMACRO{10} {0.2548} {0.9462} {1.384} {1.878} {2.550} {2.367} {2.445}
\EVENSOMEWHATSHORTERMACRO{11} {0.2482} {0.9615} {1.419} {1.942} {2.621} {2.442} {2.515}
\EVENSOMEWHATSHORTERMACRO{12} {0.2423} {0.9756} {1.451} {2.001} {2.691} {2.514} {2.582}
\EVENSOMEWHATSHORTERMACRO{13} {0.2370} {0.9887} {1.480} {2.056} {2.760} {2.582} {2.648}
\EVENSOMEWHATSHORTERMACRO{14} {0.2323} {1.001} {1.508} {2.107} {2.826} {2.648} {2.712}
\EVENSOMEWHATSHORTERMACRO{15} {0.2279} {1.012} {1.534} {2.156} {2.892} {2.712} {2.773}
\EVENSOMEWHATSHORTERMACRO{16} {0.2239} {1.023} {1.558} {2.202} {2.956} {2.773} {2.833}
\EVENSOMEWHATSHORTERMACRO{17} {0.2202} {1.033} {1.580} {2.245} {3.019} {2.832} {2.892}
\EVENSOMEWHATSHORTERMACRO{18} {0.2168} {1.043} {1.602} {2.287} {3.081} {2.890} {2.949}
\EVENSOMEWHATSHORTERMACRO{19} {0.2136} {1.052} {1.622} {2.326} {3.142} {2.946} {3.005}
\EVENSOMEWHATSHORTERMACRO{20} {0.2106} {1.061} {1.642} {2.363} {3.203} {3.001} {3.059}

\end{tabular}
\captionof{table}{\footnotesize{Effective number of digits of accuracy of each successive refined estimates for the massive macroscopic propagation function \bbd{q\ll n\uprm{\textsc{mmp}} \equiv q\ll n - q\ll n\uprm{eft}}, at the coupling \bbd{\t = {{25 i}\over{\pi}}}.  The first column, "accuracy of \bbd{e\uu{- S\lrm{WLI}}}" is the approximation
\bbd{q\ll n\uprm{\textsc{mmp}} \simeq \exp{- \sqrt{{8\pi n}\over{{\rm Im}[\s]}}}}, with \bbd{\s} denoting the infrared coupling related to the UV coupling \bbd{\t} by \bbd{e\uu{2\pi i \t} = \l[\s],} where \bbd{\l[\s]} is the modular lambda function.  The value of \bbd{\t} in this table is \bbd{\t = {{25}\over\pi}\cc i}.  The second column, "\bbd{e\uu{- S\lrm{WLI}}} with prefactor" gives the accuracy of the approximation obtained by appending the prefactor
\bbd{q\ll n\uprm{\textsc{mmp}} \simeq \big[\cc {{2\uu{13}\cc n}\over{\pi\uu 5\cc {\rm Im}[\s]}}\cc \big ]\uu{+{1\over 4}}} to the exponential.  In each successive column we add a correction of the form \bbd{-n\uu{-{p\over 2}}\cc w\ll p[\s]} to the
exponent of the exponential for \bbd{p=1,\cdots 5}, and give the resulting accuracy.  The "effective number of accurate digits" is defined here as the log of the relative error of the estimate of \bbd{q\ll n\uprm{\textsc{mmp}}} , times \bbd{-{1\over{{\rm Log}[10]}}}.
That is, the table entries are given by \bbd{- {1\over{{\rm Log}[10]}}\cc {\rm Log}\left | \cc {{q\ll n\uprm{\textsc{mmp}} - (q\ll n\uprm{\textsc{mmp}})\lrm{estimate}}\over{q\ll n\uprm{\textsc{mmp}}  }}    \cc \right | }.}}
\label{TypesOfLiouvilleAmplitudeTransformationTableSpecificParameterChoiceOfInterestToUs}
\end{center}
\end{table}

\newpage
 \vskip-1in
 \begin{table}[H]
\begin{center}
\begin{tabular}{ |c||c|c|c|c|c|c|c| } 
 \hline\hline
 ~ & ${{\rm accuracy}\atop{{\rm of~}e\uu{- S\lrm{WLI}}}}$  & ${{\rm accuracy~w/}\atop{\rm prefactor}}$ 
 & ${{\rm accuracy~w/}\atop{O(n\uu{-\hh})}}$ &${{\rm accuracy~w/}\atop{O(n\uu{-1})}}$  &${{\rm accuracy~w/}\atop{O(n\uu{-{3\over 2}})}}$ &${{\rm accuracy~w/}\atop{O(n\uu{-2})}}$ &${{\rm accuracy~w/}\atop{O(n\uu{-{5\over 2}})}}$ \\ \hline\hline

\EVENSOMEWHATSHORTERMACRO{21} {0.2078} {1.069} {1.661} {2.399} {3.263} {3.054} {3.112}
\EVENSOMEWHATSHORTERMACRO{22} {0.2052} {1.077} {1.678} {2.433} {3.322} {3.106} {3.165}
\EVENSOMEWHATSHORTERMACRO{23} {0.2027} {1.084} {1.696} {2.465} {3.381} {3.157} {3.216}
\EVENSOMEWHATSHORTERMACRO{24} {0.2004} {1.092} {1.712} {2.497} {3.439} {3.207} {3.266}
\EVENSOMEWHATSHORTERMACRO{25} {0.1981} {1.099} {1.728} {2.527} {3.498} {3.255} {3.315}
\EVENSOMEWHATSHORTERMACRO{26} {0.1960} {1.106} {1.743} {2.556} {3.556} {3.303} {3.364}
\EVENSOMEWHATSHORTERMACRO{27} {0.1940} {1.112} {1.758} {2.583} {3.614} {3.350} {3.411}
\EVENSOMEWHATSHORTERMACRO{28} {0.1921} {1.118} {1.772} {2.610} {3.672} {3.395} {3.458}
\EVENSOMEWHATSHORTERMACRO{29} {0.1903} {1.124} {1.786} {2.636} {3.730} {3.440} {3.504}
\EVENSOMEWHATSHORTERMACRO{30} {0.1885} {1.130} {1.799} {2.661} {3.789} {3.485} {3.550}
\EVENSOMEWHATSHORTERMACRO{31} {0.1868} {1.136} {1.812} {2.685} {3.848} {3.528} {3.594}
\EVENSOMEWHATSHORTERMACRO{32} {0.1852} {1.142} {1.825} {2.708} {3.908} {3.571} {3.638}
\EVENSOMEWHATSHORTERMACRO{33} {0.1837} {1.147} {1.837} {2.731} {3.969} {3.613} {3.682}
\EVENSOMEWHATSHORTERMACRO{34} {0.1822} {1.152} {1.849} {2.752} {4.030} {3.654} {3.725}
\EVENSOMEWHATSHORTERMACRO{35} {0.1808} {1.157} {1.861} {2.774} {4.093} {3.695} {3.767}
\EVENSOMEWHATSHORTERMACRO{36} {0.1794} {1.162} {1.872} {2.794} {4.158} {3.735} {3.809}
\EVENSOMEWHATSHORTERMACRO{37} {0.1780} {1.167} {1.883} {2.814} {4.225} {3.774} {3.850}
\EVENSOMEWHATSHORTERMACRO{38} {0.1768} {1.172} {1.894} {2.833} {4.294} {3.813} {3.891}
\EVENSOMEWHATSHORTERMACRO{39} {0.1755} {1.177} {1.904} {2.852} {4.366} {3.852} {3.931}
\EVENSOMEWHATSHORTERMACRO{40} {0.1743} {1.181} {1.915} {2.870} {4.442} {3.889} {3.971}
\EVENSOMEWHATSHORTERMACRO{41} {0.1731} {1.185} {1.925} {2.888} {4.522} {3.927} {4.011}
\EVENSOMEWHATSHORTERMACRO{42} {0.1720} {1.190} {1.935} {2.905} {4.609} {3.963} {4.050}
\EVENSOMEWHATSHORTERMACRO{43} {0.1709} {1.194} {1.944} {2.922} {4.704} {4.000} {4.088}
\EVENSOMEWHATSHORTERMACRO{44} {0.1698} {1.198} {1.953} {2.938} {4.811} {4.035} {4.127}
\EVENSOMEWHATSHORTERMACRO{45} {0.1688} {1.202} {1.963} {2.954} {4.934} {4.071} {4.164}

\end{tabular}
\captionof{table}{\footnotesize{Effective number of digits of accuracy of each successive refined estimates for the massive macroscopic propagation function \bbd{q\ll n\uprm{\textsc{mmp}} \equiv q\ll n - q\ll n\uprm{eft}}, at the coupling \bbd{\t = {{25 i}\over{\pi}}}.  The first column, "WLI w/o prefactor" is the approximation
\bbd{q\ll n\uprm{\textsc{mmp}} \simeq \exp{- \sqrt{{8\pi n}\over{{\rm Im}[\s]}}}}, with \bbd{\s} denoting the infrared coupling related to the UV coupling \bbd{\t} by \bbd{e\uu{2\pi i \t} = \l[\s],} where \bbd{\l[\s]} is the modular lambda function.  The value of \bbd{\t} in this table is \bbd{\t = {{25}\over\pi}\cc i}.  The second column, "\bbd{e\uu{- S\lrm{WLI}}} with prefactor" gives the accuracy of the approximation obtained by appending the prefactor
\bbd{q\ll n\uprm{\textsc{mmp}} \simeq \big[\cc {{2\uu{13}\cc n}\over{\pi\uu 5\cc {\rm Im}[\s]}}\cc \big ]\uu{+{1\over 4}}} to the exponential.  In each successive column we add a correction of the form \bbd{-n\uu{-{p\over 2}}\cc w\ll p[\s]} to the
exponent of the exponential for \bbd{p=1,\cdots 5}, and give the resulting accuracy.  The "effective number of accurate digits" is defined here as the log of the relative error of the estimate of \bbd{q\ll n\uprm{\textsc{mmp}}} , times \bbd{-{1\over{{\rm Log}[10]}}}.
That is, the table entries are given by \bbd{- {1\over{{\rm Log}[10]}}\cc {\rm Log}\left | \cc {{q\ll n\uprm{\textsc{mmp}} - (q\ll n\uprm{\textsc{mmp}})\lrm{estimate}}\over{q\ll n\uprm{\textsc{mmp}}  }}    \cc \right | }.}}
\label{TypesOfLiouvilleAmplitudeTransformationTableSpecificParameterChoiceOfInterestToUs}
\end{center}
\end{table}

\newpage
 \vskip-1in
 \begin{table}[H]
\begin{center}
\begin{tabular}{ |c||c|c|c|c|c|c|c| } 
 \hline\hline
 ~ & ${{\rm accuracy}\atop{{\rm of~}e\uu{- S\lrm{WLI}}}}$  & ${{\rm accuracy~w/}\atop{\rm prefactor}}$ 
 & ${{\rm accuracy~w/}\atop{O(n\uu{-\hh})}}$ &${{\rm accuracy~w/}\atop{O(n\uu{-1})}}$  &${{\rm accuracy~w/}\atop{O(n\uu{-{3\over 2}})}}$ &${{\rm accuracy~w/}\atop{O(n\uu{-2})}}$ &${{\rm accuracy~w/}\atop{O(n\uu{-{5\over 2}})}}$ \\ \hline\hline

\EVENSOMEWHATSHORTERMACRO{46} {0.1678} {1.206} {1.972} {2.969} {5.084} {4.105} {4.202}
\EVENSOMEWHATSHORTERMACRO{47} {0.1668} {1.210} {1.980} {2.985} {5.282} {4.140} {4.239}
\EVENSOMEWHATSHORTERMACRO{48} {0.1659} {1.214} {1.989} {2.999} {5.592} {4.174} {4.276}
\EVENSOMEWHATSHORTERMACRO{49} {0.1649} {1.217} {1.998} {3.014} {6.728} {4.207} {4.312}
\EVENSOMEWHATSHORTERMACRO{50} {0.1640} {1.221} {2.006} {3.028} {5.718} {4.240} {4.349}
\EVENSOMEWHATSHORTERMACRO{51} {0.1632} {1.225} {2.014} {3.042} {5.424} {4.273} {4.384}
\EVENSOMEWHATSHORTERMACRO{52} {0.1623} {1.228} {2.022} {3.055} {5.267} {4.305} {4.420}
\EVENSOMEWHATSHORTERMACRO{53} {0.1615} {1.232} {2.030} {3.068} {5.164} {4.337} {4.455}
\EVENSOMEWHATSHORTERMACRO{54} {0.1607} {1.235} {2.038} {3.081} {5.090} {4.368} {4.490}
\EVENSOMEWHATSHORTERMACRO{55} {0.1599} {1.238} {2.045} {3.094} {5.034} {4.399} {4.525}
\EVENSOMEWHATSHORTERMACRO{56} {0.1591} {1.242} {2.053} {3.106} {4.990} {4.430} {4.559}
\EVENSOMEWHATSHORTERMACRO{57} {0.1583} {1.245} {2.060} {3.119} {4.955} {4.460} {4.594}
\EVENSOMEWHATSHORTERMACRO{58} {0.1576} {1.248} {2.067} {3.131} {4.927} {4.490} {4.628}
\EVENSOMEWHATSHORTERMACRO{59} {0.1568} {1.251} {2.074} {3.142} {4.904} {4.520} {4.661}
\EVENSOMEWHATSHORTERMACRO{60} {0.1561} {1.254} {2.081} {3.154} {4.885} {4.549} {4.695}
\EVENSOMEWHATSHORTERMACRO{61} {0.1554} {1.257} {2.088} {3.165} {4.869} {4.578} {4.728}
\EVENSOMEWHATSHORTERMACRO{62} {0.1548} {1.260} {2.095} {3.176} {4.857} {4.606} {4.761}
\EVENSOMEWHATSHORTERMACRO{63} {0.1541} {1.263} {2.101} {3.187} {4.846} {4.635} {4.794}
\EVENSOMEWHATSHORTERMACRO{64} {0.1534} {1.266} {2.108} {3.198} {4.838} {4.662} {4.827}
\EVENSOMEWHATSHORTERMACRO{65} {0.1528} {1.269} {2.114} {3.208} {4.831} {4.690} {4.859}

\end{tabular}
\captionof{table}{\footnotesize{Effective number of digits of accuracy of each successive refined estimates for the massive macroscopic propagation function \bbd{q\ll n\uprm{\textsc{mmp}} \equiv q\ll n - q\ll n\uprm{eft}}, at the coupling \bbd{\t = {{25 i}\over{\pi}}}.  The first column, "WLI w/o prefactor" is the approximation
\bbd{q\ll n\uprm{\textsc{mmp}} \simeq \exp{- \sqrt{{8\pi n}\over{{\rm Im}[\s]}}}}, with \bbd{\s} denoting the infrared coupling related to the UV coupling \bbd{\t} by \bbd{e\uu{2\pi i \t} = \l[\s],} where \bbd{\l[\s]} is the modular lambda function.  The value of \bbd{\t} in this table is \bbd{\t = {{25}\over\pi}\cc i}.  The second column, "\bbd{e\uu{- S\lrm{WLI}}} with prefactor" gives the accuracy of the approximation obtained by appending the prefactor
\bbd{q\ll n\uprm{\textsc{mmp}} \simeq \big[\cc {{2\uu{13}\cc n}\over{\pi\uu 5\cc {\rm Im}[\s]}}\cc \big ]\uu{+{1\over 4}}} to the exponential.  In each successive column we add a correction of the form \bbd{-n\uu{-{p\over 2}}\cc w\ll p[\s]} to the
exponent of the exponential for \bbd{p=1,\cdots 5}, and give the resulting accuracy.  The "effective number of accurate digits" is defined here as the log of the relative error of the estimate of \bbd{q\ll n\uprm{\textsc{mmp}}} , times \bbd{-{1\over{{\rm Log}[10]}}}.
That is, the table entries are given by \bbd{- {1\over{{\rm Log}[10]}}\cc {\rm Log}\left | \cc {{q\ll n\uprm{\textsc{mmp}} - (q\ll n\uprm{\textsc{mmp}})\lrm{estimate}}\over{q\ll n\uprm{\textsc{mmp}}  }}    \cc \right | }.}}
\label{TypesOfLiouvilleAmplitudeTransformationTableSpecificParameterChoiceOfInterestToUs}
\end{center}
\end{table}

\newpage
 \vskip-1in
 \begin{table}[H]
\begin{center}
\begin{tabular}{ |c||c|c|c|c|c|c|c| } 
 \hline\hline
 ~ & ${{\rm accuracy}\atop{{\rm of~}e\uu{- S\lrm{WLI}}}}$  & ${{\rm accuracy~w/}\atop{\rm prefactor}}$ 
 & ${{\rm accuracy~w/}\atop{O(n\uu{-\hh})}}$ &${{\rm accuracy~w/}\atop{O(n\uu{-1})}}$  &${{\rm accuracy~w/}\atop{O(n\uu{-{3\over 2}})}}$ &${{\rm accuracy~w/}\atop{O(n\uu{-2})}}$ &${{\rm accuracy~w/}\atop{O(n\uu{-{5\over 2}})}}$ \\ \hline\hline

\EVENSOMEWHATSHORTERMACRO{66} {0.1522} {1.272} {2.121} {3.218} {4.826} {4.717} {4.892}
\EVENSOMEWHATSHORTERMACRO{67} {0.1515} {1.275} {2.127} {3.229} {4.822} {4.744} {4.924}
\EVENSOMEWHATSHORTERMACRO{68} {0.1509} {1.277} {2.133} {3.238} {4.819} {4.770} {4.956}
\EVENSOMEWHATSHORTERMACRO{69} {0.1503} {1.280} {2.139} {3.248} {4.817} {4.797} {4.987}
\EVENSOMEWHATSHORTERMACRO{70} {0.1498} {1.283} {2.145} {3.258} {4.816} {4.822} {5.019}
\EVENSOMEWHATSHORTERMACRO{71} {0.1492} {1.285} {2.151} {3.267} {4.816} {4.848} {5.050}
\EVENSOMEWHATSHORTERMACRO{72} {0.1486} {1.288} {2.157} {3.277} {4.816} {4.873} {5.082}
\EVENSOMEWHATSHORTERMACRO{73} {0.1481} {1.291} {2.163} {3.286} {4.817} {4.898} {5.113}
\EVENSOMEWHATSHORTERMACRO{74} {0.1475} {1.293} {2.169} {3.295} {4.819} {4.923} {5.144}
\EVENSOMEWHATSHORTERMACRO{75} {0.1470} {1.296} {2.174} {3.304} {4.820} {4.947} {5.175}
\EVENSOMEWHATSHORTERMACRO{76} {0.1465} {1.298} {2.180} {3.313} {4.823} {4.971} {5.206}
\EVENSOMEWHATSHORTERMACRO{77} {0.1459} {1.301} {2.185} {3.322} {4.826} {4.995} {5.236}
\EVENSOMEWHATSHORTERMACRO{78} {0.1454} {1.303} {2.191} {3.330} {4.829} {5.018} {5.267}
\EVENSOMEWHATSHORTERMACRO{79} {0.1449} {1.305} {2.196} {3.339} {4.832} {5.041} {5.297}
\EVENSOMEWHATSHORTERMACRO{80} {0.1444} {1.308} {2.201} {3.347} {4.836} {5.064} {5.327}
\EVENSOMEWHATSHORTERMACRO{81} {0.1440} {1.310} {2.206} {3.355} {4.840} {5.087} {5.358}
\EVENSOMEWHATSHORTERMACRO{82} {0.1435} {1.312} {2.212} {3.363} {4.844} {5.109} {5.388}
\EVENSOMEWHATSHORTERMACRO{83} {0.1430} {1.315} {2.217} {3.371} {4.849} {5.131} {5.418}
\EVENSOMEWHATSHORTERMACRO{84} {0.1426} {1.317} {2.222} {3.379} {4.853} {5.152} {5.448}
\EVENSOMEWHATSHORTERMACRO{85} {0.1421} {1.319} {2.227} {3.387} {4.858} {5.174} {5.477}
\EVENSOMEWHATSHORTERMACRO{86} {0.1417} {1.321} {2.232} {3.395} {4.863} {5.195} {5.507}
\EVENSOMEWHATSHORTERMACRO{87} {0.1412} {1.323} {2.236} {3.402} {4.868} {5.216} {5.537}

\end{tabular}
\captionof{table}{\footnotesize{Effective number of digits of accuracy of each successive refined estimates for the massive macroscopic propagation function \bbd{q\ll n\uprm{\textsc{mmp}} \equiv q\ll n - q\ll n\uprm{eft}}, at the coupling \bbd{\t = {{25 i}\over{\pi}}}.  The first column, "WLI w/o prefactor" is the approximation
\bbd{q\ll n\uprm{\textsc{mmp}} \simeq \exp{- \sqrt{{8\pi n}\over{{\rm Im}[\s]}}}}, with \bbd{\s} denoting the infrared coupling related to the UV coupling \bbd{\t} by \bbd{e\uu{2\pi i \t} = \l[\s],} where \bbd{\l[\s]} is the modular lambda function.  The value of \bbd{\t} in this table is \bbd{\t = {{25}\over\pi}\cc i}.  The second column, "\bbd{e\uu{- S\lrm{WLI}}} with prefactor" gives the accuracy of the approximation obtained by appending the prefactor
\bbd{q\ll n\uprm{\textsc{mmp}} \simeq \big[\cc {{2\uu{13}\cc n}\over{\pi\uu 5\cc {\rm Im}[\s]}}\cc \big ]\uu{+{1\over 4}}} to the exponential.  In each successive column we add a correction of the form \bbd{-n\uu{-{p\over 2}}\cc w\ll p[\s]} to the
exponent of the exponential for \bbd{p=1,\cdots 5}, and give the resulting accuracy.  The "effective number of accurate digits" is defined here as the log of the relative error of the estimate of \bbd{q\ll n\uprm{\textsc{mmp}}} , times \bbd{-{1\over{{\rm Log}[10]}}}.
That is, the table entries are given by \bbd{- {1\over{{\rm Log}[10]}}\cc {\rm Log}\left | \cc {{q\ll n\uprm{\textsc{mmp}} - (q\ll n\uprm{\textsc{mmp}})\lrm{estimate}}\over{q\ll n\uprm{\textsc{mmp}}  }}    \cc \right | }.}}
\label{TypesOfLiouvilleAmplitudeTransformationTableSpecificParameterChoiceOfInterestToUs}
\end{center}
\end{table}

\newpage
 \vskip-1in
 \begin{table}[H]
\begin{center}
\begin{tabular}{ |c||c|c|c|c|c|c|c| } 
 \hline\hline
 ~ & ${{\rm accuracy}\atop{{\rm of~}e\uu{- S\lrm{WLI}}}}$  & ${{\rm accuracy~w/}\atop{\rm prefactor}}$ 
 & ${{\rm accuracy~w/}\atop{O(n\uu{-\hh})}}$ &${{\rm accuracy~w/}\atop{O(n\uu{-1})}}$  &${{\rm accuracy~w/}\atop{O(n\uu{-{3\over 2}})}}$ &${{\rm accuracy~w/}\atop{O(n\uu{-2})}}$ &${{\rm accuracy~w/}\atop{O(n\uu{-{5\over 2}})}}$ \\ \hline\hline

\EVENSOMEWHATSHORTERMACRO{88} {0.1408} {1.326} {2.241} {3.410} {4.873} {5.236} {5.567}
\EVENSOMEWHATSHORTERMACRO{89} {0.1403} {1.328} {2.246} {3.417} {4.879} {5.257} {5.596}
\EVENSOMEWHATSHORTERMACRO{90} {0.1399} {1.330} {2.251} {3.425} {4.884} {5.277} {5.626}
\EVENSOMEWHATSHORTERMACRO{91} {0.1395} {1.332} {2.255} {3.432} {4.890} {5.296} {5.655}
\EVENSOMEWHATSHORTERMACRO{92} {0.1391} {1.334} {2.260} {3.439} {4.895} {5.316} {5.685}
\EVENSOMEWHATSHORTERMACRO{93} {0.1387} {1.336} {2.264} {3.446} {4.901} {5.335} {5.714}
\EVENSOMEWHATSHORTERMACRO{94} {0.1383} {1.338} {2.269} {3.453} {4.907} {5.354} {5.744}
\EVENSOMEWHATSHORTERMACRO{95} {0.1379} {1.340} {2.273} {3.460} {4.913} {5.373} {5.773}
\EVENSOMEWHATSHORTERMACRO{96} {0.1375} {1.342} {2.278} {3.467} {4.919} {5.391} {5.803}
\EVENSOMEWHATSHORTERMACRO{97} {0.1371} {1.344} {2.282} {3.474} {4.925} {5.409} {5.832}
\EVENSOMEWHATSHORTERMACRO{98} {0.1368} {1.346} {2.286} {3.480} {4.931} {5.427} {5.861}
\EVENSOMEWHATSHORTERMACRO{99} {0.1364} {1.348} {2.291} {3.487} {4.937} {5.445} {5.891}
\EVENSOMEWHATSHORTERMACRO{100} {0.1360} {1.350} {2.295} {3.494} {4.943} {5.463} {5.920}
\EVENSOMEWHATSHORTERMACRO{101} {0.1356} {1.352} {2.299} {3.500} {4.949} {5.480} {5.950}
\EVENSOMEWHATSHORTERMACRO{102} {0.1353} {1.354} {2.303} {3.507} {4.955} {5.497} {5.979}
\EVENSOMEWHATSHORTERMACRO{103} {0.1349} {1.355} {2.307} {3.513} {4.961} {5.514} {6.009}
\EVENSOMEWHATSHORTERMACRO{104} {0.1346} {1.357} {2.312} {3.519} {4.967} {5.530} {6.039}
\EVENSOMEWHATSHORTERMACRO{105} {0.1342} {1.359} {2.316} {3.525} {4.973} {5.546} {6.068}
\EVENSOMEWHATSHORTERMACRO{106} {0.1339} {1.361} {2.320} {3.532} {4.979} {5.562} {6.098}
\EVENSOMEWHATSHORTERMACRO{107} {0.1336} {1.363} {2.324} {3.538} {4.986} {5.578} {6.128}
\EVENSOMEWHATSHORTERMACRO{108} {0.1332} {1.364} {2.327} {3.544} {4.992} {5.594} {6.158}

\end{tabular}
\captionof{table}{\footnotesize{Effective number of digits of accuracy of each successive refined estimates for the massive macroscopic propagation function \bbd{q\ll n\uprm{\textsc{mmp}} \equiv q\ll n - q\ll n\uprm{eft}}, at the coupling \bbd{\t = {{25 i}\over{\pi}}}.  The first column, "WLI w/o prefactor" is the approximation
\bbd{q\ll n\uprm{\textsc{mmp}} \simeq \exp{- \sqrt{{8\pi n}\over{{\rm Im}[\s]}}}}, with \bbd{\s} denoting the infrared coupling related to the UV coupling \bbd{\t} by \bbd{e\uu{2\pi i \t} = \l[\s],} where \bbd{\l[\s]} is the modular lambda function.  The value of \bbd{\t} in this table is \bbd{\t = {{25}\over\pi}\cc i}.  The second column, "\bbd{e\uu{- S\lrm{WLI}}} with prefactor" gives the accuracy of the approximation obtained by appending the prefactor
\bbd{q\ll n\uprm{\textsc{mmp}} \simeq \big[\cc {{2\uu{13}\cc n}\over{\pi\uu 5\cc {\rm Im}[\s]}}\cc \big ]\uu{+{1\over 4}}} to the exponential.  In each successive column we add a correction of the form \bbd{-n\uu{-{p\over 2}}\cc w\ll p[\s]} to the
exponent of the exponential for \bbd{p=1,\cdots 5}, and give the resulting accuracy.  The "effective number of accurate digits" is defined here as the log of the relative error of the estimate of \bbd{q\ll n\uprm{\textsc{mmp}}} , times \bbd{-{1\over{{\rm Log}[10]}}}.
That is, the table entries are given by \bbd{- {1\over{{\rm Log}[10]}}\cc {\rm Log}\left | \cc {{q\ll n\uprm{\textsc{mmp}} - (q\ll n\uprm{\textsc{mmp}})\lrm{estimate}}\over{q\ll n\uprm{\textsc{mmp}}  }}    \cc \right | }.}}
\label{TypesOfLiouvilleAmplitudeTransformationTableSpecificParameterChoiceOfInterestToUs}
\end{center}
\end{table}

\newpage
 \vskip-1in
 \begin{table}[H]
\begin{center}
\begin{tabular}{ |c||c|c|c|c|c|c|c| } 
 \hline\hline
 ~ & ${{\rm accuracy}\atop{{\rm of~}e\uu{- S\lrm{WLI}}}}$  & ${{\rm accuracy~w/}\atop{\rm prefactor}}$ 
 & ${{\rm accuracy~w/}\atop{O(n\uu{-\hh})}}$ &${{\rm accuracy~w/}\atop{O(n\uu{-1})}}$  &${{\rm accuracy~w/}\atop{O(n\uu{-{3\over 2}})}}$ &${{\rm accuracy~w/}\atop{O(n\uu{-2})}}$ &${{\rm accuracy~w/}\atop{O(n\uu{-{5\over 2}})}}$ \\ \hline\hline

\EVENSOMEWHATSHORTERMACRO{109} {0.1329} {1.366} {2.331} {3.550} {4.998} {5.609} {6.188}
\EVENSOMEWHATSHORTERMACRO{110} {0.1326} {1.368} {2.335} {3.556} {5.004} {5.625} {6.219}
\EVENSOMEWHATSHORTERMACRO{111} {0.1322} {1.370} {2.339} {3.562} {5.010} {5.640} {6.249}
\EVENSOMEWHATSHORTERMACRO{112} {0.1319} {1.371} {2.343} {3.567} {5.016} {5.654} {6.280}
\EVENSOMEWHATSHORTERMACRO{113} {0.1316} {1.373} {2.347} {3.573} {5.023} {5.669} {6.311}
\EVENSOMEWHATSHORTERMACRO{114} {0.1313} {1.375} {2.350} {3.579} {5.029} {5.683} {6.342}
\EVENSOMEWHATSHORTERMACRO{115} {0.1310} {1.377} {2.354} {3.585} {5.035} {5.697} {6.373}
\EVENSOMEWHATSHORTERMACRO{116} {0.1307} {1.378} {2.358} {3.590} {5.041} {5.711} {6.405}
\EVENSOMEWHATSHORTERMACRO{117} {0.1304} {1.380} {2.361} {3.596} {5.047} {5.725} {6.437}
\EVENSOMEWHATSHORTERMACRO{118} {0.1301} {1.381} {2.365} {3.601} {5.053} {5.739} {6.469}
\EVENSOMEWHATSHORTERMACRO{119} {0.1298} {1.383} {2.368} {3.607} {5.059} {5.752} {6.501}
\EVENSOMEWHATSHORTERMACRO{120} {0.1295} {1.385} {2.372} {3.612} {5.065} {5.765} {6.534}
\EVENSOMEWHATSHORTERMACRO{121} {0.1292} {1.386} {2.375} {3.618} {5.071} {5.779} {6.567}
\EVENSOMEWHATSHORTERMACRO{122} {0.1289} {1.388} {2.379} {3.623} {5.077} {5.791} {6.601}
\EVENSOMEWHATSHORTERMACRO{123} {0.1286} {1.389} {2.382} {3.628} {5.083} {5.804} {6.635}
\EVENSOMEWHATSHORTERMACRO{124} {0.1284} {1.391} {2.386} {3.633} {5.089} {5.817} {6.670}
\EVENSOMEWHATSHORTERMACRO{125} {0.1281} {1.393} {2.389} {3.639} {5.095} {5.829} {6.705}
\EVENSOMEWHATSHORTERMACRO{126} {0.1278} {1.394} {2.393} {3.644} {5.101} {5.841} {6.741}
\EVENSOMEWHATSHORTERMACRO{127} {0.1275} {1.396} {2.396} {3.649} {5.107} {5.853} {6.778}
\EVENSOMEWHATSHORTERMACRO{128} {0.1273} {1.397} {2.399} {3.654} {5.113} {5.865} {6.815}
\EVENSOMEWHATSHORTERMACRO{129} {0.1270} {1.399} {2.403} {3.659} {5.119} {5.877} {6.854}
\EVENSOMEWHATSHORTERMACRO{130} {0.1267} {1.400} {2.406} {3.664} {5.125} {5.888} {6.893}

\end{tabular}
\captionof{table}{\footnotesize{Effective number of digits of accuracy of each successive refined estimates for the massive macroscopic propagation function \bbd{q\ll n\uprm{\textsc{mmp}} \equiv q\ll n - q\ll n\uprm{eft}}, at the coupling \bbd{\t = {{25 i}\over{\pi}}}.  The first column, "WLI w/o prefactor" is the approximation
\bbd{q\ll n\uprm{\textsc{mmp}} \simeq \exp{- \sqrt{{8\pi n}\over{{\rm Im}[\s]}}}}, with \bbd{\s} denoting the infrared coupling related to the UV coupling \bbd{\t} by \bbd{e\uu{2\pi i \t} = \l[\s],} where \bbd{\l[\s]} is the modular lambda function.  The value of \bbd{\t} in this table is \bbd{\t = {{25}\over\pi}\cc i}.  The second column, "\bbd{e\uu{- S\lrm{WLI}}} with prefactor" gives the accuracy of the approximation obtained by appending the prefactor
\bbd{q\ll n\uprm{\textsc{mmp}} \simeq \big[\cc {{2\uu{13}\cc n}\over{\pi\uu 5\cc {\rm Im}[\s]}}\cc \big ]\uu{+{1\over 4}}} to the exponential.  In each successive column we add a correction of the form \bbd{-n\uu{-{p\over 2}}\cc w\ll p[\s]} to the
exponent of the exponential for \bbd{p=1,\cdots 5}, and give the resulting accuracy.  The "effective number of accurate digits" is defined here as the log of the relative error of the estimate of \bbd{q\ll n\uprm{\textsc{mmp}}} , times \bbd{-{1\over{{\rm Log}[10]}}}.
That is, the table entries are given by \bbd{- {1\over{{\rm Log}[10]}}\cc {\rm Log}\left | \cc {{q\ll n\uprm{\textsc{mmp}} - (q\ll n\uprm{\textsc{mmp}})\lrm{estimate}}\over{q\ll n\uprm{\textsc{mmp}}  }}    \cc \right | }.}}
\label{TypesOfLiouvilleAmplitudeTransformationTableSpecificParameterChoiceOfInterestToUs}
\end{center}
\end{table}

\newpage
 \vskip-1in
 \begin{table}[H]
\begin{center}
\begin{tabular}{ |c||c|c|c|c|c|c|c| } 
 \hline\hline
 ~ & ${{\rm accuracy}\atop{{\rm of~}e\uu{- S\lrm{WLI}}}}$  & ${{\rm accuracy~w/}\atop{\rm prefactor}}$ 
 & ${{\rm accuracy~w/}\atop{O(n\uu{-\hh})}}$ &${{\rm accuracy~w/}\atop{O(n\uu{-1})}}$  &${{\rm accuracy~w/}\atop{O(n\uu{-{3\over 2}})}}$ &${{\rm accuracy~w/}\atop{O(n\uu{-2})}}$ &${{\rm accuracy~w/}\atop{O(n\uu{-{5\over 2}})}}$ \\ \hline\hline

\EVENSOMEWHATSHORTERMACRO{131} {0.1265} {1.402} {2.409} {3.669} {5.131} {5.900} {6.933}
\EVENSOMEWHATSHORTERMACRO{132} {0.1262} {1.403} {2.412} {3.674} {5.137} {5.911} {6.975}
\EVENSOMEWHATSHORTERMACRO{133} {0.1260} {1.405} {2.415} {3.679} {5.142} {5.922} {7.018}
\EVENSOMEWHATSHORTERMACRO{134} {0.1257} {1.406} {2.419} {3.684} {5.148} {5.933} {7.063}
\EVENSOMEWHATSHORTERMACRO{135} {0.1255} {1.407} {2.422} {3.689} {5.154} {5.944} {7.110}
\EVENSOMEWHATSHORTERMACRO{136} {0.1252} {1.409} {2.425} {3.693} {5.160} {5.955} {7.159}
\EVENSOMEWHATSHORTERMACRO{137} {0.1250} {1.410} {2.428} {3.698} {5.165} {5.965} {7.210}
\EVENSOMEWHATSHORTERMACRO{138} {0.1247} {1.412} {2.431} {3.703} {5.171} {5.975} {7.264}
\EVENSOMEWHATSHORTERMACRO{139} {0.1245} {1.413} {2.434} {3.707} {5.177} {5.986} {7.322}
\EVENSOMEWHATSHORTERMACRO{140} {0.1242} {1.414} {2.437} {3.712} {5.182} {5.996} {7.385}
\EVENSOMEWHATSHORTERMACRO{141} {0.1240} {1.416} {2.440} {3.717} {5.188} {6.006} {7.453}
\EVENSOMEWHATSHORTERMACRO{142} {0.1238} {1.417} {2.443} {3.721} {5.194} {6.016} {7.528}
\EVENSOMEWHATSHORTERMACRO{143} {0.1235} {1.419} {2.446} {3.726} {5.199} {6.026} {7.612}
\EVENSOMEWHATSHORTERMACRO{144} {0.1233} {1.420} {2.449} {3.730} {5.205} {6.035} {7.709}
\EVENSOMEWHATSHORTERMACRO{145} {0.1231} {1.421} {2.452} {3.735} {5.210} {6.045} {7.825}
\EVENSOMEWHATSHORTERMACRO{146} {0.1228} {1.423} {2.455} {3.739} {5.216} {6.054} {7.971}
\EVENSOMEWHATSHORTERMACRO{147} {0.1226} {1.424} {2.458} {3.744} {5.221} {6.064} {8.174}
\EVENSOMEWHATSHORTERMACRO{148} {0.1224} {1.425} {2.461} {3.748} {5.226} {6.073} {8.529}
\EVENSOMEWHATSHORTERMACRO{149} {0.1222} {1.427} {2.464} {3.752} {5.232} {6.082} {9.273}
\EVENSOMEWHATSHORTERMACRO{150} {0.1220} {1.428} {2.466} {3.757} {5.237} {6.091} {8.421}

\end{tabular}
\captionof{table}{\footnotesize{Effective number of digits of accuracy of each successive refined estimates for the massive macroscopic propagation function \bbd{q\ll n\uprm{\textsc{mmp}} \equiv q\ll n - q\ll n\uprm{eft}}, at the coupling \bbd{\t = {{25 i}\over{\pi}}}.  The first column, "WLI w/o prefactor" is the approximation
\bbd{q\ll n\uprm{\textsc{mmp}} \simeq \exp{- \sqrt{{8\pi n}\over{{\rm Im}[\s]}}}}, with \bbd{\s} denoting the infrared coupling related to the UV coupling \bbd{\t} by \bbd{e\uu{2\pi i \t} = \l[\s],} where \bbd{\l[\s]} is the modular lambda function.  The value of \bbd{\t} in this table is \bbd{\t = {{25}\over\pi}\cc i}.  The second column, "\bbd{e\uu{- S\lrm{WLI}}} with prefactor" gives the accuracy of the approximation obtained by appending the prefactor
\bbd{q\ll n\uprm{\textsc{mmp}} \simeq \big[\cc {{2\uu{13}\cc n}\over{\pi\uu 5\cc {\rm Im}[\s]}}\cc \big ]\uu{+{1\over 4}}} to the exponential.  In each successive column we add a correction of the form \bbd{-n\uu{-{p\over 2}}\cc w\ll p[\s]} to the
exponent of the exponential for \bbd{p=1,\cdots 5}, and give the resulting accuracy.  The "effective number of accurate digits" is defined here as the log of the relative error of the estimate of \bbd{q\ll n\uprm{\textsc{mmp}}} , times \bbd{-{1\over{{\rm Log}[10]}}}.
That is, the table entries are given by \bbd{- {1\over{{\rm Log}[10]}}\cc {\rm Log}\left | \cc {{q\ll n\uprm{\textsc{mmp}} - (q\ll n\uprm{\textsc{mmp}})\lrm{estimate}}\over{q\ll n\uprm{\textsc{mmp}}  }}    \cc \right | }.}}
\label{TypesOfLiouvilleAmplitudeTransformationTableSpecificParameterChoiceOfInterestToUs}
\end{center}
\end{table}

\newpage

\section{Subleading large-\bbd{n} corrections to the double-scaling limit}\label{AsymptoticsOfQMMPDoubleScaledLargeCharge}

\subsection{Recursion relation for the two-loop correction}

Now we want to analyze eq.  \rr{BasicMMPRecursionRelation} in the double-scaling limit rather than the fixed-\bbd{\t} large-\bbd{n} limit.
In this limit we know from the start that gauge instanton effects are absent to all orders in \bbd{n}, since the instanton factor \bbd{e\uu{-2\pi {\rm Im}[\t]} = e\uu{- {n\over{2\l}}}} is
exponentially small in \bbd{n}  at fixed double-scaling parameter \bbd{\ddsc}.  This means the correlators are independent of the UV \bbd{\th}-angle \bbd{\th = 2\pi {\rm Re}[\t]} to all orders
in \bbd{n} in the double scaling limit, depending only on \bbd{n} and \bbd{t  \equiv {\rm Im}[\t]} or equivalently on \bbd{n} and \bbd{\ddsc = {n\over{4\pi t}}}.
For a function depending only on \bbd{t \equiv {\rm Im}[\t]}  and not on \bbd{{\rm Re}[t]} we can use \bbd{\pp\ll\t\pb\ll\tb = {1\over 4}\cc \pp\ll t\sqd} to write the recursion relation for \bbd{q\ll n\uprm{\textsc{mmp}}} as
\bbb
{1\over 4}\cc e\uu{-A}\cc \pp\ll t\sqd q\ll n = (2n+ {7\over 2})(2n + {5\over 2}) \cc \biggl [ \cc
 {{Z\ll{n+1}\uprm{\textsc{mmp}} }\over{   Z\ll n\uprm{\textsc{mmp}}}}  - 1
 \cc \biggl ]
-    (2n+ {3\over 2})(2n + {1\over 2}) \cc  \biggl [ \cc
 {{Z\ll n\uprm{\textsc{mmp}} }\over{   Z\ll {n-1}\uprm{\textsc{mmp}}}}  - 1
 \cc \biggl ]\llsk
\een{MainEOVEquation}

We must give eq. \rr{MainEOVEquation} as an \textsc{eov} in terms of \bbd{n} and  \bbd{\ddsc \equiv  {n\over {4\pi t }}} rather than \bbd{n} and \bbd{t}.  The inverse
formula is \bbd{t = {n\over{4\pi\ddsc}}}. 
At fixed \bbd{n} the relationship between \bbd{\pp\ll t} and \bbd{\pp\ll \ddsc} is
\bbb
\pp\ll t = {{d\ddsc}\over{dt}}\cc \pp\ll\ddsc = ({{dt}\over{d\ddsc}})\uu{-1}\cc\pp\ll \ddsc = (- n/(4\pi\ddsc\sqd))\uu{-1}\cc\pp\ll\ddsc = - {{4\pi\ddsc\sqd}\over n}\cc\pp\ll\ddsc\ .
\eee
so then
\bbb
\pp\ll t\sqd = + {{16\cc\pi\sqd}\over{n\sqd}}\cc [\ddsc\uu 4 \pp\ll\ddsc\sqd + 2\ddsc\uu 3\pp\ll \ddsc]
\eee
and so
\bbb
\pp\ll\t\pb\ll\tb =  + {{4\pi\sqd}\over{ n\sqd}}\cc [\ddsc\uu 4 \pp\ll\ddsc\sqd + 2\ddsc\uu 3\pp\ll \ddsc]
\een{TauLaplacianInTermsOfNAndLambda}

Now we would like to write \bbd{e\uu{-A\ls\t}} in terms of \bbd{n} and \bbd{\ddsc}.  We have
\bbb
e\uu{A\ls\s} = {1\over{16\cc [{\rm Im}[\s]]\sqd}}\ .
\een{AFunctionInSigmaFrame}
So then in \bbd{\t} frame we have
\bbb
e\uu{A\ls\t} = |{{d\s}\over{d\t}}|\sqd \cc {1\over{16\cc [{\rm Im}[\s]]\sqd}}
\eee
which means
\bbb
e\uu{-A\ls\t} = 16\cc [{\rm Im}[\s]]\sqd\cc |{{d\t}\over{d\s}}|\sqd
\eee
The perturbative relationship between \bbd{\s} and \bbd{\t} is
\bbb
\t\simeq {\s\over 2} - {{2i}\over\pi}\cc {\tt log}[2]\ , \llsk
\s\simeq 2\t + {{4i}\over{\pi}}\cc {\tt log}[2]\ ,
\eee
with the \bbd{\simeq}s indicating equality up to exponentially small corrections.
So
\bbb
e\uu{- A\ls\t} \simeq 16\cc [{\rm Im}[\t] + {2\over\pi} \cc {\tt log}[2] ]\sqd
\eee
Then using
\bbb
\ddsc 
= {{  n}\over{4\pi\cc {\rm Im}[\t]}}
\eee
we have
\bbb
{\rm Im}[\t] = {n\over{4\pi \ddsc}}
\eee
which means
\bbb
e\uu{- A\ls\t} \simeq 16\cc [{n\over{4\pi \ddsc}}+ {2\over\pi}\cc {\tt log}[2] ]\sqd = {{ n\sqd}\over{\pi\sqd\ddsc\sqd}}\cc [1 + {{8\ddsc}\over n} \cc {\tt log}[2] ]\sqd
\eee
Then, using expression \rr{TauLaplacianInTermsOfNAndLambda} for \bbd{\pp\ll\t\bar{\pp}\ll{\bar{\t}}} in terms of \bbd{n} and \bbd{\ddsc}, which we recap here,
\bbb
\pp\ll\t\pb\ll\tb =  + {{4\pi\sqd}\over{ n\sqd}}\cc [\ddsc\uu 4 \pp\ll\ddsc\sqd + 2\ddsc\uu 3\pp\ll \ddsc]
\een{TauLaplacianInTermsOfNAndLambdaRecap}
we have
\bbb
e\uu{-A}\pp\pb \simeq 4\cc [1 + {{8\ddsc}\over n }\cc {\tt log}[2] ]\sqd \cc [\ddsc\uu 2 \pp\ll\ddsc\sqd + 2\ddsc\pp\ll \ddsc]
\eee
where the \bbd{\simeq} indicates the omission of exponentially small corrections only.

Now expand in \bbd{n} at fixed \bbd{\ddsc} using the ansatz
\bbb
q\ll n\uprm{\textsc{mmp}} = \MacroFLPLower 0[\ddsc] + n\uu{-1}\cc \MacroFLPLower 1[\ddsc] + n\uu{-2}\cc \MacroFLPLower 2[\ddsc] + \cdots
\eee
Defining
\bbb
[{\rm RHS}] \equiv  (2n+ {7\over 2})(2n + {5\over 2}) \cc \biggl [ \cc
 {{Z\ll{n+1}\uprm{\textsc{mmp}} }\over{   Z\ll n\uprm{\textsc{mmp}}}}  - 1
 \cc \biggl ]
-    (2n+ {3\over 2})(2n + {1\over 2}) \cc  \biggl [ \cc
 {{Z\ll n\uprm{\textsc{mmp}} }\over{   Z\ll {n-1}\uprm{\textsc{mmp}}}}  - 1
 \cc \biggl ]\llsk
\een{RecursionRelationDoubleScaledContext}
and expanding at large \bbd{n} and fixed \bbd{\l}, we find that the \bbd{n\uu{+2}} and \bbd{n\uu{+1}} terms vanish, and we have
\bbb
[{\rm RHS}] = [{\rm RHS}]\ll 0 + n\uu{-1}\cc [{\rm RHS}]\ll 1 + n\uu{-2}\cc [{\rm RHS}]\ll 2 + \cdots
\eee
with
\bbb
[{\rm RHS}]\ll 0 = 4\ddsc\sqd \MacroFLPLower 0\prpr[\ddsc] + 8\ddsc \MacroFLPLower 0\pr[\ddsc]
\xxnn
[{\rm RHS}]\ll 1 = 4\ddsc\sqd \MacroFLPLower 1\prpr[\ddsc] + {{\cal G}}[\ddsc]\ ,
\xxnn
{{\cal G}}[\ddsc] \equiv 4\ddsc \cc (\ddsc \MacroFLPLower 0\pr[\ddsc] + 2)(\ddsc \MacroFLPLower 0\prpr[\ddsc] + \MacroFLPLower 0\pr[\ddsc])
\xxnn
= 8\ddsc\cc \pp\ll\ddsc\cc \bigg [ \cc 
{{\ddsc\sqd}\over 4}\cc (\MacroFLPLower 0\pr[\ddsc])\sqd + \ddsc\cc \MacroFLPLower 0\pr[\ddsc]
\cc \bigg ]
\eee
and so forth.

We can expand the LHS similarly:
\bbb
[{\rm LHS}]\ll 0 = 4\ddsc\sqd \MacroFLPLower 0\prpr[\ddsc] + 8\ddsc \MacroFLPLower 0\pr[\ddsc]
\xxnn
[{\rm LHS}]\ll 1 = 4\ddsc\sqd \MacroFLPLower 1\prpr[\ddsc] + 8\ddsc \MacroFLPLower 1\pr[\ddsc] + 64\cc {\rm log}[2] \cc \ddsc\cc \bigg [ \cc
\ddsc\sqd\cc \MacroFLPLower 0\prpr[\ddsc] + 2 \ddsc\cc \MacroFLPLower 0\pr[\ddsc]
\cc\bigg ]
\eee
and so on.

\subsection{Solving the recursion relation for the two-loop correction}

Now we can constrain the functions \bbd{\MacroFLPLower k} appearing in the double-scaled large-R-charge expansion of \bbd{q\ll n\uprm{\textsc{mmp}}}, by
setting equal each \bbd{[{\rm LHS}]\ll k} to each \bbd{[{\rm RHS}]\ll k}.

The functions \bbd{[{\rm LHS}]\ll 0} and \bbd{[{\rm RHS}]\ll 0} are equal identically, for any form of \bbd{\MacroFLPLower 0[\ddsc],} so the recursion relation at order
\bbd{n\uu 0} gives no information at all.  At order \bbd{n\uu{-1}} we can divide by \bbd{8\ddsc} and we find
\bbb
0 = {1\over{8\ddsc}}\big [ \cc [{\rm LHS}]\ll 0 - [{\rm RHS}]\ll 0 \cc \big ]
\xxnn
= \MacroFLPLower 1\pr[\ddsc] 
+8\cc {\rm log}[2]\cc  \bigg [ \cc
\ddsc\sqd\cc \MacroFLPLower 0\prpr[\ddsc] + 2 \ddsc\cc \MacroFLPLower 0\pr[\ddsc] \cc \bigg ] - {{{{\cal G}}[\ddsc]}\over{8\ddsc}}
\xxnn
= \MacroFLPLower 1\pr[\ddsc] 
+8\cc {\rm log}[2]\cc  \bigg [ \cc
\ddsc\sqd\cc \MacroFLPLower 0\prpr[\ddsc] + 2 \ddsc\cc \MacroFLPLower 0\pr[\ddsc] \cc \bigg ] -\pp\ll\ddsc\cc \bigg [ \cc 
{{\ddsc\sqd}\over 4}\cc (\MacroFLPLower 0\pr[\ddsc])\sqd + \ddsc\cc \MacroFLPLower 0\pr[\ddsc]
\cc \bigg ]
\xxnn
 = \MacroFLPLower 1\pr[\ddsc] 
+8\cc {\rm log}[2]\cc \cc \pp\ll\ddsc  \bigg [ \cc
\ddsc\sqd\cc \MacroFLPLower 0\pr[\ddsc] \cc \bigg ] 
 -\pp\ll\ddsc\cc \bigg [ \cc 
{{\ddsc\sqd}\over 4}\cc (\MacroFLPLower 0\pr[\ddsc])\sqd + \ddsc\cc \MacroFLPLower 0\pr[\ddsc]
\cc \bigg ]
\eee
So the content of the recursion relation at order \bbd{n\uu{-1}} is
\bbb
 \MacroFLPLower 1\pr[\ddsc]  = \pp\ll\ddsc\bigg [\cc
 {{\ddsc\sqd}\over 4}\cc (\MacroFLPLower 0\pr[\ddsc])\sqd + \ddsc\cc \MacroFLPLower 0\pr[\ddsc] - 8\cc {\rm log}[2]\cc \ddsc\sqd\cc \MacroFLPLower 0\pr[\ddsc] 
 \cc \bigg ]
\eee
so the general solution of the recursion relation is
\bbb
\MacroFLPLower 1[\ddsc] = {{\ddsc\sqd}\over 4}\cc (\MacroFLPLower 0\pr[\ddsc])\sqd + \ddsc\cc \MacroFLPLower 0\pr[\ddsc] - 8\cc {\rm log}[2]\cc \ddsc\sqd\cc \MacroFLPLower 0\pr[\ddsc] + ({\rm constant})\ ,
\eee
where the constant is independent of \bbd{\ddsc} and \bbd{n}.

The function \bbd{\MacroFLPLower 0} vanishes exponentially with \bbd{\ddsc\uu\hh} at large \bbd{\ddsc}, so every term on the RHS other than the constant
vanishes at large \bbd{\ddsc}; on physical grounds we should expect the two-loop correction \bbd{\MacroFLPLower 1 [\ddsc]} to the connected MMP function
to vanish exponentially as well, since every contribution to it contains at least one macroscopic massive worldline; so the constant on the RHS must
vanish, and we have a unique solution consistent with the expected physical behavior of the MMP function,
\bbb
\MacroFLPLower 1[\ddsc] = {{\ddsc\sqd}\over 4}\cc (\MacroFLPLower 0\pr[\ddsc])\sqd + \ddsc\cc \MacroFLPLower 0\pr[\ddsc] - 8\cc {\rm log}[2]\cc \ddsc\sqd\cc \MacroFLPLower 0\pr[\ddsc] \ .
\een{TwoLoopConnectedMMPAmplitude}

\subsection{Physical interpretation of the double-scaled MMP term
at order \bbd{n\uu{-1}}}\label{TwoLoopPhysicalInterpretation}

The three terms in \rr{TwoLoopConnectedMMPAmplitude} can each be identified with three types of two-loop corrections to the connected macroscopic massive propagation amplitude.  The first term, \bbd{{{\ddsc\sqd}\over 4}\cc (\MacroFLPLower 0\pr[\ddsc])\sqd } is quadratic in \bbd{F\ls 0\pr[\ddsc]} and therefore has a magnitude at least twice as exponentially suppressed as a single electric hypermultiplet worldline instanton, in the
regime of large \bbd{\ddsc}.  This term should be thought of as a sum of terms with two macroscopic massive hypermultiplet worldlines connected by a massless propagator.
The second and third terms are both linear in \bbd{F\ls 0\pr[\ddsc]} and have a single massive macroscopic worldline with a massless propagator beginning and ending on it.  The second term  \bbd{\ddsc\cc \MacroFLPLower 0\pr[\ddsc] } is the piece of this diagram that is sub-extensive along
the macroscopic massive worldline.  The last term \bbd{- 8\cc {\rm log}[2]\cc \ddsc\sqd\cc \MacroFLPLower 0\pr[\ddsc] } is extensive along the worldline, and gives a finite mass renormalization to the BPS particle,
corresponding to the one-loop threshold correction to the effective prepotential of the \bbd{U(1)} vector multiplet.  Now we will comment
briefly on the meaning of these terms, particularly
the mass renormalization.

\subsubsection{Breakdown of the double-scaling regime at ultra-large \bbd{n}
}\label{MainDSCBreakdownSectionToMergeTheLargerSubsectionInto}

The relationship between the two- and one-loop MMP results contains a clue as to the distinction between the large-\bbd{\ddsc} limit of the double-scaled large-\bbd{n} 
limit, on the one hand, and the fixed-coupling large-\bbd{n} limit on the other hand.

The double-scaled large-\bbd{n} expansion of the MMP amplitude \bbd{q\ll n\uprm{\textsc{mmp}},} breaks down at large \bbd{n} and fixed \bbd{\t}.  This can be seen
in the ratio of the two-loop term to the one-loop term, \bbd{{{\MacroFLPLower 1[\ddsc]}\over{n\cc \MacroFLPLower 0[\ddsc]}}}.
In the double-scaled laege-\bbd{n} limit, this goes as \bbd{n\uu{-1},} but at fixed \bbd{\t} and large \bbd{n}, the coupling \bbd{\ddsc} goes to infinity, and each term
\bbd{F\ls{0,1}[\ddsc]} is replaced by its strong-\bbd{\ddsc} limit.  At strong coupling \bbd{\MacroFLPLower 0} goes as \bbd{[{\rm const.}]\times \ddsc\uu{{1\over 4}}\times e\uu{-4\pi \sqrt{\ddsc}}}.
There are three terms in \bbd{\MacroFLPLower 1[\ddsc]}, of which 
the first has twice the exponential suppression and we can ignore it.  The other two factors have only
one macroscopic massive propagator and go as \bbd{e\uu{-4\pi\ddsc\uu\hh}} times powers of \bbd{\ddsc}.  One has an enhancement of \bbd{\ddsc\uu{+\hh}} over \bbd{\MacroFLPLower 0[\ddsc]} at large \bbd{\ddsc} and the other has an enhancement of
\bbd{\ddsc\uu{+{3\over 2}}}.  Dividing by \bbd{n \MacroFLPLower 0[\ddsc]} one of these terms goes as \bbd{\ddsc\uu{+\hh} / n \propto n\uu{-\hh} [{\rm Im}[\t]]\uu{-\hh}}
and the other goes as \bbd{\ddsc\uu{+{3\over 2}} / n \propto n\uu{+\hh} [{\rm Im}[\t]]\uu{-{3\over 2}}}.  The term in the ratio that grows with \bbd{n} at fixed coupling \bbd{t = {\rm Im}[\t]}, is
\bbb
{{\MacroFLPLower 1[\ddsc]}\over{n\cc \MacroFLPLower 0[\ddsc]}} \cc \bigg |\ll{\ddsc\uu {+{3\over 2}}\cc n\uu{-1}{\rm ~term}} = + {{16\pi\cc\ddsc\uu{{3\over 2}}\cc {\rm Log}[2]}\over n} = + {{2\cc {\rm Log}[2]}\over{\sqrt{\pi}}}\cc {{n\uu{+\hh}\over{t\uu{{3\over 2}}}}}
\een{FirstStatement}

So we find that at large \bbd{\ddsc,} the ratio of the two-loop MMP contribution to the one-loop MMP contribution is
\bbb
\MacroFLPLower 1 / (n \MacroFLPLower 0) \sim  
 16\pi\cc{\rm log}[2]\cc n\uu{-1}\cc \ddsc\uu{+{3\over 2}}
= 
2\cc\pi\uu{-{3\over 2}} \cc {\rm log}[2]\cc n\uu{+\hh}\cc {\rm Im}[\t]\uu{-{1\over 2}}
\een{SecondStatement}
This behavior means that double-scaled perturbation expansion breaks down altogether -- \emm{even while \bbd{n} is large and the gauge coupling is weak}  -- when
\bbb
\ddsc \cc \muchgreaterthan \cc 
~ n\uu{+{2\over 3}}
\eee
or equivalently when
\bbb
n \cc \lesssim \cc  
 ~ \ddsc\uu{+{3\over 2}}
\eee
or equivalently when
\bbb
n\cc \muchgreaterthan \cc
 ~ {\rm Im}[\t]\uu 3
\eee
or equivalently when
\bbb
{\rm Im}[\t] \cc \muchlessthan \cc
~ n\uu{+{1\over 3}}
\eee

So the condition for the validity of the double-scaled large-charge expansion -- beyond the minimal necessary condition \bbd{n \geq 1} -- can be expressed in four equivalent ways,
\bbb
{\rm Im}[\t] \cc \muchgreaterthan \cc 
 ~ n\uu{+{1\over 3}}
 \xxnn
 n\cc \muchlessthan \cc
 ~ {\rm Im}[\t]\uu 3
 \xxnn
 n \cc \muchgreaterthan \cc  
 ~ \ddsc\uu{+{3\over 2}}
 \xxnn
 \ddsc \cc\cc \muchlessthan \cc 
~ n\uu{+{2\over 3}}
\eee

In summary, double-scaled perturbation theory is applicable only for
\bbb
1 \muchlessthan n \muchlessthan {\rm Im}[\t]\uu 3\ .
\eee
This condition forces the gauge coupling to be \emm{at least somewhat} weak and separates the regime of fixed-\bbd{\l} large charge from the regime
of fixed \bbd{\t} large charge.

\subsubsection{Physical mechanism for the breakdown of
double-scaled perturbation theory}

The significance of the threshold correction term \bbd{- 8\cc {\rm log}[2]\cc \ddsc\sqd\cc \MacroFLPLower 0\pr[\ddsc]}{, the third term in the connected two-loop double-scaled MMP amplitude \rr{TwoLoopConnectedMMPAmplitude}, becomes clearer when we recall the exact BPS formula for the worldline instanton action controlling the exponential falloff of the MMP function \bbd{q\ll n} at large \bbd{n} and fixed \bbd{\t},
expression \rr{ExactBPSWLI}.  The leading behavior at large charge and
fixed \bbd{\t} is
\bbb
q\ll n\uprm{\textsc{mmp}} \simeq ({\rm constant}) \times ({\rm power~law})\times \exp{- \sqrt{{{8\pi n}\over{{\rm Im}[\s]}}}}
\eee
in the range of \bbd{\s} in which the electric hypermultiplet is the lightest massive BPS particle.

If we expand this expression at large \bbd{n} and fixed \bbd{\ddsc}, 
retaining the subleading large-\bbd{n} corrections,
the expression for \bbd{\s} in terms of \bbd{\t} is replaced with its
weak-coupling expansion, 
\bbb
\s \sim 2\t + {{4 i\cc {\rm Log}[2]}\over\pi}\ , \llsk\llsk 
{\rm Im}[\s] = 2\cc {\rm Im}[\t] + {{4\cc {\rm Log}[2]}\over\pi}
\eee
So the expected behavior at large \bbd{n} is
\bbb

q\ll n\uprm{\textsc{mmp}} \simeq ({\rm constant}) \times ({\rm power~law})\times \exp{- \sqrt{{{4\pi n}\over{{\rm Im}[\t] + {{2\cc {\rm Log}[2]}\over\pi} + O(e\uu{-2\pi\cc{\rm Im}[\t]}) }}}}
\eee
Re-expressing this in terms of \bbd{n} and \bbd{\ddsc} we have
\bbb
q\ll n\uprm{\textsc{mmp}} \simeq 
 ({\rm constant}) \times ({\rm power~law})\times \exp{- 4\pi\cc \sqrt{{\ddsc\over{ 1 + {{8\cc {\rm Log}[2]\cc \ddsc}\over n} + O({{\ddsc}\over n}\cc e\uu{- {n\over{2\ddsc}}}) }}}}
\eee

\subsubsection{Higher order in \bbd{n}}

We have not computed beyond two loops so far, 
but in sec. \ref{HigherMMPThroughNToMinusFive} we will.  Let us first discuss what
our interpretation of the threshold correction implies for the large-\bbd{\ddsc}
limit of the higher \bbd{{1\over n}} corrections to the double-scaling
limit of \bbd{q\ll n\uprm{\textsc{mmp}}}.

Expanding at large \bbd{n,} we expect the terms with the maximal number of threshold corrections to the \bbd{{1\over n}} series in the double-scaling limit, to be given by
the series
\bbb
\bbsk
q\ll n\uprm{\textsc{mmp}} \cc \biggl |\lrm{ {{maximal~number~of}\atop{threshold~corrections}}} = 
 ({\rm constant}) \times ({\rm power~law})\times e\uu{-4\pi \ddsc\uu\hh}\times \exp{  4\pi \ddsc\uu\hh\times \bigg [ \cc 
 1-\big (\cc
  1 + {{8\cc {\rm Log}[2]\cc \ddsc}\over n} + O({{\ddsc}\over n}\cc e\uu{- {n\over{2\ddsc}}})
 \cc \big )\uu{-\hh}
 \cc \bigg ]}
 \llsk
 \xxnn
 \bbsk
\simeq
q\ll n\uprm{\textsc{mmp}} \cc \biggl |\lrm{ {{maximal~number~of}\atop{threshold~corrections}}} = q\ll n\uprm{\textsc{mmp}} \cc \biggl |\lrm{ terms~of~relative~size~{{\ddsc\uu{{{3p}\over 2}}}\over{n\uu p}}}
\xxnn
\bbsk
= \sum\
 ({\rm constant}) \times ({\rm power~law})\times e\uu{-4\pi \ddsc\uu\hh}\times \exp{  4\pi \ddsc\uu\hh\times \bigg [ \cc 
 1-\big (\cc
  1 - {{4\cc {\rm Log}[2]\cc \ddsc}\over n} + O({{\ddsc\sqd}\over{n\sqd}}) \cc \big )
 \cc \bigg ]} \cc \biggl |\lrm{ terms~of~relative~size~{{\ddsc\uu{{{3p}\over 2}}}\over{n\uu p}}}
 \llsk\llsk
 \xxnn
 \bbsk
= \sum\cc
 ({\rm constant}) \times ({\rm power~law})\times e\uu{-4\pi \ddsc\uu\hh}\times \exp{ + {{16\pi\cc {\rm Log}[2]\cc \ddsc\uu{+{3\over 2}}}\over n} }
 \eee
So, the interpretation of the largest terms in the subleading large-\bbd{n} corrections at double-scaling, as the expansion of the one-loop mass correction
to the BPS formula due to the one-loop threshold correction to the holomorphic prepotential, predicts that the largest term in \bbd{n\uu{-p}\cc \MacroFLPLower p[\ddsc] \in q\ll n\uprm{\textsc{mmp}}} is given by
\bbb
n\uu{-p}\cc \MacroFLPLower p[\ddsc]\big |\lrm{ {{largest}\atop{term}}} = 
 ({\rm constant}) \times ({\rm power~law})\times e\uu{-4\pi \ddsc\uu\hh}\times \bigg [ \cc \exp{ + {{16\pi\cc {\rm Log}[2]\cc \ddsc\uu{+{3\over 2}}}\over n} }
 \cc \bigg ]\ll{n\uu{-p}\cc{\rm term}}
 \eee
 Taking the ratio with \bbd{\MacroFLPLower 0[\ddsc]} to cancel out the power law and constant prefactors and the \bbd{n\uu 0} exponential term, we have
 \bbb
\bbsk
 {{n\uu{-p}\cc \MacroFLPLower p[\ddsc]}\over{\MacroFLPLower 0[\ddsc]}}\cc \bigg |\lrm{{{largest~term}\atop{at~large~\ddsc}}} 
 = \bigg [ \cc \exp{ + {{16\pi\cc {\rm Log}[2]\cc \ddsc\uu{+{3\over 2}}}\over n} }
 \cc \bigg ]\ll{n\uu{-p}\cc{\rm term}} = (16\pi\cc {\rm Log}[2])\uu p\cc \ddsc\uu{{3p}\over 2}\cc n\uu{-p}\times {1\over{p!}}\ . \llsk
 \eee

 Filling in the constant and power law from the exact
 solution of \cite{Grassi:2019txd},
 \bbb
 ({\rm constant})\times ({\rm power~law}) = {{8 \sqrt{2}}\over{\pi}}\cc \ddsc\uu{+{1\over 4}}\cc
 \eee
  we have the prediction
\bbb
n\uu{-p}\cc \MacroFLPLower p[\ddsc]\big |\lrm{ {{largest}\atop{term}}} = 
{{8 \sqrt{2}}\over{\pi}}\cc \ddsc\uu{+{1\over 4}}\cc e\uu{-4\pi \ddsc\uu\hh}\times(16\pi\cc {\rm Log}[2])\uu p\cc \ddsc\uu{{3p}\over 2}\cc n\uu{-p}\times {1\over{p!}}
\xxnn
= {1\over{p!}}\cc 2\uu{4p + {7\over 2}}\cc \pi\uu{p-1}\cc ({\rm Log}[2])\uu p\cc \ddsc\uu{{{3p}\over 2} + {1\over 4}}\cc n\uu{-p}\cc  e\uu{-4\pi \ddsc\uu\hh}
 \eee

 So in particular we predict
 \bbb
 {{n\uu{-1}\cc \MacroFLPLower 1[\ddsc]}\over{\MacroFLPLower 0[\ddsc]}}\cc \bigg |\lrm{{{largest~term}\atop{at~large~\ddsc}}} 
 \qeqA  16\pi\cc {\rm Log}[2] \cc \ddsc\uu{{3\over 2}}\cc n\uu{-1}
 \xxnn
  {{n\uu{-2}\cc \MacroFLPLower 2[\ddsc]}\over{\MacroFLPLower 0[\ddsc]}}\cc \bigg |\lrm{{{largest~term}\atop{at~large~\ddsc}}} 
 \qeqA  128\cc\pi\sqd\cc ({\rm Log}[2])\uu 2\cc \ddsc\uu 3\cc n\uu{-2} 
  \xxnn
  {{n\uu{-3}\cc \MacroFLPLower 3[\ddsc]}\over{\MacroFLPLower 0[\ddsc]}}\cc \bigg |\lrm{{{largest~term}\atop{at~large~\ddsc}}} 
 \qeqA  
 {{2048}\over 3}\cc \pi\uu 3\cc ({\rm Log}[2])\uu 3\cc \ddsc\uu{{9\over 2}}\cc n\uu{-3}
  \xxnn
  {{n\uu{-4}\cc \MacroFLPLower 4[\ddsc]}\over{\MacroFLPLower 0[\ddsc]}}\cc \bigg |\lrm{{{largest~term}\atop{at~large~\ddsc}}} 
 \qeqA   {{8192}\over 3}\cc \pi\uu 4\cc ({\rm Log}[2])\uu 4\cc \ddsc\uu 6\cc n\uu{-4}
  \xxnn
  {{n\uu{-5}\cc \MacroFLPLower 2[\ddsc]}\over{\MacroFLPLower 5[\ddsc]}}\cc \bigg |\lrm{{{largest~term}\atop{at~large~\ddsc}}} 
 \qeqA  {{2\uu{17}}\over{15}}\cc \pi\uu 5 ({\rm Log}[2])\uu 5\cc \ddsc\uu{{15}\over 2}\cc n\uu{-5}
 \eee

 Or, in absolute terms,
 \bbb
  n\uu{-p}\cc \MacroFLPLower p[\ddsc]  \bigg |\lrm{{{largest~term}\atop{at~large~\ddsc}}}   \qeqA 
  2\uu{{{15}\over 2}} {\rm Log}[2] \cc \ddsc\uu{{7\over 4}}\cc n\uu{-1}\cc  e\uu{-4\pi \ddsc\uu\hh}
  \xxnn
   n\uu{-2}\cc \MacroFLPLower 2[\ddsc]  \bigg |\lrm{{{largest~term}\atop{at~large~\ddsc}}}   \qeqA 
    2\uu{{{21}\over 2}}\cc \pi\cc ({\rm Log}[2])\uu 2\cc \ddsc\uu{{{13}\over 4}}\cc n\uu{-2}\cc  e\uu{-4\pi \ddsc\uu\hh}
   \xxnn
   n\uu{-3}\cc \MacroFLPLower 3[\ddsc]  \bigg |\lrm{{{largest~term}\atop{at~large~\ddsc}}}   \qeqA 
   {{ 2\uu{ {{29}\over 2}  }}\over 3}\cc \pi\sqd \cc ({\rm Log}[2])\uu 3\cc \ddsc\uu{ {{19}\over 4}}\cc n\uu{-3}\cc  e\uu{-4\pi \ddsc\uu\hh}
   \xxnn
   n\uu{-4}\cc \MacroFLPLower 4[\ddsc]  \bigg |\lrm{{{largest~term}\atop{at~large~\ddsc}}}   \qeqA  {{2\uu{{{33}\over 2}}}\over 3}\cc \pi\uu 3\cc ({\rm Log}[2])\uu 4\cc \ddsc\uu {{{25}\over 4}}\cc n\uu{-4}\cc  e\uu{-4\pi \ddsc\uu\hh}
  \xxnn
   n\uu{-5}\cc \MacroFLPLower 5[\ddsc]  \bigg |\lrm{{{largest~term}\atop{at~large~\ddsc}}}   \qeqA {{2\uu{{{41}\over 2}}}\over {15}}\cc \pi\uu 4\cc ({\rm Log}[2])\uu 5\cc \ddsc\uu {{{31}\over 4}}\cc n\uu{-5}\cc  e\uu{-4\pi \ddsc\uu
   \hh}
 \eee
 and so forth.

In sec. \ref{HigherMMPThroughNToMinusFive} we have computed the functional forms of
 \bbd{\MacroFLPLower p[\ddsc]} through \bbd{p=5}.  Expanding at large \bbd{\ddsc} we will find
 \bbb
\bbsk
 \MacroFLPLower 1[\ddsc] = - 8\cc {\rm log}[2]\cc \ddsc\sqd\cc \MacroFLPLower 0\pr[\ddsc] + ({\rm relative~}O(\ddsc\uu{-\hh}) = + 16\pi\cc {\rm Log}[2]\cc \ddsc\uu {+{3\over 2}}\cc \MacroFLPLower 0[\ddsc] 
 + ({\rm relative~}O(\ddsc\uu{-\hh}) \llsk
 \eee
 and
 \bbb
 \bbsk
 \bbsk
 \MacroFLPLower 2[\ddsc] = 32\cc\ddsc\uu 4\cc ({\rm Log}[2])\sqd\cc \MacroFLPLower 0\prpr[\ddsc] + ({\rm relative~}O(\ddsc\uu{-\hh})
  = + 128\cc \pi\sqd\cc ({\rm Log}[2])\sqd\cc \ddsc\uu 3 \cc \MacroFLPLower 0[\ddsc] + ({\rm relative~}O(\ddsc\uu{-\hh})
  \llsk
 \eee
 and
 \bbb
 \bbsk\bbsk
 \MacroFLPLower 3[\ddsc] = - {{256}\over 3}\cc ({\rm Log}[2])\uu 3\cc \ddsc\uu 6\cc \MacroFLPLower 0\prprpr[\ddsc] + ({\rm relative~}O(\ddsc\uu{-\hh})) = + {{2\uu{11}}\over 3}\cc \pi\uu 3\cc ({\rm Log}[2])\uu 3\cc \ddsc\uu {{9\over 2}}\cc \MacroFLPLower 0[\ddsc] 
  + ({\rm relative~}O(\ddsc\uu{-\hh}))
 \llsk
 \eee
 and
 \bbb
\bbsk\bbsk
 \MacroFLPLower 4[\ddsc] = {{512}\over 3}\cc ({\rm Log}[2])\uu 4\cc \ddsc\uu 8\cc \MacroFLPLower 0\prprprpr[\ddsc] + ({\rm relative~}O(\ddsc\uu{-\hh})) = {{2\uu{13}}\over 3}\cc \pi\uu 4\cc ({\rm Log}[2])\uu 4\cc \ddsc\uu 6\cc \MacroFLPLower 0[\ddsc]
 + ({\rm relative~}O(\ddsc\uu{-\hh}))
  \llsk
 \eee
 and
 \bbb
 \bbsk
 \bbsk
 \MacroFLPLower 5[\ddsc] = -{{4096}\over {15}}\cc ({\rm Log}[2])\uu 5\cc \ddsc\uu {10}\cc \MacroFLPLower 0\prprprprpr[\ddsc] 
  + ({\rm relative~}O(\ddsc\uu{-\hh}))
 = + {{2\uu{17}}\over{15}}\cc \pi\uu 5\cc  ({\rm Log}[2])\uu 5\cc \ddsc\uu{{{15}\over 2}}\cc \MacroFLPLower 0[\ddsc]
  + ({\rm relative~}O(\ddsc\uu{-\hh}))
  \llsk
 \eee

So, at least for \bbd{0 \leq p \leq 5} we have the pattern
\bbb
\MacroFLPLower p[\ddsc] =   {1\over{p!}}\cc (-8\cc {\rm Log}[2])\uu p\cc \ddsc\uu{2p}\cc \big (\cc {d\over{d\ddsc}}\cc \big )\uu p\cc \MacroFLPLower 0[\ddsc] + ({\rm smaller})\ ,
\een{AsymptoticsOfLargestTermAtPPlus1LoopsDoubleScaled}
where
\bbb
({\rm smaller}) \ni ({\rm exponentially~smaller}) + ({\rm smaller~by~powers~of~}\ddsc)\ ,
\eee
where
\bbb
({\rm exponentially~smaller}) \ni \ddsc\uu a \prod\ll{i = 1}\uu w\cc \bigg [\cc \big (\cc {d\over{d\ddsc}}\cc \big )\uu {r\ll i}\cc \MacroFLPLower 0[\ddsc]\cc \bigg ]\ , \llsk\llsk w\geq 2\ ,
\xxnn
 ({\rm smaller~by~powers~of~}\ddsc) \ni  \ddsc\uu a \cc \big (\cc {d\over{d\ddsc}}\cc \big )\uu r\cc \MacroFLPLower 0[\ddsc] \ , \llsk\llsk a - {r\over 2} < {{3p}\over 2}\ .
\eee
The exponentially smaller terms go as \bbd{\ddsc\uu {a - \hh\cc \sum\ll i r\ll i}\cc e\uu{- 4\pi w \ddsc\uu{+\hh}}} and the subleading-by-power-law terms go as
\bbd{\ddsc\uu{a - {r\over 2}}\cc e\uu{- 4\pi \ddsc\uu\hh}} which is smaller than \bbd{\ddsc\uu {{{3p}}\over 2}\cc e\uu{- 4\pi \ddsc\uu\hh}} at large \bbd{\ddsc}.

In conclusion, the analysis of the large-\bbd{\ddsc} limit of each \bbd{F\ls k[\ddsc]}
supports the interpretation of the breakdown of the double-scaling regime
as attributable to the threshold correction to the BPS bound controlling the mass of the lightest massive BPS particle.  This analysis has focused
on a particular term in \bbd{F\ls k[\ddsc]}, whose full functional form we have not yet given.  In the next subsection, sec. \ref{HigherMMPThroughNToMinusFive}, we
will present the full functional form of the double-scaled higher loop corrections  \bbd{n\uu{-k}\cc F\ls k[\ddsc]} to the MMP function.

\subsection{Order \bbd{n\uu{-2}} through \bbd{n\uu{-5}} corrections to the MMP function at fixed \bbd{\ddsc}}\label{HigherMMPThroughNToMinusFive}

We can carry out the same algorithm to derive the higher-order corrections \bbd{n\uu{-k}\cc \MacroFLPLower k[\ddsc]} to the MMP function \bbd{q\ll n\uprm{\textsc{mmp}}} as exact functions of \bbd{\ddsc} at higher order in \bbd{n}.
Using {\it Mathematica} to organize the algebraic calculations, we can write the recursion relation \rr{RecursionRelationDoubleScaledContext} as an asymptotic series in \bbd{n\uu{-1}}.  At order \bbd{n\uu{-k}} we obtain
a first-order linear inhomogeneous ODE for \bbd{\MacroFLPLower k[\ddsc]} that can be integrated in closed form for a general functional form for \bbd{\MacroFLPLower 0[\ddsc]}; that is, we obtain 
the general solution for each \bbd{\MacroFLPLower k[\ddsc]} in closed form as a polynomial in \bbd{\ddsc} and the derivatives of \bbd{\MacroFLPLower 0[\ddsc]}.  The single integration constant is fixed by
the condition that the \bbd{\MacroFLPLower k[\ddsc]} should fall off exponentially at large \bbd{\ddsc}.  

We present the results here up to and including \bbd{k=5,} the term of order \bbd{n\uu{-5}} at fixed \bbd{\ddsc}.

\def\MacroForFNought{\MacroFLPLower 0}
\begin{doublespace}
\bbb
\text{}\\
\MacroFLPLower 1[\ddsc ] \text{=}\ddsc  \MacroForFNought'[\ddsc ]-8 \ddsc ^2 \text{Log}[2] \MacroForFNought'[\ddsc ]+\frac{1}{4} \ddsc ^2 \MacroForFNought'[\ddsc
]^2\\
\xxx
\ordinary{\MacroFLPLower 2[\ddsc ] \text{=} -8 \ddsc ^2 \text{Log}[2] \MacroForFNought'[\ddsc ]+64 \ddsc ^3 \text{Log}[2]^2 \MacroForFNought'[\ddsc ]+\frac{1}{4}
\ddsc ^2 \MacroForFNought'[\ddsc ]^2-4 \ddsc ^3 \text{Log}[2] \MacroForFNought'[\ddsc ]^2+}\\
\ordinary{\frac{1}{12} \ddsc ^3 \MacroForFNought'[\ddsc ]^3+\frac{17}{32} \ddsc ^2 \MacroForFNought''[\ddsc ]-8 \ddsc ^3 \text{Log}[2] \MacroForFNought''[\ddsc
]+32 \ddsc ^4 \text{Log}[2]^2 \MacroForFNought''[\ddsc ]+}\\
\ordinary{\frac{1}{2} \ddsc ^3 \MacroForFNought'[\ddsc ] \MacroForFNought''[\ddsc ]-4 \ddsc ^4 \text{Log}[2] \MacroForFNought'[\ddsc ] \MacroForFNought''[\ddsc ]+\frac{1}{8}
\ddsc ^4 \MacroForFNought'[\ddsc ]^2 \MacroForFNought''[\ddsc ]+\frac{1}{48} \ddsc ^3 \MacroForFNought^{(3)}[\ddsc ]}\\
\xxx
\ordinary{\MacroFLPLower 3[\ddsc ] \text{=} 64 \ddsc ^3 \text{Log}[2]^2 \MacroForFNought'[\ddsc ]-512 \ddsc ^4 \text{Log}[2]^3 \MacroForFNought'[\ddsc ]-4
\ddsc ^3 \text{Log}[2] \MacroForFNought'[\ddsc ]^2+}\\
\ordinary{48 \ddsc ^4 \text{Log}[2]^2 \MacroForFNought'[\ddsc ]^2+\frac{1}{12} \ddsc ^3 \MacroForFNought'[\ddsc ]^3-2 \ddsc ^4 \text{Log}[2] \MacroForFNought'[\ddsc
]^3+\frac{1}{32} \ddsc ^4 \MacroForFNought'[\ddsc ]^4-}\\
\ordinary{\frac{17}{2} \ddsc ^3 \text{Log}[2] \MacroForFNought''[\ddsc ]+128 \ddsc ^4 \text{Log}[2]^2 \MacroForFNought''[\ddsc ]-512 \ddsc ^5 \text{Log}[2]^3
\MacroForFNought''[\ddsc ]+}\\
\ordinary{\frac{17}{32} \ddsc ^3 \MacroForFNought'[\ddsc ] \MacroForFNought''[\ddsc ]-16 \ddsc ^4 \text{Log}[2] \MacroForFNought'[\ddsc ] \MacroForFNought''[\ddsc
]+96 \ddsc ^5 \text{Log}[2]^2 \MacroForFNought'[\ddsc ] \MacroForFNought''[\ddsc ]+}\\
\ordinary{\frac{1}{2} \ddsc ^4 \MacroForFNought'[\ddsc ]^2 \MacroForFNought''[\ddsc ]-6 \ddsc ^5 \text{Log}[2] \MacroForFNought'[\ddsc ]^2 \MacroForFNought''[\ddsc
]+\frac{1}{8} \ddsc ^5 \MacroForFNought'[\ddsc ]^3 \MacroForFNought''[\ddsc ]+}\\
\ordinary{\frac{55}{192} \ddsc ^4 \MacroForFNought''[\ddsc ]^2-4 \ddsc ^5 \text{Log}[2] \MacroForFNought''[\ddsc ]^2+16 \ddsc ^6 \text{Log}[2]^2 \MacroForFNought''[\ddsc
]^2+\frac{1}{4} \ddsc ^5 \MacroForFNought'[\ddsc ] \MacroForFNought''[\ddsc ]^2-}\\
\ordinary{2 \ddsc ^6 \text{Log}[2] \MacroForFNought'[\ddsc ] \MacroForFNought''[\ddsc ]^2+\frac{1}{16} \ddsc ^6 \MacroForFNought'[\ddsc ]^2 \MacroForFNought''[\ddsc
]^2+\frac{7}{32} \ddsc ^3 \MacroForFNought^{(3)}[\ddsc ]-}\\
\ordinary{\frac{19}{4} \ddsc ^4 \text{Log}[2] \MacroForFNought^{(3)}[\ddsc ]+32 \ddsc ^5 \text{Log}[2]^2 \MacroForFNought^{(3)}[\ddsc ]-\frac{256}{3} \ddsc
^6 \text{Log}[2]^3 \MacroForFNought^{(3)}[\ddsc ]+}\\
\ordinary{\frac{19}{64} \ddsc ^4 \MacroForFNought'[\ddsc ] \MacroForFNought^{(3)}[\ddsc ]-4 \ddsc ^5 \text{Log}[2] \MacroForFNought'[\ddsc ] \MacroForFNought^{(3)}[\ddsc
]+16 \ddsc ^6 \text{Log}[2]^2 \MacroForFNought'[\ddsc ] \MacroForFNought^{(3)}[\ddsc ]+}\\
\ordinary{\frac{1}{8} \ddsc ^5 \MacroForFNought'[\ddsc ]^2 \MacroForFNought^{(3)}[\ddsc ]-\ddsc ^6 \text{Log}[2] \MacroForFNought'[\ddsc ]^2 \MacroForFNought^{(3)}[\ddsc
]+\frac{1}{48} \ddsc ^6 \MacroForFNought'[\ddsc ]^3 \MacroForFNought^{(3)}[\ddsc ]+}\\
\ordinary{\frac{1}{48} \ddsc ^5 \MacroForFNought''[\ddsc ] \MacroForFNought^{(3)}[\ddsc ]+\frac{1}{48} \ddsc ^4 \MacroForFNought^{(4)}[\ddsc ]-\frac{1}{6} \ddsc
^5 \text{Log}[2] \MacroForFNought^{(4)}[\ddsc ]+\frac{1}{96} \ddsc ^5 \MacroForFNought'[\ddsc ] \MacroForFNought^{(4)}[\ddsc ]}
\eee

\bbb
\ordinary{\MacroFLPLower 4[\ddsc ] \text{=} -512 \ddsc ^4 \text{Log}[2]^3 \MacroForFNought'[\ddsc ]+4096 \ddsc ^5 \text{Log}[2]^4 \MacroForFNought'[\ddsc ]+48
\ddsc ^4 \text{Log}[2]^2 \MacroForFNought'[\ddsc ]^2-}\\
\ordinary{512 \ddsc ^5 \text{Log}[2]^3 \MacroForFNought'[\ddsc ]^2-2 \ddsc ^4 \text{Log}[2] \MacroForFNought'[\ddsc ]^3+32 \ddsc ^5 \text{Log}[2]^2 \MacroForFNought'[\ddsc
]^3+\frac{1}{32} \ddsc ^4 \MacroForFNought'[\ddsc ]^4-}\\
\ordinary{\ddsc ^5 \text{Log}[2] \MacroForFNought'[\ddsc ]^4+\frac{1}{80} \ddsc ^5 \MacroForFNought'[\ddsc ]^5+102 \ddsc ^4 \text{Log}[2]^2 \MacroForFNought''[\ddsc
]-1536 \ddsc ^5 \text{Log}[2]^3 \MacroForFNought''[\ddsc ]+}\\
\ordinary{6144 \ddsc ^6 \text{Log}[2]^4 \MacroForFNought''[\ddsc ]-\frac{51}{4} \ddsc ^4 \text{Log}[2] \MacroForFNought'[\ddsc ] \MacroForFNought''[\ddsc ]+288
\ddsc ^5 \text{Log}[2]^2 \MacroForFNought'[\ddsc ] \MacroForFNought''[\ddsc ]-}\\
\ordinary{1536 \ddsc ^6 \text{Log}[2]^3 \MacroForFNought'[\ddsc ] \MacroForFNought''[\ddsc ]+\frac{51}{128} \ddsc ^4 \MacroForFNought'[\ddsc ]^2 \MacroForFNought''[\ddsc
]-18 \ddsc ^5 \text{Log}[2] \MacroForFNought'[\ddsc ]^2 \MacroForFNought''[\ddsc ]+}\\
\ordinary{144 \ddsc ^6 \text{Log}[2]^2 \MacroForFNought'[\ddsc ]^2 \MacroForFNought''[\ddsc ]+\frac{3}{8} \ddsc ^5 \MacroForFNought'[\ddsc ]^3 \MacroForFNought''[\ddsc
]-6 \ddsc ^6 \text{Log}[2] \MacroForFNought'[\ddsc ]^3 \MacroForFNought''[\ddsc ]+}\\
\ordinary{\frac{3}{32} \ddsc ^6 \MacroForFNought'[\ddsc ]^4 \MacroForFNought''[\ddsc ]+\frac{61}{192} \ddsc ^4 \MacroForFNought''[\ddsc ]^2-\frac{161}{12} \ddsc
^5 \text{Log}[2] \MacroForFNought''[\ddsc ]^2+}\\
\ordinary{144 \ddsc ^6 \text{Log}[2]^2 \MacroForFNought''[\ddsc ]^2-512 \ddsc ^7 \text{Log}[2]^3 \MacroForFNought''[\ddsc ]^2+\frac{161}{192} \ddsc ^5 \MacroForFNought'[\ddsc
] \MacroForFNought''[\ddsc ]^2-}
\\
\ordinary{18 \ddsc ^6 \text{Log}[2] \MacroForFNought'[\ddsc ] \MacroForFNought''[\ddsc ]^2+96 \ddsc ^7 \text{Log}[2]^2 \MacroForFNought'[\ddsc ] \MacroForFNought''[\ddsc
]^2+\frac{9}{16} \ddsc ^6 \MacroForFNought'[\ddsc ]^2 \MacroForFNought''[\ddsc ]^2-}\\
\ordinary{6 \ddsc ^7 \text{Log}[2] \MacroForFNought'[\ddsc ]^2 \MacroForFNought''[\ddsc ]^2+\frac{1}{8} \ddsc ^7 \MacroForFNought'[\ddsc ]^3 \MacroForFNought''[\ddsc
]^2+\frac{59}{384} \ddsc ^6 \MacroForFNought''[\ddsc ]^3-}\\
\ordinary{2 \ddsc ^7 \text{Log}[2] \MacroForFNought''[\ddsc ]^3+8 \ddsc ^8 \text{Log}[2]^2 \MacroForFNought''[\ddsc ]^3+\frac{1}{8} \ddsc ^7 \MacroForFNought'[\ddsc
] \MacroForFNought''[\ddsc ]^3-}\\
\ordinary{\ddsc ^8 \text{Log}[2] \MacroForFNought'[\ddsc ] \MacroForFNought''[\ddsc ]^3+\frac{1}{32} \ddsc ^8 \MacroForFNought'[\ddsc ]^2 \MacroForFNought''[\ddsc
]^3-\frac{21}{4} \ddsc ^4 \text{Log}[2] \MacroForFNought^{(3)}[\ddsc ]+}\\
\ordinary{110 \ddsc ^5 \text{Log}[2]^2 \MacroForFNought^{(3)}[\ddsc ]-768 \ddsc ^6 \text{Log}[2]^3 \MacroForFNought^{(3)}[\ddsc ]+2048 \ddsc ^7 \text{Log}[2]^4
\MacroForFNought^{(3)}[\ddsc ]+}\\
\ordinary{\frac{21}{64} \ddsc ^4 \MacroForFNought'[\ddsc ] \MacroForFNought^{(3)}[\ddsc ]-\frac{55}{4} \ddsc ^5 \text{Log}[2] \MacroForFNought'[\ddsc ] \MacroForFNought^{(3)}[\ddsc
]+144 \ddsc ^6 \text{Log}[2]^2 \MacroForFNought'[\ddsc ] \MacroForFNought^{(3)}[\ddsc ]-}\\
\ordinary{512 \ddsc ^7 \text{Log}[2]^3 \MacroForFNought'[\ddsc ] \MacroForFNought^{(3)}[\ddsc ]+\frac{55}{128} \ddsc ^5 \MacroForFNought'[\ddsc ]^2 \MacroForFNought^{(3)}[\ddsc
]-9 \ddsc ^6 \text{Log}[2] \MacroForFNought'[\ddsc ]^2 \MacroForFNought^{(3)}[\ddsc ]+}\\
\ordinary{48 \ddsc ^7 \text{Log}[2]^2 \MacroForFNought'[\ddsc ]^2 \MacroForFNought^{(3)}[\ddsc ]+\frac{3}{16} \ddsc ^6 \MacroForFNought'[\ddsc ]^3 \MacroForFNought^{(3)}[\ddsc
]-2 \ddsc ^7 \text{Log}[2] \MacroForFNought'[\ddsc ]^3 \MacroForFNought^{(3)}[\ddsc ]+}\\
\ordinary{\frac{1}{32} \ddsc ^7 \MacroForFNought'[\ddsc ]^4 \MacroForFNought^{(3)}[\ddsc ]+\frac{79}{192} \ddsc ^5 \MacroForFNought''[\ddsc ] \MacroForFNought^{(3)}[\ddsc
]-\frac{187}{24} \ddsc ^6 \text{Log}[2] \MacroForFNought''[\ddsc ] \MacroForFNought^{(3)}[\ddsc ]+}\\
\ordinary{48 \ddsc ^7 \text{Log}[2]^2 \MacroForFNought''[\ddsc ] \MacroForFNought^{(3)}[\ddsc ]-128 \ddsc ^8 \text{Log}[2]^3 \MacroForFNought''[\ddsc ] \MacroForFNought^{(3)}[\ddsc
]+}\\
\ordinary{\frac{187}{384} \ddsc ^6 \MacroForFNought'[\ddsc ] \MacroForFNought''[\ddsc ] \MacroForFNought^{(3)}[\ddsc ]-6 \ddsc ^7 \text{Log}[2] \MacroForFNought'[\ddsc
] \MacroForFNought''[\ddsc ] \MacroForFNought^{(3)}[\ddsc ]+
\blue{\rightarrow\rightarrow\rightarrow\downarrow\downarrow\downarrow}}\\ \text{{\rm (continued~on~next~page,~sorry!)}
}
\xxnn
 \text{{\rm (second page of the expression for \bbd{\MacroFLPLower 4[\ddsc]}, continued~from~previous~page)}}\\
\ordinary{\blue{\uparrow}\blue{\uparrow}\blue{\uparrow}\blue{\leftarrow}\blue{\leftarrow}\blue{\leftarrow}+24 \ddsc ^8 \text{Log}[2]^2 \MacroForFNought'[\ddsc ] \MacroForFNought''[\ddsc ] \MacroForFNought^{(3)}[\ddsc ]+\frac{3}{16} \ddsc ^7 \MacroForFNought'[\ddsc
]^2 \MacroForFNought''[\ddsc ] \MacroForFNought^{(3)}[\ddsc ]-}\\
\ordinary{\frac{3}{2} \ddsc ^8 \text{Log}[2] \MacroForFNought'[\ddsc ]^2 \MacroForFNought''[\ddsc ] \MacroForFNought^{(3)}[\ddsc ]+\frac{1}{32} \ddsc ^8 \MacroForFNought'[\ddsc
]^3 \MacroForFNought''[\ddsc ] \MacroForFNought^{(3)}[\ddsc ]+}\\
\ordinary{\frac{1}{64} \ddsc ^7 \MacroForFNought''[\ddsc ]^2 \MacroForFNought^{(3)}[\ddsc ]+\frac{1}{48} \ddsc ^6 \MacroForFNought^{(3)}[\ddsc ]^2-\frac{1}{6}
\ddsc ^7 \text{Log}[2] \MacroForFNought^{(3)}[\ddsc ]^2+}\\
\ordinary{\frac{1}{96} \ddsc ^7 \MacroForFNought'[\ddsc ] \MacroForFNought^{(3)}[\ddsc ]^2+\frac{163 \ddsc ^4 \MacroForFNought^{(4)}[\ddsc ]}{2048}-\frac{29}{12}
\ddsc ^5 \text{Log}[2] \MacroForFNought^{(4)}[\ddsc ]+}\\
\ordinary{\frac{67}{3} \ddsc ^6 \text{Log}[2]^2 \MacroForFNought^{(4)}[\ddsc ]-\frac{256}{3} \ddsc ^7 \text{Log}[2]^3 \MacroForFNought^{(4)}[\ddsc ]+\frac{512}{3}
\ddsc ^8 \text{Log}[2]^4 \MacroForFNought^{(4)}[\ddsc ]+
}
\\
\ordinary{+\frac{29}{192} \ddsc ^5 \MacroForFNought'[\ddsc ] \MacroForFNought^{(4)}[\ddsc ]-
\frac{67}{24} \ddsc ^6 \text{Log}[2] \MacroForFNought'[\ddsc ] \MacroForFNought^{(4)}[\ddsc
]+16 \ddsc ^7 \text{Log}[2]^2 \MacroForFNought'[\ddsc ] \MacroForFNought^{(4)}[\ddsc ]- } \\
\ordinary{\frac{128}{3} \ddsc ^8 \text{Log}[2]^3 \MacroForFNought'[\ddsc ] \MacroForFNought^{(4)}[\ddsc ]+\frac{67}{768} \ddsc ^6 \MacroForFNought'[\ddsc ]^2
\MacroForFNought^{(4)}[\ddsc ]-\ddsc ^7 \text{Log}[2] \MacroForFNought'[\ddsc ]^2 \MacroForFNought^{(4)}[\ddsc ]+}\\
\ordinary{4 \ddsc ^8 \text{Log}[2]^2 \MacroForFNought'[\ddsc ]^2 \MacroForFNought^{(4)}[\ddsc ]+\frac{1}{48} \ddsc ^7 \MacroForFNought'[\ddsc ]^3 \MacroForFNought^{(4)}[\ddsc
]-\frac{1}{6} \ddsc ^8 \text{Log}[2] \MacroForFNought'[\ddsc ]^3 \MacroForFNought^{(4)}[\ddsc ]+}\\
\ordinary{\frac{1}{384} \ddsc ^8 \MacroForFNought'[\ddsc ]^4 \MacroForFNought^{(4)}[\ddsc ]+\frac{1}{32} \ddsc ^6 \MacroForFNought''[\ddsc ] \MacroForFNought^{(4)}[\ddsc
]-\frac{1}{4} \ddsc ^7 \text{Log}[2] \MacroForFNought''[\ddsc ] \MacroForFNought^{(4)}[\ddsc ]+}\\
\ordinary{\frac{1}{64} \ddsc ^7 \MacroForFNought'[\ddsc ] \MacroForFNought''[\ddsc ] \MacroForFNought^{(4)}[\ddsc ]+\frac{91 \ddsc ^5 \MacroForFNought^{(5)}[\ddsc
]}{7680}-\frac{1}{6} \ddsc ^6 \text{Log}[2] \MacroForFNought^{(5)}[\ddsc ]+}\\
\ordinary{\frac{2}{3} \ddsc ^7 \text{Log}[2]^2 \MacroForFNought^{(5)}[\ddsc ]+\frac{1}{96} \ddsc ^6 \MacroForFNought'[\ddsc ] \MacroForFNought^{(5)}[\ddsc ]-\frac{1}{12}
\ddsc ^7 \text{Log}[2] \MacroForFNought'[\ddsc ] \MacroForFNought^{(5)}[\ddsc ]+}\\
\ordinary{\frac{1}{384} \ddsc ^7 \MacroForFNought'[\ddsc ]^2 \MacroForFNought^{(5)}[\ddsc ]+\frac{\ddsc ^6 \MacroForFNought^{(6)}[\ddsc ]}{4608}} 
\eenn
\end{doublespace}
\pagebreak
The formula for the coupling-dependence \bbd{\MacroFLPLower 5[\ddsc]} of the \bbd{n\uu{-5}} term in the exponent of the MMP amplitude, is lengthy
and cannot be fit easily on fewer than five pages.  We give the expression here but this is the last order we're going to compute.  For the
function \bbd{\MacroFLPLower 5[\ddsc]} we have:
\begin{doublespace}
\bbb
\ordinary{\MacroFLPLower 5[\ddsc ] \text{=} 4096 \ddsc ^5 \text{Log}[2]^4 \MacroForFNought'[\ddsc ]-32768 \ddsc ^6 \text{Log}[2]^5 \MacroForFNought'[\ddsc
]-512 \ddsc ^5 \text{Log}[2]^3 \MacroForFNought'[\ddsc ]^2+}\\
\ordinary{5120 \ddsc ^6 \text{Log}[2]^4 \MacroForFNought'[\ddsc ]^2+32 \ddsc ^5 \text{Log}[2]^2 \MacroForFNought'[\ddsc ]^3-\frac{1280}{3} \ddsc ^6 \text{Log}[2]^3
\MacroForFNought'[\ddsc ]^3-}\\
\ordinary{\ddsc ^5 \text{Log}[2] \MacroForFNought'[\ddsc ]^4+20 \ddsc ^6 \text{Log}[2]^2 \MacroForFNought'[\ddsc ]^4+\frac{1}{80} \ddsc ^5 \MacroForFNought'[\ddsc
]^5-\frac{1}{2} \ddsc ^6 \text{Log}[2] \MacroForFNought'[\ddsc ]^5+}\\
\ordinary{\frac{1}{192} \ddsc ^6 \MacroForFNought'[\ddsc ]^6-1088 \ddsc ^5 \text{Log}[2]^3 \MacroForFNought''[\ddsc ]+16384 \ddsc ^6 \text{Log}[2]^4 \MacroForFNought''[\ddsc
]-}\\
\ordinary{65536 \ddsc ^7 \text{Log}[2]^5 \MacroForFNought''[\ddsc ]+204 \ddsc ^5 \text{Log}[2]^2 \MacroForFNought'[\ddsc ] \MacroForFNought''[\ddsc ]-4096 \ddsc
^6 \text{Log}[2]^3 \MacroForFNought'[\ddsc ] \MacroForFNought''[\ddsc ]+}\\
\ordinary{20480 \ddsc ^7 \text{Log}[2]^4 \MacroForFNought'[\ddsc ] \MacroForFNought''[\ddsc ]-\frac{51}{4} \ddsc ^5 \text{Log}[2] \MacroForFNought'[\ddsc ]^2
\MacroForFNought''[\ddsc ]+}\\
\ordinary{384 \ddsc ^6 \text{Log}[2]^2 \MacroForFNought'[\ddsc ]^2 \MacroForFNought''[\ddsc ]-2560 \ddsc ^7 \text{Log}[2]^3 \MacroForFNought'[\ddsc ]^2 \MacroForFNought''[\ddsc
]+}\\
\ordinary{\frac{17}{64} \ddsc ^5 \MacroForFNought'[\ddsc ]^3 \MacroForFNought''[\ddsc ]-16 \ddsc ^6 \text{Log}[2] \MacroForFNought'[\ddsc ]^3 \MacroForFNought''[\ddsc
]+160 \ddsc ^7 \text{Log}[2]^2 \MacroForFNought'[\ddsc ]^3 \MacroForFNought''[\ddsc ]+}\\
\ordinary{\frac{1}{4} \ddsc ^6 \MacroForFNought'[\ddsc ]^4 \MacroForFNought''[\ddsc ]-5 \ddsc ^7 \text{Log}[2] \MacroForFNought'[\ddsc ]^4 \MacroForFNought''[\ddsc
]+\frac{1}{16} \ddsc ^7 \MacroForFNought'[\ddsc ]^5 \MacroForFNought''[\ddsc ]-}\\
\ordinary{\frac{61}{6} \ddsc ^5 \text{Log}[2] \MacroForFNought''[\ddsc ]^2+\frac{958}{3} \ddsc ^6 \text{Log}[2]^2 \MacroForFNought''[\ddsc ]^2-3072 \ddsc
^7 \text{Log}[2]^3 \MacroForFNought''[\ddsc ]^2+
\\
10240 \ddsc ^8 \text{Log}[2]^4 \MacroForFNought''[\ddsc ]^2+\frac{61}{96} \ddsc ^5 \MacroForFNought'[\ddsc ] \MacroForFNought''[\ddsc ]^2-\frac{479}{12}
\ddsc ^6 \text{Log}[2] \MacroForFNought'[\ddsc ] \MacroForFNought''[\ddsc ]^2+}\\
\ordinary{576 \ddsc ^7 \text{Log}[2]^2 \MacroForFNought'[\ddsc ] \MacroForFNought''[\ddsc ]^2-2560 \ddsc ^8 \text{Log}[2]^3 \MacroForFNought'[\ddsc ] \MacroForFNought''[\ddsc
]^2+ \blue{\rightarrow\rightarrow\rightarrow\downarrow\downarrow\downarrow}}\\ \text{{\rm (continued~on~next~page,~sorry!)}} 
\eee
\pagebreak 
\bbb
 \text{{\rm (second page of the expression for \bbd{\MacroFLPLower 5[\ddsc]}, continued~from~previous~page)}}\\
\ordinary{\blue{\uparrow}\blue{\uparrow}\blue{\uparrow}\blue{\leftarrow}\blue{\leftarrow}\blue{\leftarrow}+\frac{479}{384} \ddsc ^6 \MacroForFNought'[\ddsc ]^2 \MacroForFNought''[\ddsc ]^2-36 \ddsc ^7 \text{Log}[2] \MacroForFNought'[\ddsc ]^2 \MacroForFNought''[\ddsc
]^2+240 \ddsc ^8 \text{Log}[2]^2 \MacroForFNought'[\ddsc ]^2 \MacroForFNought''[\ddsc ]^2+}\\
\ordinary{\frac{3}{4} \ddsc ^7 \MacroForFNought'[\ddsc ]^3 \MacroForFNought''[\ddsc ]^2-10 \ddsc ^8 \text{Log}[2] \MacroForFNought'[\ddsc ]^3 \MacroForFNought''[\ddsc
]^2+\frac{5}{32} \ddsc ^8 \MacroForFNought'[\ddsc ]^4 \MacroForFNought''[\ddsc ]^2+}\\
\ordinary{\frac{23}{48} \ddsc ^6 \MacroForFNought''[\ddsc ]^3-\frac{169}{12} \ddsc ^7 \text{Log}[2] \MacroForFNought''[\ddsc ]^3+128 \ddsc ^8 \text{Log}[2]^2
\MacroForFNought''[\ddsc ]^3-}\\
\ordinary{\frac{1280}{3} \ddsc ^9 \text{Log}[2]^3 \MacroForFNought''[\ddsc ]^3+\frac{169}{192} \ddsc ^7 \MacroForFNought'[\ddsc ] \MacroForFNought''[\ddsc ]^3-16
\ddsc ^8 \text{Log}[2] \MacroForFNought'[\ddsc ] \MacroForFNought''[\ddsc ]^3+}\\
\ordinary{80 \ddsc ^9 \text{Log}[2]^2 \MacroForFNought'[\ddsc ] \MacroForFNought''[\ddsc ]^3+\frac{1}{2} \ddsc ^8 \MacroForFNought'[\ddsc ]^2 \MacroForFNought''[\ddsc
]^3-5 \ddsc ^9 \text{Log}[2] \MacroForFNought'[\ddsc ]^2 \MacroForFNought''[\ddsc ]^3+}\\
\ordinary{\frac{5}{48} \ddsc ^9 \MacroForFNought'[\ddsc ]^3 \MacroForFNought''[\ddsc ]^3+\frac{21}{256} \ddsc ^8 \MacroForFNought''[\ddsc ]^4-\ddsc ^9 \text{Log}[2]
\MacroForFNought''[\ddsc ]^4+4 \ddsc ^{10} \text{Log}[2]^2 \MacroForFNought''[\ddsc ]^4+}\\
\ordinary{\frac{1}{16} \ddsc ^9 \MacroForFNought'[\ddsc ] \MacroForFNought''[\ddsc ]^4-\frac{1}{2} \ddsc ^{10} \text{Log}[2] \MacroForFNought'[\ddsc ] \MacroForFNought''[\ddsc
]^4+\frac{1}{64} \ddsc ^{10} \MacroForFNought'[\ddsc ]^2 \MacroForFNought''[\ddsc ]^4+}\\
\ordinary{84 \ddsc ^5 \text{Log}[2]^2 \MacroForFNought^{(3)}[\ddsc ]-\frac{5216}{3} \ddsc ^6 \text{Log}[2]^3 \MacroForFNought^{(3)}[\ddsc ]+12288 \ddsc
^7 \text{Log}[2]^4 \MacroForFNought^{(3)}[\ddsc ]-}\\
\ordinary{32768 \ddsc ^8 \text{Log}[2]^5 \MacroForFNought^{(3)}[\ddsc ]-\frac{21}{2} \ddsc ^5 \text{Log}[2] \MacroForFNought'[\ddsc ] \MacroForFNought^{(3)}[\ddsc
]+326 \ddsc ^6 \text{Log}[2]^2 \MacroForFNought'[\ddsc ] \MacroForFNought^{(3)}[\ddsc ]-}\\
\ordinary{3072 \ddsc ^7 \text{Log}[2]^3 \MacroForFNought'[\ddsc ] \MacroForFNought^{(3)}[\ddsc ]+10240 \ddsc ^8 \text{Log}[2]^4 \MacroForFNought'[\ddsc ] \MacroForFNought^{(3)}[\ddsc
]+}\\
\ordinary{\frac{21}{64} \ddsc ^5 \MacroForFNought'[\ddsc ]^2 \MacroForFNought^{(3)}[\ddsc ]-\frac{163}{8} \ddsc ^6 \text{Log}[2] \MacroForFNought'[\ddsc ]^2 \MacroForFNought^{(3)}[\ddsc
]+}\\
\ordinary{288 \ddsc ^7 \text{Log}[2]^2 \MacroForFNought'[\ddsc ]^2 \MacroForFNought^{(3)}[\ddsc ]-1280 \ddsc ^8 \text{Log}[2]^3 \MacroForFNought'[\ddsc ]^2 \MacroForFNought^{(3)}[\ddsc
]+}\\
\ordinary{\frac{163}{384} \ddsc ^6 \MacroForFNought'[\ddsc ]^3 \MacroForFNought^{(3)}[\ddsc ]-12 \ddsc ^7 \text{Log}[2] \MacroForFNought'[\ddsc ]^3 \MacroForFNought^{(3)}[\ddsc
]+80 \ddsc ^8 \text{Log}[2]^2 \MacroForFNought'[\ddsc ]^3 \MacroForFNought^{(3)}[\ddsc ]+}\\
\ordinary{\frac{3}{16} \ddsc ^7 \MacroForFNought'[\ddsc ]^4 \MacroForFNought^{(3)}[\ddsc ]-\frac{5}{2} \ddsc ^8 \text{Log}[2] \MacroForFNought'[\ddsc ]^4 \MacroForFNought^{(3)}[\ddsc
]+\frac{1}{32} \ddsc ^8 \MacroForFNought'[\ddsc ]^5 \MacroForFNought^{(3)}[\ddsc ]+}\\
\ordinary{\frac{455 \ddsc ^5 \MacroForFNought''[\ddsc ] \MacroForFNought^{(3)}[\ddsc ]}{1024}-\frac{145}{6} \ddsc ^6 \text{Log}[2] \MacroForFNought''[\ddsc ]
\MacroForFNought^{(3)}[\ddsc ]+}\\
\ordinary{\frac{1048}{3} \ddsc ^7 \text{Log}[2]^2 \MacroForFNought''[\ddsc ] \MacroForFNought^{(3)}[\ddsc ]-2048 \ddsc ^8 \text{Log}[2]^3 \MacroForFNought''[\ddsc
] \MacroForFNought^{(3)}[\ddsc ]+}\\
\ordinary{5120 \ddsc ^9 \text{Log}[2]^4 \MacroForFNought''[\ddsc ] \MacroForFNought^{(3)}[\ddsc ]+\frac{145}{96} \ddsc ^6 \MacroForFNought'[\ddsc ] \MacroForFNought''[\ddsc
] \MacroForFNought^{(3)}[\ddsc ]-}\\
\ordinary{\frac{131}{3} \ddsc ^7 \text{Log}[2] \MacroForFNought'[\ddsc ] \MacroForFNought''[\ddsc ] \MacroForFNought^{(3)}[\ddsc ]+384 \ddsc ^8 \text{Log}[2]^2
\MacroForFNought'[\ddsc ] \MacroForFNought''[\ddsc ] \MacroForFNought^{(3)}[\ddsc ]-}\\
\ordinary{1280 \ddsc ^9 \text{Log}[2]^3 \MacroForFNought'[\ddsc ] \MacroForFNought''[\ddsc ] \MacroForFNought^{(3)}[\ddsc ]+\frac{131}{96} \ddsc ^7 \MacroForFNought'[\ddsc
]^2 \MacroForFNought''[\ddsc ] \MacroForFNought^{(3)}[\ddsc ]-}\\
\ordinary{24 \ddsc ^8 \text{Log}[2] \MacroForFNought'[\ddsc ]^2 \MacroForFNought''[\ddsc ] \MacroForFNought^{(3)}[\ddsc ]+120 \ddsc ^9 \text{Log}[2]^2 \MacroForFNought'[\ddsc
]^2 \MacroForFNought''[\ddsc ] \MacroForFNought^{(3)}[\ddsc ] +
\blue{\rightarrow\rightarrow\rightarrow\downarrow\downarrow\downarrow}}\\ \text{{\rm (continued~on~next~page,~sorry!)}} 
\xxnn
\pagebreak \\ \text{{\rm (third page of the expression for \bbd{\MacroFLPLower 5[\ddsc]}, continued~from~previous~page)}}\\
\ordinary{\blue{\uparrow}\blue{\uparrow}\blue{\uparrow}\blue{\leftarrow}\blue{\leftarrow}\blue{\leftarrow}+ \frac{1}{2} \ddsc ^8 \MacroForFNought'[\ddsc ]^3 \MacroForFNought''[\ddsc ] \MacroForFNought^{(3)}[\ddsc ]-5 \ddsc ^9 \text{Log}[2] \MacroForFNought'[\ddsc
]^3 \MacroForFNought''[\ddsc ] \MacroForFNought^{(3)}[\ddsc ]+}\\
\ordinary{\frac{5}{64} \ddsc ^9 \MacroForFNought'[\ddsc ]^4 \MacroForFNought''[\ddsc ] \MacroForFNought^{(3)}[\ddsc ]+\frac{95}{192} \ddsc ^7 \MacroForFNought''[\ddsc
]^2 \MacroForFNought^{(3)}[\ddsc ]-}\\
\ordinary{\frac{203}{24} \ddsc ^8 \text{Log}[2] \MacroForFNought''[\ddsc ]^2 \MacroForFNought^{(3)}[\ddsc ]+48 \ddsc ^9 \text{Log}[2]^2 \MacroForFNought''[\ddsc
]^2 \MacroForFNought^{(3)}[\ddsc ]-}\\
\ordinary{128 \ddsc ^{10} \text{Log}[2]^3 \MacroForFNought''[\ddsc ]^2 \MacroForFNought^{(3)}[\ddsc ]+\frac{203}{384} \ddsc ^8 \MacroForFNought'[\ddsc ] \MacroForFNought''[\ddsc
]^2 \MacroForFNought^{(3)}[\ddsc ]-}\\
\ordinary{6 \ddsc ^9 \text{Log}[2] \MacroForFNought'[\ddsc ] \MacroForFNought''[\ddsc ]^2 \MacroForFNought^{(3)}[\ddsc ]+24 \ddsc ^{10} \text{Log}[2]^2 \MacroForFNought'[\ddsc
] \MacroForFNought''[\ddsc ]^2 \MacroForFNought^{(3)}[\ddsc ]+}\\
\ordinary{\frac{3}{16} \ddsc ^9 \MacroForFNought'[\ddsc ]^2 \MacroForFNought''[\ddsc ]^2 \MacroForFNought^{(3)}[\ddsc ]-\frac{3}{2} \ddsc ^{10} \text{Log}[2]
\MacroForFNought'[\ddsc ]^2 \MacroForFNought''[\ddsc ]^2 \MacroForFNought^{(3)}[\ddsc ]+}\\
\ordinary{\frac{1}{32} \ddsc ^{10} \MacroForFNought'[\ddsc ]^3 \MacroForFNought''[\ddsc ]^2 \MacroForFNought^{(3)}[\ddsc ]+\frac{1}{96} \ddsc ^9 \MacroForFNought''[\ddsc
]^3 \MacroForFNought^{(3)}[\ddsc ]+\frac{10631 \ddsc ^6 \MacroForFNought^{(3)}[\ddsc ]^2}{61440}-}\\
\ordinary{\frac{103}{24} \ddsc ^7 \text{Log}[2] \MacroForFNought^{(3)}[\ddsc ]^2+\frac{215}{6} \ddsc ^8 \text{Log}[2]^2 \MacroForFNought^{(3)}[\ddsc ]^2-128
\ddsc ^9 \text{Log}[2]^3 \MacroForFNought^{(3)}[\ddsc ]^2+}\\
\ordinary{256 \ddsc ^{10} \text{Log}[2]^4 \MacroForFNought^{(3)}[\ddsc ]^2+\frac{103}{384} \ddsc ^7 \MacroForFNought'[\ddsc ] \MacroForFNought^{(3)}[\ddsc ]^2-\frac{215}{48}
\ddsc ^8 \text{Log}[2] \MacroForFNought'[\ddsc ] \MacroForFNought^{(3)}[\ddsc ]^2+}\\
\ordinary{24 \ddsc ^9 \text{Log}[2]^2 \MacroForFNought'[\ddsc ] \MacroForFNought^{(3)}[\ddsc ]^2-64 \ddsc ^{10} \text{Log}[2]^3 \MacroForFNought'[\ddsc ] \MacroForFNought^{(3)}[\ddsc
]^2+}\\
\ordinary{\frac{215 \ddsc ^8 \MacroForFNought'[\ddsc ]^2 \MacroForFNought^{(3)}[\ddsc ]^2}{1536}-\frac{3}{2} \ddsc ^9 \text{Log}[2] \MacroForFNought'[\ddsc ]^2
\MacroForFNought^{(3)}[\ddsc ]^2+}\\
\ordinary{6 \ddsc ^{10} \text{Log}[2]^2 \MacroForFNought'[\ddsc ]^2 \MacroForFNought^{(3)}[\ddsc ]^2+\frac{1}{32} \ddsc ^9 \MacroForFNought'[\ddsc ]^3 \MacroForFNought^{(3)}[\ddsc
]^2-}\\
\ordinary{\frac{1}{4} \ddsc ^{10} \text{Log}[2] \MacroForFNought'[\ddsc ]^3 \MacroForFNought^{(3)}[\ddsc ]^2+\frac{1}{256} \ddsc ^{10} \MacroForFNought'[\ddsc
]^4 \MacroForFNought^{(3)}[\ddsc ]^2+\frac{1}{24} \ddsc ^8 \MacroForFNought''[\ddsc ] \MacroForFNought^{(3)}[\ddsc ]^2-}\\
\ordinary{\frac{1}{3} \ddsc ^9 \text{Log}[2] \MacroForFNought''[\ddsc ] \MacroForFNought^{(3)}[\ddsc ]^2+\frac{1}{48} \ddsc ^9 \MacroForFNought'[\ddsc ] \MacroForFNought''[\ddsc
] \MacroForFNought^{(3)}[\ddsc ]^2-}\\
\ordinary{\frac{163}{64} \ddsc ^5 \text{Log}[2] \MacroForFNought^{(4)}[\ddsc ]+\frac{208}{3} \ddsc ^6 \text{Log}[2]^2 \MacroForFNought^{(4)}[\ddsc ]-\frac{1952}{3}
\ddsc ^7 \text{Log}[2]^3 \MacroForFNought^{(4)}[\ddsc ]+}\\
\ordinary{\frac{8192}{3} \ddsc ^8 \text{Log}[2]^4 \MacroForFNought^{(4)}[\ddsc ]-\frac{16384}{3} \ddsc ^9 \text{Log}[2]^5 \MacroForFNought^{(4)}[\ddsc ]+\frac{163
\ddsc ^5 \MacroForFNought'[\ddsc ] \MacroForFNought^{(4)}[\ddsc ]}{1024}-}\\
\ordinary{\frac{26}{3} \ddsc ^6 \text{Log}[2] \MacroForFNought'[\ddsc ] \MacroForFNought^{(4)}[\ddsc ]+122 \ddsc ^7 \text{Log}[2]^2 \MacroForFNought'[\ddsc ]
\MacroForFNought^{(4)}[\ddsc ]-}\\
\ordinary{\frac{2048}{3} \ddsc ^8 \text{Log}[2]^3 \MacroForFNought'[\ddsc ] \MacroForFNought^{(4)}[\ddsc ]+\frac{5120}{3} \ddsc ^9 \text{Log}[2]^4 \MacroForFNought'[\ddsc
] \MacroForFNought^{(4)}[\ddsc ] +
\blue{\rightarrow\rightarrow\rightarrow\downarrow\downarrow\downarrow}}\\ \text{{\rm (continued~on~next~page,~sorry!)}} 
\xxnn
\pagebreak \\ \text{{\rm (fourth page of the expression for \bbd{\MacroFLPLower 5[\ddsc]}, continued~from~previous~page)}}\\
\ordinary{\blue{\uparrow}\blue{\uparrow}\blue{\uparrow}\blue{\leftarrow}\blue{\leftarrow}\blue{\leftarrow}+\frac{13}{48} \ddsc ^6 \MacroForFNought'[\ddsc ]^2 \MacroForFNought^{(4)}[\ddsc ]-\frac{61}{8} \ddsc ^7 \text{Log}[2] \MacroForFNought'[\ddsc ]^2 \MacroForFNought^{(4)}[\ddsc
]}\\
\ordinary{ +64 \ddsc ^8 \text{Log}[2]^2 \MacroForFNought'[\ddsc ]^2 \MacroForFNought^{(4)}[\ddsc ]- \frac{640}{3} \ddsc ^9 \text{Log}[2]^3 \MacroForFNought'[\ddsc ]^2 \MacroForFNought^{(4)}[\ddsc ]+\frac{61}{384} \ddsc ^7 \MacroForFNought'[\ddsc ]^3
\MacroForFNought^{(4)}[\ddsc ]-}\\
\ordinary{\frac{8}{3} \ddsc ^8 \text{Log}[2] \MacroForFNought'[\ddsc ]^3 \MacroForFNought^{(4)}[\ddsc ]+\frac{40}{3} \ddsc ^9 \text{Log}[2]^2 \MacroForFNought'[\ddsc
]^3 \MacroForFNought^{(4)}[\ddsc ]+\frac{1}{24} \ddsc ^8 \MacroForFNought'[\ddsc ]^4 \MacroForFNought^{(4)}[\ddsc ]-}\\
\ordinary{\frac{5}{12} \ddsc ^9 \text{Log}[2] \MacroForFNought'[\ddsc ]^4 \MacroForFNought^{(4)}[\ddsc ]+\frac{1}{192} \ddsc ^9 \MacroForFNought'[\ddsc ]^5 \MacroForFNought^{(4)}[\ddsc
]+\frac{743 \ddsc ^6 \MacroForFNought''[\ddsc ] \MacroForFNought^{(4)}[\ddsc ]}{3072}-}\\
\ordinary{6 \ddsc ^7 \text{Log}[2] \MacroForFNought''[\ddsc ] \MacroForFNought^{(4)}[\ddsc ]+\frac{148}{3} \ddsc ^8 \text{Log}[2]^2 \MacroForFNought''[\ddsc ]
\MacroForFNought^{(4)}[\ddsc ]-}\\
\ordinary{\frac{512}{3} \ddsc ^9 \text{Log}[2]^3 \MacroForFNought''[\ddsc ] \MacroForFNought^{(4)}[\ddsc ]+\frac{1024}{3} \ddsc ^{10} \text{Log}[2]^4 \MacroForFNought''[\ddsc
] \MacroForFNought^{(4)}[\ddsc ]+}\\
\ordinary{\frac{3}{8} \ddsc ^7 \MacroForFNought'[\ddsc ] \MacroForFNought''[\ddsc ] \MacroForFNought^{(4)}[\ddsc ]-\frac{37}{6} \ddsc ^8 \text{Log}[2] \MacroForFNought'[\ddsc
] \MacroForFNought''[\ddsc ] \MacroForFNought^{(4)}[\ddsc ]+}\\
\ordinary{32 \ddsc ^9 \text{Log}[2]^2 \MacroForFNought'[\ddsc ] \MacroForFNought''[\ddsc ] \MacroForFNought^{(4)}[\ddsc ]-\frac{256}{3} \ddsc ^{10} \text{Log}[2]^3
\MacroForFNought'[\ddsc ] \MacroForFNought''[\ddsc ] \MacroForFNought^{(4)}[\ddsc ]+}\\
\ordinary{\frac{37}{192} \ddsc ^8 \MacroForFNought'[\ddsc ]^2 \MacroForFNought''[\ddsc ] \MacroForFNought^{(4)}[\ddsc ]-2 \ddsc ^9 \text{Log}[2] \MacroForFNought'[\ddsc
]^2 \MacroForFNought''[\ddsc ] \MacroForFNought^{(4)}[\ddsc ]+}\\
\ordinary{8 \ddsc ^{10} \text{Log}[2]^2 \MacroForFNought'[\ddsc ]^2 \MacroForFNought''[\ddsc ] \MacroForFNought^{(4)}[\ddsc ]+\frac{1}{24} \ddsc ^9 \MacroForFNought'[\ddsc
]^3 \MacroForFNought''[\ddsc ] \MacroForFNought^{(4)}[\ddsc ]-}\\
\ordinary{\frac{1}{3} \ddsc ^{10} \text{Log}[2] \MacroForFNought'[\ddsc ]^3 \MacroForFNought''[\ddsc ] \MacroForFNought^{(4)}[\ddsc ]+\frac{1}{192} \ddsc ^{10}
\MacroForFNought'[\ddsc ]^4 \MacroForFNought''[\ddsc ] \MacroForFNought^{(4)}[\ddsc ]+}\\
\ordinary{\frac{1}{32} \ddsc ^8 \MacroForFNought''[\ddsc ]^2 \MacroForFNought^{(4)}[\ddsc ]-\frac{1}{4} \ddsc ^9 \text{Log}[2] \MacroForFNought''[\ddsc ]^2 \MacroForFNought^{(4)}[\ddsc
]+\frac{1}{64} \ddsc ^9 \MacroForFNought'[\ddsc ] \MacroForFNought''[\ddsc ]^2 \MacroForFNought^{(4)}[\ddsc ]+}\\
\ordinary{\frac{231 \ddsc ^7 \MacroForFNought^{(3)}[\ddsc ] \MacroForFNought^{(4)}[\ddsc ]}{5120}-\frac{7}{12} \ddsc ^8 \text{Log}[2] \MacroForFNought^{(3)}[\ddsc
] \MacroForFNought^{(4)}[\ddsc ]+}\\
\ordinary{\frac{7}{3} \ddsc ^9 \text{Log}[2]^2 \MacroForFNought^{(3)}[\ddsc ] \MacroForFNought^{(4)}[\ddsc ]+\frac{7}{192} \ddsc ^8 \MacroForFNought'[\ddsc ]
\MacroForFNought^{(3)}[\ddsc ] \MacroForFNought^{(4)}[\ddsc ]-}\\
\ordinary{\frac{7}{24} \ddsc ^9 \text{Log}[2] \MacroForFNought'[\ddsc ] \MacroForFNought^{(3)}[\ddsc ] \MacroForFNought^{(4)}[\ddsc ]+\frac{7}{768} \ddsc ^9 \MacroForFNought'[\ddsc
]^2 \MacroForFNought^{(3)}[\ddsc ] \MacroForFNought^{(4)}[\ddsc ]+}\\
\ordinary{\frac{29 \ddsc ^8 \MacroForFNought^{(4)}[\ddsc ]^2}{46080}+\frac{55 \ddsc ^5 \MacroForFNought^{(5)}[\ddsc ]}{2048}-\frac{853}{768} \ddsc ^6 \text{Log}[2]
\MacroForFNought^{(5)}[\ddsc ]+\frac{41}{3} \ddsc ^7 \text{Log}[2]^2 \MacroForFNought^{(5)}[\ddsc ]-}\\
\ordinary{72 \ddsc ^8 \text{Log}[2]^3 \MacroForFNought^{(5)}[\ddsc ]+\frac{512}{3} \ddsc ^9 \text{Log}[2]^4 \MacroForFNought^{(5)}[\ddsc ]-\frac{4096}{15}
\ddsc ^{10} \text{Log}[2]^5 \MacroForFNought^{(5)}[\ddsc ]+}\\
\ordinary{\frac{853 \ddsc ^6 \MacroForFNought'[\ddsc ] \MacroForFNought^{(5)}[\ddsc ]}{12288}-\frac{41}{24} \ddsc ^7 \text{Log}[2] \MacroForFNought'[\ddsc ] \MacroForFNought^{(5)}[\ddsc
]+\frac{27}{2} \ddsc ^8 \text{Log}[2]^2 \MacroForFNought'[\ddsc ] \MacroForFNought^{(5)}[\ddsc ]-}\\
\ordinary{\frac{128}{3} \ddsc ^9 \text{Log}[2]^3 \MacroForFNought'[\ddsc ] \MacroForFNought^{(5)}[\ddsc ]+\frac{256}{3} \ddsc ^{10} \text{Log}[2]^4 \MacroForFNought'[\ddsc
] \MacroForFNought^{(5)}[\ddsc ]+
\blue{\rightarrow\rightarrow\rightarrow\downarrow\downarrow\downarrow}}\\ \text{{\rm (continued~on~next~page,~sorry!)}} 
\xxnn
\pagebreak \\ \text{{\rm (fifth page of the expression for \bbd{\MacroFLPLower 5[\ddsc]}, continued~from~previous~page)}}\\
\ordinary{\blue{\uparrow}\blue{\uparrow}\blue{\uparrow}\blue{\leftarrow}\blue{\leftarrow}\blue{\leftarrow}
+\frac{41}{768} \ddsc ^7 \MacroForFNought'[\ddsc ]^2 \MacroForFNought^{(5)}[\ddsc ]-\frac{27}{32} \ddsc ^8 \text{Log}[2] \MacroForFNought'[\ddsc ]^2
\MacroForFNought^{(5)}[\ddsc ]+4 \ddsc ^9 \text{Log}[2]^2 \MacroForFNought'[\ddsc ]^2 \MacroForFNought^{(5)}[\ddsc ]-}\\
\ordinary{\frac{32}{3} \ddsc ^{10} \text{Log}[2]^3 \MacroForFNought'[\ddsc ]^2 \MacroForFNought^{(5)}[\ddsc ]+\frac{9}{512} \ddsc ^8 \MacroForFNought'[\ddsc ]^3
\MacroForFNought^{(5)}[\ddsc ]-\frac{1}{6} \ddsc ^9 \text{Log}[2] \MacroForFNought'[\ddsc ]^3 \MacroForFNought^{(5)}[\ddsc ]+}
\\
\ordinary{\frac{2}{3} \ddsc ^{10} \text{Log}[2]^2 \MacroForFNought'[\ddsc ]^3 \MacroForFNought^{(5)}[\ddsc ]+\frac{1}{384} \ddsc ^9 \MacroForFNought'[\ddsc ]^4
\MacroForFNought^{(5)}[\ddsc ]-}\\
\ordinary{\frac{1}{48} \ddsc ^{10} \text{Log}[2] \MacroForFNought'[\ddsc ]^4 \MacroForFNought^{(5)}[\ddsc ]+\frac{\ddsc ^{10} \MacroForFNought'[\ddsc ]^5 \MacroForFNought^{(5)}[\ddsc
]}{3840}+\frac{59 \ddsc ^7 \MacroForFNought''[\ddsc ] \MacroForFNought^{(5)}[\ddsc ]}{2304}-}\\
\ordinary{\frac{1}{3} \ddsc ^8 \text{Log}[2] \MacroForFNought''[\ddsc ] \MacroForFNought^{(5)}[\ddsc ]+\frac{4}{3} \ddsc ^9 \text{Log}[2]^2 \MacroForFNought''[\ddsc
] \MacroForFNought^{(5)}[\ddsc ]+}\\
\ordinary{\frac{1}{48} \ddsc ^8 \MacroForFNought'[\ddsc ] \MacroForFNought''[\ddsc ] \MacroForFNought^{(5)}[\ddsc ]-\frac{1}{6} \ddsc ^9 \text{Log}[2] \MacroForFNought'[\ddsc
] \MacroForFNought''[\ddsc ] \MacroForFNought^{(5)}[\ddsc ]}
\\
\ordinary{+ \frac{1}{192} \ddsc ^9 \MacroForFNought'[\ddsc ]^2 \MacroForFNought''[\ddsc ] \MacroForFNought^{(5)}[\ddsc ]+\frac{11 \ddsc ^8 \MacroForFNought^{(3)}[\ddsc
] \MacroForFNought^{(5)}[\ddsc ]}{11520}
\\
+\frac{41 \ddsc ^6 \MacroForFNought^{(6)}[\ddsc ]}{7680}-\frac{101}{960} \ddsc ^7 \text{Log}[2] \MacroForFNought^{(6)}[\ddsc ]+\frac{2}{3} \ddsc ^8 \text{Log}[2]^2 \MacroForFNought^{(6)}[\ddsc ]-\frac{16}{9}
\ddsc ^9 \text{Log}[2]^3 \MacroForFNought^{(6)}[\ddsc ]+}\\
\ordinary{\frac{101 \ddsc ^7 \MacroForFNought'[\ddsc ] \MacroForFNought^{(6)}[\ddsc ]}{15360}-\frac{1}{12} \ddsc ^8 \text{Log}[2] \MacroForFNought'[\ddsc ] \MacroForFNought^{(6)}[\ddsc
]+\frac{1}{3} \ddsc ^9 \text{Log}[2]^2 \MacroForFNought'[\ddsc ] \MacroForFNought^{(6)}[\ddsc ]+}\\
\ordinary{\frac{1}{384} \ddsc ^8 \MacroForFNought'[\ddsc ]^2 \MacroForFNought^{(6)}[\ddsc ]-\frac{1}{48} \ddsc ^9 \text{Log}[2] \MacroForFNought'[\ddsc ]^2 \MacroForFNought^{(6)}[\ddsc
]+\frac{\ddsc ^9 \MacroForFNought'[\ddsc ]^3 \MacroForFNought^{(6)}[\ddsc ]}{2304}+}\\
\ordinary{\frac{\ddsc ^8 \MacroForFNought''[\ddsc ] \MacroForFNought^{(6)}[\ddsc ]}{2304}+\frac{\ddsc ^7 \MacroForFNought^{(7)}[\ddsc ]}{4608}-\frac{1}{576} \ddsc
^8 \text{Log}[2] \MacroForFNought^{(7)}[\ddsc ]+\frac{\ddsc ^8 \MacroForFNought'[\ddsc ] \MacroForFNought^{(7)}[\ddsc ]}{9216}}
\eenn
\end{doublespace}

\newpage

\section{Accuracy of the asymptotic large-charge expansion at large \bbd{n} and fixed double-scaled coupling \bbd{\ddsc}}\label{DoubleScaledEstimatesVsExact}

\subsection{Limitations on the accuracy of agreement between the asymptotic series and the localization calculation}

Many sources of error are the same as this discussed in 
sec. \ref{FixedCouplingSourcesOfErrorSec},
in the context of the accuracy of the fixed-\bbd{\t} large-charge
asymptotic expansions.  In particular, we are comparing with
the same numerical localization calculations, so the numerical imprecision of the localization calculations beyond ten significant
digits in the MMP function is equally a limiting factor here, in
the comparison with the double-scaled estimates.  We will
now mention some of the issues concerning sources of
error for the double-scaled estimates that differ from the issues
for the fixed-coupling estimates.

\heading{Higher-winding effects not a source of error for
the double-scaled estimates}
For the fixed-\bbd{\t} large-charge estimates of the MMP function,
only the single-equator-winding worldline instanton
of the lightest BPS particle (which for
\bbd{\t ={{25}\over\pi}\cc i} is an electrically charged hypermultiplet)
contributes to the MMP function to any order in \bbd{n\uu{-\hh}};
as a result, multiple-winding \textsc{wli}s and/or WLIs of heavier
BPS particles ({\it e.g.} massive vector multiplets, which have
twice the mass of the electric hyper) are absent from the fixed-\bbd{\t} large-\bbd{n} asymptotic
series and their omission is a source of error of order 
\bbd{n\uu{-\hh}\cc \exp{- 2\sqrt{{8\pi n}\over{{\rm Im}[\s]}}}}.  In
the double-scaled large-charge expansion, by contrast, these effects are
explicitly included in the asymptotic estaimates.  Peculiarly,
the order \bbd{n\uu 0}
connected double-scaled MMP amplitude \bbd{F\ls 0[\ddsc] = 
F\urm{inst}\lp{\smgkt}[\ddsc]} contains only \textsc{wli} contributions
of \emm{odd} winding number about the equator; clearly
from the explicit formula \rr{FormulaForF0} there are no even-winding 
terms in the order \bbd{n\uu 0} contribution to \bbd{q\ll n\uprm{\textsc{mmp}}}.
As remarked in \cite{Grassi:2019txd}, this is plausibly due to a
cancellation between macroscopic massive hypermultiplet trajectories
of winding \bbd{2} and massive vector-multiplet trajectories
of winding \bbd{1} contributing with the same magnitude and 
opposite signs, though no explicit field-theoretic computation has
yet been done to check this.  Already at next order in \bbd{n}, however,
connected macroscopic massive hypermultiplet contributions with total winding \bbd{2} \emm{do} appear
in \bbd{F\ls 1[\ddsc]}.  These come from the term
\bbd{{{\ddsc\sqd}\over 4}\cc (\MacroFLPLower 0\pr[\ddsc])\sqd}     in expression \rr{TwoLoopConnectedMMPAmplitude} for \bbd{F\ls 1[\ddsc]}, and 
correspond to two separate macroscopic massive hypermultiplet
propagators of winding \bbd{1}, connected by a single massless
vector multiplet propagator.

\heading{Gauge instanton effects and the large-order behavior of the
asymptotic series}

Though we have chosen a value \bbd{\t = {{25}\over\pi}\cc i} 
at which gauge instanton effects are far too small to contribute
at the order of precision we have attained, it is worth considering
their implications for the precision of the double-scaled
asymptotic estimates of the macroscopic massive propagation
function.

The double-scaled coupling \bbd{\ddsc} is independent of
the ultraviolet \bbd{\th}-angle \bbd{\th\lrm{UV} \equiv 2\pi{\rm Re}[\t]}, as are all the finite-order corrections \bbd{n\uu{-k}\cc F\ls k[\ddsc]} to the MMP function.  However instanton effects do contribute to the MMP 
function at large charge, through the dependence of
the exact BPS mass of the electric hyper on \bbd{{\rm Im}[\s]}.  
The best accuracy we can hope to attain from the double-scaled
large charge expansion -- at least without some
kind of hyperasymptotic extension of the series --
would be of relative size \bbd{e\uu{- 2\pi {\rm Im}[\t]} = e\uu{- {n\over{2\ddsc}}}}.  It is unclear whether or not this inaccuracy would
be announced by any internal signal within the double-scaled
large-\bbd{n} asymptotic
series itself, such as an upturn in the magnitudes of the
series coefficients \bbd{n\uu{-k} F\ls k[\ddsc]} past the optimal
truncation point of the expansion, or whether the series would
continue to converge beyond the optimally accurate point while
no longer giving accurate results.  Both cases can occur
in asymptotic series, with the latter case being particularly
well-known in supersymmetric examples.  It would
be very interesting to answer this question using tools
from resurgence theory (see {\it e.g.} \cite{Lipatov:1976ny, Argyres:2012vv, Argyres:2012ka, Dorigoni:2014hea, Aniceto:2018bis} 
and references therein\footnote{On the mathematical side, the 
key work in the subject is \cite{Ecalle}.  For the mathematical
reader interested in a category-theoretic approach to the
resurgence in quantum theory, a reference is \cite{WKBSheaves}.}) a subject which has been explored previously
in the context of supersymmetric localization \cite{Honda:2016mvg}
and recently applied to the large-quantum-number expansion
in the Wilson-Fisher \bbd{O(n)} models \cite{Dondi:2021buw}.

\heading{Proliferating complexity of the terms in the series}

The algebraic expressions for the \bbd{F\ls k[\ddsc]} grow
long and complicated very quickly, even when expressed as
compactly as possibly in terms
of \bbd{F\ls 0[\ddsc]} and its derivatives.  The \bbd{\ddsc-}dependence
\bbd{F\ls 5[\ddsc]} of the order \bbd{n\uu{-5}} term, cannot be
fit easily onto fewer than five pages.  At this order,
the numerical evaluation of the estimates themselves takes
an appreciable amount of time; evaluated on {\it Mathematica} running on
a 2013 Macbook Air with a 1.4 GHz Intel Core i5 processor, 
evaluating \bbd{F\ls 5[\ddsc]} takes
somewhere between four and a half and ten minutes\footnote{We don't know whether this is supposed to happen or
whether the battery is just wearing out or something, but the factor of two
seems to depend on whether our laptop is plugged in or not.} for each
individual value of \bbd{\ddsc}.  The quickly increasing
complexity of the terms made it impractical to include the
\bbd{n\uu{-6} F\ls 6[\ddsc]} term (or higher terms) in the asymptotic series.  This is unfortunate, since there is no indication at all
the truncation of the double-scaled exapnsion
at order \bbd{n\uu{-5}} is anywhere near optimally accurate,
and the inclusion of higher terms would likely increase the accuracy of
the estimate by another one or two significant digits at each additional
order.

It may be that the double-scaled large-charge expansion could become
more manageable in the strong-coupling expansion, truncating 
the terms to some maximum winding number depending on
the values of \bbd{n} and \bbd{\ddsc}.  This may be a useful
direction to explore to further increase the accuracy of the large-R-charge
estimates.

\subsection{Plotting the log of relative error of the double-scaled estimates through order \bbd{n\uu{-5}} at fixed \bbd{\ddsc,} evaluated at \bbd{\t = {{25}\over \pi}, 1 \leq n \leq 150}}

\centpicWidthHeightAngleFileCaptionFiglab{120mm}{80mm}{0}
{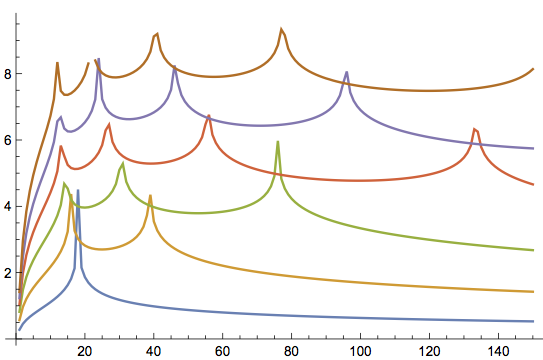}
{Plot of the giving the accuracy of the double-scaled estimates of the MMP function through N${}\uu 5$LO.  The quantity
being plotted is \bbd{- {1\over{{\rm Log}[10]}}} the \emm{logarithm} of the relative error in the estimate of the MMP function.  The 
horizontal axis is \bbd{n}, and the vertical axis is 
\bbd{- {1\over{{\rm Log}[10]}}\cc {\rm Log} \left [ \cc {{| q_n\uprm{\textsc{mmp}} - (q_n\uprm{\textsc{mmp}})\lrm{estimate} | }\over{ |q_n\uprm{\textsc{mmp}}|}} \cc  \right ]}  The LO, NLO, N${}\sqd$LO, N${}\uu 3$LO, N${}\uu 4$LO and N${}\uu 5$LO double-scaled estimates
are given by the blue, yellow, green, red, and purple, and brown dots respectively.   The
upward spikes represent "accidental accuracies" in which an estimate transitions between 
slightly overestimating and slightly underestimating the exact result as \bbd{n} is varied, generating an atypically precise agreement between a given estimate and the exact amplitude, in some small range of \bbd{n}. 
}
{SecondFigure} 

\centpicWidthHeightAngleFileCaption{120mm}{80mm}{0}
{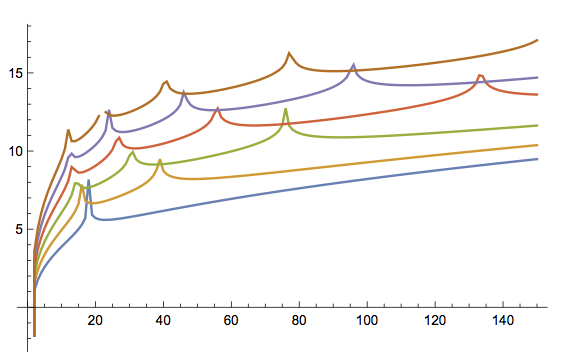}
{Plot of the giving the accuracy of the double-scaled estimates of the full log of the scheme-independent correlator  \bbd{\tilde{G}\ll{2n} \equiv {{G\ll{2n} }\over{ (G\ll 2)\uu n}} = e\uu{ q\ll n - n\cc q\ll 1 + (n-1) q\ll 0}}, through N${}\uu 5$LO.   The quantity
being plotted is \bbd{- {1\over{{\rm Log}[10]}}} the \emm{logarithm} of the relative error in the estimate of the MMP function.  The 
horizontal axis is \bbd{n}, and the vertical axis is \bbd{- {1\over{{\rm Log}[10]}}\cc {\rm Log} \left [ \cc {{| q_n - (q_n)\lrm{estimate} | }\over{ |q_n- n\cc q\ll 1 + (n-1) q\ll 0 |}} \cc  \right ]}  The leading-order, NLO, N${}\sqd$LO, N${}\uu 3$LO, N${}\uu 4$LO and N${}\uu 5$LO estimates
are given by the blue, yellow, green, red, and purple, and brown dots respectively.   The
upward spikes represent "accidental accuracies" in which an estimate transitions between 
slightly overestimating and slightly underestimating the exact result as \bbd{n} is varied, generating an atypically precise agreement between a given estimate and the exact amplitude, in some small range of \bbd{n}.}

\newpage
\subsection{Table of values of the double-scaled ({\it i.e.\rm } fixed-\bbd{\ddsc}) large-charge asymptotic estimates at \bbd{\t = {{25}\over \pi}\cc i} up to \bbd{n=150}}

 \begin{table}[H]
\begin{center}
\resizebox{\columnwidth}{!}{%
\begin{tabular}{ |c||c|c|c|c|c|c|c| } 
 \hline\hline
 ~ & $\MacroFLPLower 0[\ddsc]$  
 &${{\rm estimate~incl.}\atop{n\uu{-1}\MacroFLPLower 1[\ddsc]}}$  &${{\rm estimate~incl.}\atop{n\uu{-2}\MacroFLPLower 2[\ddsc]}}$ & ${{\rm estimate~incl.}\atop{n\uu{-3}\MacroFLPLower 3[\ddsc]}}$& ${{\rm estimate~incl.}\atop{n\uu{-4}\MacroFLPLower 4[\ddsc]}}$ & ${{\rm estimate~incl.}\atop{n\uu{-5}\MacroFLPLower 5[\ddsc]}}$ & $(q\ll n\uprm{\textsc{mmp}})\lrm{exact}$ \\ \hline\hline

\NEWERDSCMACRO {1} {0.6515789571} {0.3106159884} {0.4912085899} {0.3869269511} {0.4513455344} {0.4099885449} {0.4263073863} 
\NEWERDSCMACRO{2} {0.3946733263} {0.2608832307} {0.3020903723} {0.2894178806} {0.2934159952} {0.2921249755} {0.2924432054} 
\NEWERDSCMACRO{3} {0.2711221920} {0.2007750251} {0.2169181342} {0.2133789929} {0.2141499819} {0.2139812813} {0.2140116919} 
\NEWERDSCMACRO{4} {0.1980089293} {0.1562620119} {0.1640670070} {0.1627032722} {0.1629353694} {0.1628962899} {0.1629019007} 
\NEWERDSCMACRO{5} {0.1502153430} {0.1237476447} {0.1279557008} {0.1273368696} {0.1274245447} {0.1274123721} {0.1274138410} 
\NEWERDSCMACRO{6} {0.1170270973} {0.09954831105} {0.1019669053} {0.1016586081} {0.1016962711} {0.1016917835} {0.1016922546} 
\NEWERDSCMACRO{7} {0.09300855834} {0.08115719284} {0.08260296089} {0.08244065014} {0.08245807861} {0.08245625608} {0.08245642599} 
\NEWERDSCMACRO{8} {0.07508946762} {0.06691661367} {0.06780122735} {0.06771312872} {0.06772151249} {0.06772073565} {0.06772080050} 
\NEWERDSCMACRO{9} {0.06140276891} {0.05571131925} {0.05625864419} {0.05621029365} {0.05621436529} {0.05621403196} {0.05621405675} 
\NEWERDSCMACRO{10} {0.05074985468} {0.04677219216} {0.04711083791} {0.04708452067} {0.04708645350} {0.04708631681} {0.04708632558} 
\NEWERDSCMACRO{11} {0.04232850769} {0.03955582885} {0.03976277450} {0.03974892486} {0.03974977625} {0.03974972807} {0.03974973039} 
\NEWERDSCMACRO{12} {0.03558371710} {0.03366981364} {0.03379259457} {0.03378585902} {0.03378616421} {0.03378615512} {0.03378615497} 
\NEWERDSCMACRO{13} {0.03012096484} {0.02882524012} {0.02889391825} {0.02889122382} {0.02889126049} {0.02889126744} {0.02889126648} 
\NEWERDSCMACRO{14} {0.02565341289} {0.02480576447} {0.02483969766} {0.02483925094} {0.02483916477} {0.02483917702} {0.02483917594} 
\NEWERDSCMACRO{15} {0.02196848039} {0.02144693696} {0.02145875613} {0.02145950165} {0.02145936873} {0.02145938143} {0.02145938050} 
\NEWERDSCMACRO{16} {0.01890600199} {0.01862208600} {0.01862010631} {0.01862142301} {0.01862128213} {0.01862129330} {0.01862129258} 
\NEWERDSCMACRO{17} {0.01634355964} {0.01623248103} {0.01622220237} {0.01622372867} {0.01622359848} {0.01622360749} {0.01622360696} 
\NEWERDSCMACRO{18} {0.01418639766} {0.01420035157} {0.01418541289} {0.01418694359} {0.01418683189} {0.01418683872} {0.01418683837} 
\NEWERDSCMACRO{19} {0.01236034544} {0.01246385154} {0.01244664912} {0.01244807353} {0.01244798241} {0.01244798732} {0.01244798710} 
\NEWERDSCMACRO{20} {0.01080676066} {0.01097337293} {0.01095546710} {0.01095673146} {0.01095666011} {0.01095666344} {0.01095666332} 
\NEWERDSCMACRO{21} {0.009478857881} {0.009688810997} {0.009671197203} {0.009672281587} {0.009672227894} {0.009672229973} {0.009672229925} 
\NEWERDSCMACRO{22} {0.008339004755} {0.008577510675} {0.008560802010} {0.008561706111} {0.008561667481} {0.008561668609} {0.008561668609} 
\NEWERDSCMACRO{23} {0.007356704985} {0.007612707402} {0.007597258612} {0.007597992954} {0.007597966770} {0.007597967198} {0.007597967228} 
\NEWERDSCMACRO{24} {0.006507076091} {0.006772331309} {0.006758323571} {0.006758904129} {0.006758887969} {0.006758887898} {0.006758887946} 
\NEWERDSCMACRO{25} {0.005769688366} {0.006038081569} {0.006025580719} {0.006026025648} {0.006026017381} {0.006026016969} {0.006026017027} 
\NEWERDSCMACRO{26} {0.005127670735} {0.005394703710} {0.005383700338} {0.005384028043} {0.005384025851} {0.005384025221} {0.005384025281} 
\NEWERDSCMACRO{27} {0.004567016031} {0.004829420924} {0.004819857988} {0.004820086027} {0.004820088395} {0.004820087638} {0.004820087697} 
\NEWERDSCMACRO{28} {0.004076036781} {0.004331483276} {0.004323275076} {0.004323419587} {0.004323425275} {0.004323424459} {0.004323424514} 
\NEWERDSCMACRO{29} {0.003644935681} {0.003891807929} {0.003884853075} {0.003884928518} {0.003884936528} {0.003884935700} {0.003884935751} 
\NEWERDSCMACRO{30} {0.003265464184} {0.003502690183} {0.003496880366} {0.003496899468} {0.003496909006} {0.003496908200} {0.003496908245} 
\NEWERDSCMACRO{31} {0.002930649381} {0.003157570017} {0.003152795854} {0.003152769661} {0.003152780103} {0.003152779343} {0.003152779381} 
\NEWERDSCMACRO{32} {0.002634574151} {0.002850842411} {0.002846997247} {0.002846935240} {0.002846946105} {0.002846945404} {0.002846945436} 
\NEWERDSCMACRO{33} {0.002372199210} {0.002577702455} {0.002574684734} {0.002574594972} {0.002574605896} {0.002574605261} {0.002574605288} 
\NEWERDSCMACRO{34} {0.002139218286} {0.002334018223} {0.002331732873} {0.002331622153} {0.002331632862} {0.002331632298} {0.002331632319} 
\NEWERDSCMACRO{35} {0.001931939664} {0.002116225963} {0.002114585104} {0.002114459111} {0.002114469409} {0.002114468915} {0.002114468932}

\end{tabular}
}
\captionof{table}{The successive refined estimates for the massive macroscopic propagation function \bbd{q\ll n\uprm{\textsc{mmp}} \equiv q\ll n - q\ll n\uprm{eft}}, at the coupling \bbd{\t = {{25 i}\over{\pi}}}.  The first column represents the strict infinite-\bbd{n} limit at fixed double-scaling parameter \bbd{\ddsc},
\bbd{q\ll n\uprm{\textsc{mmp}} \simeq \MacroFLPLower 0[\ddsc] = F\urm{inst}\lp{\smgkt}[\ddsc]}.  The next five columns represent this approximation corrected by adding successive terms \bbd{n\uu{-k}\cc \MacroFLPLower k[\ddsc]} for \bbd{k = 1,\cdots 5}.  The last column is the exact
result as computed \cite{LocData} following the method of \cite{Gerchkovitz:2016gxx}.}
\label{TypesOfLiouvilleAmplitudeTransformationTableSpecificParameterChoiceOfInterestToUs}
\end{center}
\end{table}

\newpage
 \vskip-1in
 \begin{table}[H]
\begin{center}
\resizebox{\columnwidth}{!}{%
\begin{tabular}{ |c||c|c|c|c|c|c|c| } 
 \hline\hline
 ~ & $\MacroFLPLower 0[\ddsc]$  
 &${{\rm estimate~incl.}\atop{n\uu{-1}\MacroFLPLower 1[\ddsc]}}$  &${{\rm estimate~incl.}\atop{n\uu{-2}\MacroFLPLower 2[\ddsc]}}$ & ${{\rm estimate~incl.}\atop{n\uu{-3}\MacroFLPLower 3[\ddsc]}}$& ${{\rm estimate~incl.}\atop{n\uu{-4}\MacroFLPLower 4[\ddsc]}}$ & ${{\rm estimate~incl.}\atop{n\uu{-5}\MacroFLPLower 5[\ddsc]}}$ & $(q\ll n\uprm{\textsc{mmp}})\lrm{exact}$ \\ \hline\hline 

\NEWERDSCMACRO{36} {0.001747188853} {0.001921243312} {0.001920166501} {0.001920029953} {0.001920039702} {0.001920039277} {0.001920039289} 
\NEWERDSCMACRO{37} {0.001582228233} {0.001746397130} {0.001745811295} {0.001745668079} {0.001745677189} {0.001745676828} {0.001745676837} 
\NEWERDSCMACRO{38} {0.001434690459} {0.001589363259} {0.001589202444} {0.001589055732} {0.001589064149} {0.001589063848} {0.001589063854} 
\NEWERDSCMACRO{39} {0.001302523033} {0.001448116056} {0.001448321035} {0.001448173393} {0.001448181090} {0.001448180845} {0.001448180848} 
\NEWERDSCMACRO{40} {0.001183942002} {0.001320885953} {0.001321403782} {0.001321257261} {0.001321264234} {0.001321264039} {0.001321264040} 
\NEWERDSCMACRO{41} {0.001077393142} {0.001206123664} {0.001206907196} {0.001206763414} {0.001206769674} {0.001206769523} {0.001206769522} 
\NEWERDSCMACRO{42} {0.0009815192976} {0.001102469881} {0.001103477277} {0.001103337490} {0.001103343060} {0.001103342948} {0.001103342946} 
\NEWERDSCMACRO{43} {0.0008951328279} {0.001008729558} {0.001009923809} {0.001009788968} {0.001009793878} {0.001009793802} {0.001009793798} 
\NEWERDSCMACRO{44} {0.0008171922687} {0.0009238500137} {0.0009251984792} {0.0009250692846} {0.0009250735723} {0.0009250735254} {0.0009250735212} 
\NEWERDSCMACRO{45} {0.0007467825211} {0.0008469022370} {0.0008483762128} {0.0008482531567} {0.0008482568617} {0.0008482568405} {0.0008482568358} 
\NEWERDSCMACRO{46} {0.0006830979808} {0.0007770648850} {0.0007786391999} {0.0007785226037} {0.0007785257677} {0.0007785257685} {0.0007785257633} 
\NEWERDSCMACRO{47} {0.0006254281327} {0.0007136105490} {0.0007152631935} {0.0007151532393} {0.0007151559046} {0.0007151559239} {0.0007151559185} 
\NEWERDSCMACRO{48} {0.0005731452189} {0.0006558939394} {0.0006576057257} {0.0006575024833} {0.0006575046917} {0.0006575047264} {0.0006575047208} 
\NEWERDSCMACRO{49} {0.0005256936560} {0.0006033416976} {0.0006050959505} {0.0006049993989} {0.0006050011913} {0.0006050012386} {0.0006050012329} 
\NEWERDSCMACRO{50} {0.0004825809307} {0.0005554435896} {0.0005572258663} {0.0005571359133} {0.0005571373289} {0.0005571373863} {0.0005571373807} 
\NEWERDSCMACRO{51} {0.0004433697498} {0.0005117448790} {0.0005135427166} {0.0005134592139} {0.0005134602901} {0.0005134603555} {0.0005134603500} 
\NEWERDSCMACRO{52} {0.0004076712568} {0.0004718397065} {0.0004736423948} {0.0004735651513} {0.0004735659233} {0.0004735659948} {0.0004735659895} 
\NEWERDSCMACRO{53} {0.0003751391572} {0.0004353653317} {0.0004371637095} {0.0004370925022} {0.0004370930031} {0.0004370930790} {0.0004370930739} 
\NEWERDSCMACRO{54} {0.0003454646213} {0.0004019971155} {0.0004037833887} {0.0004037179712} {0.0004037182317} {0.0004037183106} {0.0004037183057} 
\NEWERDSCMACRO{55} {0.0003183718512} {0.0003714441394} {0.0003732117176} {0.0003731518279} {0.0003731518766} {0.0003731519572} {0.0003731519527} 
\NEWERDSCMACRO{56} {0.0002936142202} {0.0003434453736} {0.0003451887254} {0.0003451340919} {0.0003451339549} {0.0003451340363} {0.0003451340321} 
\NEWERDSCMACRO{57} {0.0002709709024} {0.0003177663198} {0.0003194808436} {0.0003194311898} {0.0003194308912} {0.0003194309725} {0.0003194309685} 
\NEWERDSCMACRO{58} {0.0002502439267} {0.0002941960646} {0.0002958779737} {0.0002958330222} {0.0002958325840} {0.0002958326644} {0.0002958326607} 
\NEWERDSCMACRO{59} {0.0002312555960} {0.0002725446892} {0.0002741909098} {0.0002741503855} {0.0002741498277} {0.0002741499066} {0.0002741499032} 
\NEWERDSCMACRO{60} {0.0002138462236} {0.0002526409886} {0.0002542490695} {0.0002542127019} {0.0002542120425} {0.0002542121195} {0.0002542121164} 
\NEWERDSCMACRO{61} {0.0001978721442} {0.0002343304601} {0.0002358984921} {0.0002358660175} {0.0002358652730} {0.0002358653476} {0.0002358653448} 
\NEWERDSCMACRO{62} {0.0001832039628} {0.0002174735268} {0.0002190000722} {0.0002189712350} {0.0002189704201} {0.0002189704921} {0.0002189704896} 
\NEWERDSCMACRO{63} {0.0001697250120} {0.0002019439669} {0.0002034279965} {0.0002034025502} {0.0002034016782} {0.0002034017473} {0.0002034017450} 
\NEWERDSCMACRO{64} {0.0001573299900} {0.0001876275216} {0.0001890683593} {0.0001890460675} {0.0001890451501} {0.0001890452162} {0.0001890452141} 
\NEWERDSCMACRO{65} {0.0001459237564} {0.0001744206612} {0.0001758179350} {0.0001757985716} {0.0001757976195} {0.0001757976824} {0.0001757976806} 
\NEWERDSCMACRO{66} {0.0001354202667} {0.0001622294889} {0.0001635830878} {0.0001635664377} {0.0001635654603} {0.0001635655198} {0.0001635655183} 
\NEWERDSCMACRO{67} {0.0001257416277} {0.0001509687666} {0.0001522788025} {0.0001522646613} {0.0001522636668} {0.0001522637231} {0.0001522637217} 
\NEWERDSCMACRO{68} {0.0001168172587} {0.0001405610465} {0.0001418278208} {0.0001418159951} {0.0001418149908} {0.0001418150437} {0.0001418150425} 
\NEWERDSCMACRO{69} {0.0001085831469} {0.0001309358982} {0.0001321598721} {0.0001321501793} {0.0001321491716} {0.0001321492212} {0.0001321492202} 
\NEWERDSCMACRO{70} {0.0001009811840} {0.0001220292180} {0.0001232109870} {0.0001232032549} {0.0001232022494} {0.0001232022959} {0.0001232022950}

\end{tabular}
}
\captionof{table}{The successive refined estimates for the massive macroscopic propagation function \bbd{q\ll n\uprm{\textsc{mmp}} \equiv q\ll n - q\ll n\uprm{eft}}, at the coupling \bbd{\t = {{25 i}\over{\pi}}}.  The first column represents the strict infinite-\bbd{n} limit at fixed double-scaling parameter \bbd{\ddsc},
\bbd{q\ll n\uprm{\textsc{mmp}} \simeq \MacroFLPLower 0[\ddsc] = F\urm{inst}\lp{\smgkt}[\ddsc]}.  The next five columns represent this approximation corrected by adding successive terms \bbd{n\uu{-k}\cc \MacroFLPLower k[\ddsc]} for \bbd{k = 1,\cdots 5}.  The last column is the exact
result as computed \cite{LocData} following the method of \cite{Gerchkovitz:2016gxx}.}
\label{TypesOfLiouvilleAmplitudeTransformationTableSpecificParameterChoiceOfInterestToUs}
\end{center}
\end{table}

\newpage
 \vskip-1in
 \begin{table}[H]
\begin{center}
\resizebox{\columnwidth}{!}{%
\begin{tabular}{ |c||c|c|c|c|c|c|c| } 
 \hline\hline
 ~ & $\MacroFLPLower 0[\ddsc]$  
 &${{\rm estimate~incl.}\atop{n\uu{-1}\MacroFLPLower 1[\ddsc]}}$  &${{\rm estimate~incl.}\atop{n\uu{-2}\MacroFLPLower 2[\ddsc]}}$ & ${{\rm estimate~incl.}\atop{n\uu{-3}\MacroFLPLower 3[\ddsc]}}$& ${{\rm estimate~incl.}\atop{n\uu{-4}\MacroFLPLower 4[\ddsc]}}$ & ${{\rm estimate~incl.}\atop{n\uu{-5}\MacroFLPLower 5[\ddsc]}}$ & $(q\ll n\uprm{\textsc{mmp}})\lrm{exact}$ \\ \hline\hline

\NEWERDSCMACRO{71} {0.00009395857627} {0.0001137826120} {0.0001149228833} {0.0001149169501} {0.0001149159516} {0.0001149159950} {0.0001149159943} 
\NEWERDSCMACRO{72} {0.00008746731725} {0.0001061428449} {0.0001072424172} {0.0001072381311} {0.0001072371438} {0.0001072371840} {0.0001072371835} 
\NEWERDSCMACRO{73} {0.00008146371776} {0.00009906134541} {0.0001001210918} {0.0001001183104} {0.0001001173379} {0.0001001173752} {0.0001001173747} 
\NEWERDSCMACRO{74} {0.00007590798546} {0.00009249376341} {0.00009351461614} {0.00009351320646} {0.00009351225181} {0.00009351228623} {0.00009351228591} 
\NEWERDSCMACRO{75} {0.00007076384866} {0.00008639957275} {0.00008738251023} {0.00008738234799} {0.00008738141374} {0.00008738144540} {0.00008738144519} 
\NEWERDSCMACRO{76} {0.00006599821932} {0.00008074171402} {0.00008168774925} {0.00008168871860} {0.00008168780685} {0.00008168783588} {0.00008168783576} 
\NEWERDSCMACRO{77} {0.00006158089089} {0.00007548627365} {0.00007639644430} {0.00007639843736} {0.00007639754985} {0.00007639757635} {0.00007639757633} 
\NEWERDSCMACRO{78} {0.00005748426706} {0.00007060219491} {0.00007147755480} {0.00007148047123} {0.00007147960935} {0.00007147963346} {0.00007147963351} 
\NEWERDSCMACRO{79} {0.00005368311801} {0.00006606101768} {0.00006690262936} {0.00006690637599} {0.00006690554082} {0.00006690556266} {0.00006690556278} 
\NEWERDSCMACRO{80} {0.00005015436127} {0.00006183664370} {0.00006264557213} {0.00006265006256} {0.00006264925490} {0.00006264927459} {0.00006264927477} 
\NEWERDSCMACRO{81} {0.00004687686437} {0.00005790512479} {0.00005868243193} {0.00005868758610} {0.00005868680653} {0.00005868682419} {0.00005868682442} 
\NEWERDSCMACRO{82} {0.00004383126706} {0.00005424447148} {0.00005499121170} {0.00005499695553} {0.00005499620441} {0.00005499622015} {0.00005499622043} 
\NEWERDSCMACRO{83} {0.00004099982096} {0.00005083448000} {0.00005155169620} {0.00005155796124} {0.00005155723874} {0.00005155725269} {0.00005155725300} 
\NEWERDSCMACRO{84} {0.00003836624486} {0.00004765657580} {0.00004834529612} {0.00004835201920} {0.00004835132533} {0.00004835133759} {0.00004835133795} 
\NEWERDSCMACRO{85} {0.00003591559396} {0.00004469367174} {0.00004535490689} {0.00004536202976} {0.00004536136441} {0.00004536137509} {0.00004536137547} 
\NEWERDSCMACRO{86} {0.00003363414166} {0.00004193003964} {0.00004256478065} {0.00004257224970} {0.00004257161261} {0.00004257162182} {0.00004257162223} 
\NEWERDSCMACRO{87} {0.00003150927258} {0.00003935119372} {0.00003996041018} {0.00003996817610} {0.00003996756692} {0.00003996757476} {0.00003996757518} 
\NEWERDSCMACRO{88} {0.00002952938572} {0.00003694378487} {0.00003752842343} {0.00003753644094} {0.00003753585923} {0.00003753586580} {0.00003753586624} 
\NEWERDSCMACRO{89} {0.00002768380662} {0.00003469550447} {0.00003525648777} {0.00003526471537} {0.00003526416061} {0.00003526416600} {0.00003526416645} 
\NEWERDSCMACRO{90} {0.00002596270774} {0.00003259499712} {0.00003313322295} {0.00003314162260} {0.00003314109423} {0.00003314109852} {0.00003314109898} 
\NEWERDSCMACRO{91} {0.00002435703620} {0.00003063178108} {0.00003114812176} {0.00003115665871} {0.00003115615610} {0.00003115615938} {0.00003115615984} 
\NEWERDSCMACRO{92} {0.00002285844809} {0.00002879617598} {0.00002929147799} {0.00002930012051} {0.00002929964298} {0.00002929964533} {0.00002929964580} 
\NEWERDSCMACRO{93} {0.00002145924875} {0.00002707923686} {0.00002755432063} {0.00002756303979} {0.00002756258666} {0.00002756258815} {0.00002756258862} 
\NEWERDSCMACRO{94} {0.00002015233856} {0.00002547269406} {0.00002592835392} {0.00002593712343} {0.00002593669396} {0.00002593669467} {0.00002593669514} 
\NEWERDSCMACRO{95} {0.00001893116342} {0.00002396889846} {0.00002440590270} {0.00002441469868} {0.00002441429212} {0.00002441429211} {0.00002441429258} 
\NEWERDSCMACRO{96} {0.00001778966979} {0.00002256077135} {0.00002297986244} {0.00002298866325} {0.00002298827885} {0.00002298827819} {0.00002298827865} 
\NEWERDSCMACRO{97} {0.00001672226361} {0.00002124175876} {0.00002164365363} {0.00002165243972} {0.00002165207670} {0.00002165207545} {0.00002165207590} 
\NEWERDSCMACRO{98} {0.00001572377287} {0.00002000578969} {0.00002039118006} {0.00002039993379} {0.00002039959140} {0.00002039958960} {0.00002039959005} 
\NEWERDSCMACRO{99} {0.00001478941348} {0.00001884723783} {0.00001921679069} {0.00001922549621} {0.00001922517367} {0.00001922517139} {0.00001922517183} 
\NEWERDSCMACRO{100} {0.00001391475800} {0.00001776088663} {0.00001811524470} {0.00001812388779} {0.00001812358435} {0.00001812358164} {0.00001812358207} 
\NEWERDSCMACRO{101} {0.00001309570717} {0.00001674189727} {0.00001708167952} {0.00001709024751} {0.00001708996240} {0.00001708995929} {0.00001708995971} 
\NEWERDSCMACRO{102} {0.00001232846384} {0.00001578577929} {0.00001611158153} {0.00001612006311} {0.00001611979559} {0.00001611979214} {0.00001611979255} 
\NEWERDSCMACRO{103} {0.00001160950905} {0.00001488836370} {0.00001520075917} {0.00001520914432} {0.00001520889367} {0.00001520888991} {0.00001520889031} 
\NEWERDSCMACRO{104} {0.00001093558026} {0.00001404577830} {0.00001434531825} {0.00001435359817} {0.00001435336365} {0.00001435335962} {0.00001435336001} 
\NEWERDSCMACRO{105} {0.00001030365130} {0.00001325442503} {0.00001354163939} {0.00001354980634} {0.00001354958725} {0.00001354958299} {0.00001354958336} 
\NEWERDSCMACRO{106} {9.710914034$\times 10\uu{-6}$} {0.00001251095916} {0.00001278635714} {0.00001279440439} {0.00001279420005} {0.00001279419559} {0.00001279419595} 
\NEWERDSCMACRO{107} {9.154761542$\times 10\uu{-6}$} {0.00001181227019} {0.00001207634095} {0.00001208426269} {0.00001208407243} {0.00001208406779} {0.00001208406814} 
\NEWERDSCMACRO{108} {8.632772702$\times 10\uu{-6}$} {0.00001115546427} {0.00001140867755} {0.00001141646882} {0.00001141629198} {0.00001141628720} {0.00001141628753} 
\NEWERDSCMACRO{109} {8.142698011$\times 10\uu{-6}$} {0.00001053784805} {0.00001078065480} {0.00001078831140} {0.00001078814736} {0.00001078814245} {0.00001078814278} 
\NEWERDSCMACRO{110} {7.682446573$\times 10\uu{-6}$} {9.956913799$\times 10\uu{-6}$} {0.00001018974683} {0.00001019726528} {0.00001019711342} {0.00001019710841} {0.00001019710872} 

\end{tabular}
}
\captionof{table}{The successive refined estimates for the massive macroscopic propagation function \bbd{q\ll n\uprm{\textsc{mmp}} \equiv q\ll n - q\ll n\uprm{eft}}, at the coupling \bbd{\t = {{25 i}\over{\pi}}}.  The first column represents the strict infinite-\bbd{n} limit at fixed double-scaling parameter \bbd{\ddsc},
\bbd{q\ll n\uprm{\textsc{mmp}} \simeq \MacroFLPLower 0[\ddsc] = F\urm{inst}\lp{\smgkt}[\ddsc]}.  The next five columns represent this approximation corrected by adding successive terms \bbd{n\uu{-k}\cc \MacroFLPLower k[\ddsc]} for \bbd{k = 1,\cdots 5}.  The last column is the exact
result as computed \cite{LocData} following the method of \cite{Gerchkovitz:2016gxx}.}
\label{TypesOfLiouvilleAmplitudeTransformationTableSpecificParameterChoiceOfInterestToUs}
\end{center}
\end{table}

 \vskip-1in
 \begin{table}[H]
\begin{center}
\resizebox{\columnwidth}{!}{%
\begin{tabular}{ |c||c|c|c|c|c|c|c| } 
 \hline\hline
 ~ & $\MacroFLPLower 0[\ddsc]$  
 &${{\rm estimate~incl.}\atop{n\uu{-1}\MacroFLPLower 1[\ddsc]}}$  &${{\rm estimate~incl.}\atop{n\uu{-2}\MacroFLPLower 2[\ddsc]}}$ & ${{\rm estimate~incl.}\atop{n\uu{-3}\MacroFLPLower 3[\ddsc]}}$& ${{\rm estimate~incl.}\atop{n\uu{-4}\MacroFLPLower 4[\ddsc]}}$ & ${{\rm estimate~incl.}\atop{n\uu{-5}\MacroFLPLower 5[\ddsc]}}$ & $(q\ll n\uprm{\textsc{mmp}})\lrm{exact}$ \\ \hline\hline 

\NEWERDSCMACRO{111} {7.250074141$\times 10\uu{-6}$} {9.410325706$\times 10\uu{-6}$} {9.633600332$\times 10\uu{-6}$} {9.640977777$\times 10\uu{-6}$} {9.640837502$\times 10\uu{-6}$} {9.640832419$\times 10\uu{-6}$} {9.640832718$\times 10\uu{-6}$} 
\NEWERDSCMACRO{112} {6.843772122$\times 10\uu{-6}$} {8.895907290$\times 10\uu{-6}$} {9.110021916$\times 10\uu{-6}$} {9.117256104$\times 10\uu{-6}$} {9.117126842$\times 10\uu{-6}$} {9.117121703$\times 10\uu{-6}$} {9.117121989$\times 10\uu{-6}$} 
\NEWERDSCMACRO{113} {6.461857457$\times 10\uu{-6}$} {8.411629758$\times 10\uu{-6}$} {8.616966505$\times 10\uu{-6}$} {8.624055717$\times 10\uu{-6}$} {8.623936916$\times 10\uu{-6}$} {8.623931737$\times 10\uu{-6}$} {8.623932012$\times 10\uu{-6}$} 
\NEWERDSCMACRO{114} {6.102763311$\times 10\uu{-6}$} {7.955601288$\times 10\uu{-6}$} {8.152526593$\times 10\uu{-6}$} {8.159469597$\times 10\uu{-6}$} {8.159360721$\times 10\uu{-6}$} {8.159355520$\times 10\uu{-6}$} {8.159355782$\times 10\uu{-6}$} 
\NEWERDSCMACRO{115} {5.765030502$\times 10\uu{-6}$} {7.526057145$\times 10\uu{-6}$} {7.714922340$\times 10\uu{-6}$} {7.721718345$\times 10\uu{-6}$} {7.721618881$\times 10\uu{-6}$} {7.721613674$\times 10\uu{-6}$} {7.721613924$\times 10\uu{-6}$} 
\NEWERDSCMACRO{116} {5.447299596$\times 10\uu{-6}$} {7.121350554$\times 10\uu{-6}$} {7.302492439$\times 10\uu{-6}$} {7.309141055$\times 10\uu{-6}$} {7.309050508$\times 10\uu{-6}$} {7.309045308$\times 10\uu{-6}$} {7.309045546$\times 10\uu{-6}$} 
\NEWERDSCMACRO{117} {5.148303631$\times 10\uu{-6}$} {6.739944272$\times 10\uu{-6}$} {6.913685667$\times 10\uu{-6}$} {6.920186869$\times 10\uu{-6}$} {6.920104763$\times 10\uu{-6}$} {6.920099582$\times 10\uu{-6}$} {6.920099809$\times 10\uu{-6}$} 
\NEWERDSCMACRO{118} {4.866861398$\times 10\uu{-6}$} {6.380402806$\times 10\uu{-6}$} {6.547053095$\times 10\uu{-6}$} {6.553407181$\times 10\uu{-6}$} {6.553333061$\times 10\uu{-6}$} {6.553327910$\times 10\uu{-6}$} {6.553328125$\times 10\uu{-6}$} 
\NEWERDSCMACRO{119} {4.601871247$\times 10\uu{-6}$} {6.041385219$\times 10\uu{-6}$} {6.201240871$\times 10\uu{-6}$} {6.207448435$\times 10\uu{-6}$} {6.207381862$\times 10\uu{-6}$} {6.207376753$\times 10\uu{-6}$} {6.207376957$\times 10\uu{-6}$} 
\NEWERDSCMACRO{120} {4.352305373$\times 10\uu{-6}$} {5.721638475$\times 10\uu{-6}$} {5.874983557$\times 10\uu{-6}$} {5.881045458$\times 10\uu{-6}$} {5.880986013$\times 10\uu{-6}$} {5.880980953$\times 10\uu{-6}$} {5.880981146$\times 10\uu{-6}$} 
\NEWERDSCMACRO{121} {4.117204528$\times 10\uu{-6}$} {5.419991292$\times 10\uu{-6}$} {5.567097960$\times 10\uu{-6}$} {5.573015291$\times 10\uu{-6}$} {5.572962572$\times 10\uu{-6}$} {5.572957570$\times 10\uu{-6}$} {5.572957753$\times 10\uu{-6}$} 
\NEWERDSCMACRO{122} {3.895673153$\times 10\uu{-6}$} {5.135348446$\times 10\uu{-6}$} {5.276477422$\times 10\uu{-6}$} {5.282251488$\times 10\uu{-6}$} {5.282205110$\times 10\uu{-6}$} {5.282200173$\times 10\uu{-6}$} {5.282200346$\times 10\uu{-6}$} 
\NEWERDSCMACRO{123} {3.686874861$\times 10\uu{-6}$} {4.866685498$\times 10\uu{-6}$} {5.002086533$\times 10\uu{-6}$} {5.007718826$\times 10\uu{-6}$} {5.007678421$\times 10\uu{-6}$} {5.007673556$\times 10\uu{-6}$} {5.007673719$\times 10\uu{-6}$} 
\NEWERDSCMACRO{124} {3.490028274$\times 10\uu{-6}$} {4.613043919$\times 10\uu{-6}$} {4.742956236$\times 10\uu{-6}$} {4.748448413$\times 10\uu{-6}$} {4.748413630$\times 10\uu{-6}$} {4.748408842$\times 10\uu{-6}$} {4.748408996$\times 10\uu{-6}$} 
\NEWERDSCMACRO{125} {3.304403163$\times 10\uu{-6}$} {4.373526560$\times 10\uu{-6}$} {4.498179287$\times 10\uu{-6}$} {4.503533150$\times 10\uu{-6}$} {4.503503654$\times 10\uu{-6}$} {4.503498948$\times 10\uu{-6}$} {4.503499093$\times 10\uu{-6}$} 
\NEWERDSCMACRO{126} {3.129316885$\times 10\uu{-6}$} {4.147293465$\times 10\uu{-6}$} {4.266906049$\times 10\uu{-6}$} {4.272123528$\times 10\uu{-6}$} {4.272098998$\times 10\uu{-6}$} {4.272094378$\times 10\uu{-6}$} {4.272094514$\times 10\uu{-6}$} 
\NEWERDSCMACRO{127} {2.964131077$\times 10\uu{-6}$} {3.933557980$\times 10\uu{-6}$} {4.048340586$\times 10\uu{-6}$} {4.053423721$\times 10\uu{-6}$} {4.053403852$\times 10\uu{-6}$} {4.053399322$\times 10\uu{-6}$} {4.053399450$\times 10\uu{-6}$} 
\NEWERDSCMACRO{128} {2.808248593$\times 10\uu{-6}$} {3.731583150$\times 10\uu{-6}$} {3.841737048$\times 10\uu{-6}$} {3.846687974$\times 10\uu{-6}$} {3.846672474$\times 10\uu{-6}$} {3.846668038$\times 10\uu{-6}$} {3.846668157$\times 10\uu{-6}$} 
\NEWERDSCMACRO{129} {2.661110677$\times 10\uu{-6}$} {3.540678374$\times 10\uu{-6}$} {3.646396307$\times 10\uu{-6}$} {3.651217239$\times 10\uu{-6}$} {3.651205831$\times 10\uu{-6}$} {3.651201491$\times 10\uu{-6}$} {3.651201602$\times 10\uu{-6}$} 
\NEWERDSCMACRO{130} {2.522194331$\times 10\uu{-6}$} {3.360196292$\times 10\uu{-6}$} {3.461662834$\times 10\uu{-6}$} {3.466356057$\times 10\uu{-6}$} {3.466348478$\times 10\uu{-6}$} {3.466344235$\times 10\uu{-6}$} {3.466344339$\times 10\uu{-6}$} 
\NEWERDSCMACRO{131} {2.391009879$\times 10\uu{-6}$} {3.189529909$\times 10\uu{-6}$} {3.286921808$\times 10\uu{-6}$} {3.291489661$\times 10\uu{-6}$} {3.291485659$\times 10\uu{-6}$} {3.291481516$\times 10\uu{-6}$} {3.291481613$\times 10\uu{-6}$} 
\NEWERDSCMACRO{132} {2.267098709$\times 10\uu{-6}$} {3.028109907$\times 10\uu{-6}$} {3.121596413$\times 10\uu{-6}$} {3.126041283$\times 10\uu{-6}$} {3.126040619$\times 10\uu{-6}$} {3.126036577$\times 10\uu{-6}$} {3.126036667$\times 10\uu{-6}$} 
\NEWERDSCMACRO{133} {2.150031176$\times 10\uu{-6}$} {2.875402160$\times 10\uu{-6}$} {2.965145344$\times 10\uu{-6}$} {2.969469651$\times 10\uu{-6}$} {2.969472099$\times 10\uu{-6}$} {2.969468159$\times 10\uu{-6}$} {2.969468243$\times 10\uu{-6}$} 
\NEWERDSCMACRO{134} {2.039404656$\times 10\uu{-6}$} {2.730905416$\times 10\uu{-6}$} {2.817060471$\times 10\uu{-6}$} {2.821266664$\times 10\uu{-6}$} {2.821272008$\times 10\uu{-6}$} {2.821268170$\times 10\uu{-6}$} {2.821268247$\times 10\uu{-6}$} 
\NEWERDSCMACRO{135} {1.934841737$\times 10\uu{-6}$} {2.594149141$\times 10\uu{-6}$} {2.676864676$\times 10\uu{-6}$} {2.680955221$\times 10\uu{-6}$} {2.680963258$\times 10\uu{-6}$} {2.680959523$\times 10\uu{-6}$} {2.680959594$\times 10\uu{-6}$} 
\NEWERDSCMACRO{136} {1.835988542$\times 10\uu{-6}$} {2.464691520$\times 10\uu{-6}$} {2.544109839$\times 10\uu{-6}$} {2.548087215$\times 10\uu{-6}$} {2.548097750$\times 10\uu{-6}$} {2.548094118$\times 10\uu{-6}$} {2.548094183$\times 10\uu{-6}$} 
\NEWERDSCMACRO{137} {1.742513168$\times 10\uu{-6}$} {2.342117589$\times 10\uu{-6}$} {2.418374960$\times 10\uu{-6}$} {2.422241650$\times 10\uu{-6}$} {2.422254499$\times 10\uu{-6}$} {2.422250970$\times 10\uu{-6}$} {2.422251029$\times 10\uu{-6}$} 
\NEWERDSCMACRO{138} {1.654104234$\times 10\uu{-6}$} {2.226037502$\times 10\uu{-6}$} {2.299264412$\times 10\uu{-6}$} {2.303022899$\times 10\uu{-6}$} {2.303037888$\times 10\uu{-6}$} {2.303034460$\times 10\uu{-6}$} {2.303034515$\times 10\uu{-6}$} 
\NEWERDSCMACRO{139} {1.570469537$\times 10\uu{-6}$} {2.116084908$\times 10\uu{-6}$} {2.186406315$\times 10\uu{-6}$} {2.190059074$\times 10\uu{-6}$} {2.190076038$\times 10\uu{-6}$} {2.190072712$\times 10\uu{-6}$} {2.190072761$\times 10\uu{-6}$} 
\NEWERDSCMACRO{140} {1.491334790$\times 10\uu{-6}$} {2.011915454$\times 10\uu{-6}$} {2.079451018$\times 10\uu{-6}$} {2.083000514$\times 10\uu{-6}$} {2.083019296$\times 10\uu{-6}$} {2.083016070$\times 10\uu{-6}$} {2.083016115$\times 10\uu{-6}$} 
\NEWERDSCMACRO{141} {1.416442463$\times 10\uu{-6}$} {1.913205372$\times 10\uu{-6}$} {1.978069686$\times 10\uu{-6}$} {1.981518367$\times 10\uu{-6}$} {1.981538820$\times 10\uu{-6}$} {1.981535694$\times 10\uu{-6}$} {1.981535734$\times 10\uu{-6}$} 
\NEWERDSCMACRO{142} {1.345550688$\times 10\uu{-6}$} {1.819650176$\times 10\uu{-6}$} {1.881952982$\times 10\uu{-6}$} {1.885303277$\times 10\uu{-6}$} {1.885325262$\times 10\uu{-6}$} {1.885322234$\times 10\uu{-6}$} {1.885322270$\times 10\uu{-6}$} 
\NEWERDSCMACRO{143} {1.278432254$\times 10\uu{-6}$} {1.730963443$\times 10\uu{-6}$} {1.790809839$\times 10\uu{-6}$} {1.794064153$\times 10\uu{-6}$} {1.794087538$\times 10\uu{-6}$} {1.794084607$\times 10\uu{-6}$} {1.794084639$\times 10\uu{-6}$} 
\NEWERDSCMACRO{144} {1.214873661$\times 10\uu{-6}$} {1.646875668$\times 10\uu{-6}$} {1.704366308$\times 10\uu{-6}$} {1.707527023$\times 10\uu{-6}$} {1.707551682$\times 10\uu{-6}$} {1.707548848$\times 10\uu{-6}$} {1.707548876$\times 10\uu{-6}$} 
\NEWERDSCMACRO{145} {1.154674239$\times 10\uu{-6}$} {1.567133209$\times 10\uu{-6}$} {1.622364496$\times 10\uu{-6}$} {1.625433961$\times 10\uu{-6}$} {1.625459778$\times 10\uu{-6}$} {1.625457038$\times 10\uu{-6}$} {1.625457062$\times 10\uu{-6}$} 
\NEWERDSCMACRO{146} {1.097645335$\times 10\uu{-6}$} {1.491497291$\times 10\uu{-6}$} {1.544561558$\times 10\uu{-6}$} {1.547542093$\times 10\uu{-6}$} {1.547568957$\times 10\uu{-6}$} {1.547566309$\times 10\uu{-6}$} {1.547566330$\times 10\uu{-6}$} 
\NEWERDSCMACRO{147} {1.043609546$\times 10\uu{-6}$} {1.419743085$\times 10\uu{-6}$} {1.470728767$\times 10\uu{-6}$} {1.473622660$\times 10\uu{-6}$} {1.473650465$\times 10\uu{-6}$} {1.473647908$\times 10\uu{-6}$} {1.473647926$\times 10\uu{-6}$} 
\NEWERDSCMACRO{148} {9.924000085$\times 10\uu{-7}$} {1.351658836$\times 10\uu{-6}$} {1.400650644$\times 10\uu{-6}$} {1.403460146$\times 10\uu{-6}$} {1.403488795$\times 10\uu{-6}$} {1.403486327$\times 10\uu{-6}$} {1.403486342$\times 10\uu{-6}$} 
\NEWERDSCMACRO{149} {9.438597330$\times 10\uu{-7}$} {1.287045064$\times 10\uu{-6}$} {1.334124140$\times 10\uu{-6}$} {1.336851466$\times 10\uu{-6}$} {1.336880866$\times 10\uu{-6}$} {1.336878486$\times 10\uu{-6}$} {1.336878498$\times 10\uu{-6}$} 
\NEWERDSCMACRO{150} {8.978409857$\times 10\uu{-7}$} {1.225713802$\times 10\uu{-6}$} {1.270957877$\times 10\uu{-6}$} {1.273605203$\times 10\uu{-6}$} {1.273635268$\times 10\uu{-6}$} {1.273632973$\times 10\uu{-6}$} {1.273632982$\times 10\uu{-6}$} 

\end{tabular}
}
\captionof{table}{The successive refined estimates for the massive macroscopic propagation function \bbd{q\ll n\uprm{\textsc{mmp}} \equiv q\ll n - q\ll n\uprm{eft}}, at the coupling \bbd{\t = {{25 i}\over{\pi}}}.  The first column represents the strict infinite-\bbd{n} limit at fixed double-scaling parameter \bbd{\ddsc},
\bbd{q\ll n\uprm{\textsc{mmp}} \simeq \MacroFLPLower 0[\ddsc] = F\urm{inst}\lp{\smgkt}[\ddsc]}.  The next five columns represent this approximation corrected by adding successive terms \bbd{n\uu{-k}\cc \MacroFLPLower k[\ddsc]} for \bbd{k = 1,\cdots 5}.  The last column is the exact
result as computed \cite{LocData} following the method of \cite{Gerchkovitz:2016gxx}.}
\nopagebreak
\label{TypesOfLiouvilleAmplitudeTransformationTableSpecificParameterChoiceOfInterestToUs}
\nopagebreak
\filbreak
\end{center}
\nopagebreak
\filbreak
\end{table}
\filbreak
\nopagebreak[61]
\nopagebreak[62]
\nopagebreak[63]
\filbreak
\subsection{Tables listing the accuracy of the double-scaled ({\it i.e.\rm } fixed-\bbd{\ddsc}) large-charge asymptotic estimates  at \bbd{\t = {{25}\over \pi}\cc i}, up to \bbd{n=150}}

 \begin{table}[H]
\begin{center}
\begin{tabular}{ |c||c|c|c|c|c|c|c| } 
 \hline\hline
 ~ & ${{{\rm accuracy~of}}\atop{\MacroFLPLower 0[\ddsc] 
 }}$  
 &${{\rm accuracy~incl.}\atop{n\uu{-1}\MacroFLPLower 1[\ddsc]}}$  &${{\rm accuracy~incl.}\atop{n\uu{-2}\MacroFLPLower 2[\ddsc]}}$ & ${{\rm accuracy~incl.}\atop{n\uu{-3}\MacroFLPLower 3[\ddsc]}}$& ${{\rm accuracy~incl.}\atop{n\uu{-4}\MacroFLPLower 4[\ddsc]}}$ & ${{\rm accuracy~incl.}\atop{n\uu{-5}\MacroFLPLower 5[\ddsc]}}$
    \\ \hline\hline

\NEWDSCMACRO {1} {0.2770} {0.5664} {0.8175} {1.034} {1.231} {1.417}

\NEWDSCMACRO{2} {0.4565} {0.9669} {1.482} {1.985} {2.478} {2.963}
\NEWDSCMACRO{3} {0.5737} {1.209} {1.867} {2.529} {3.190} {3.847} 
\NEWDSCMACRO{4} {0.6665} {1.390} {2.146} {2.914} {3.687} {4.463} 
\NEWDSCMACRO{5} {0.7473} {1.541} {2.371} {3.219} {4.076} {4.938} 
\NEWDSCMACRO{6} {0.8216} {1.676} {2.569} {3.480} {4.403} {5.334} 
\NEWDSCMACRO{7} {0.8929} {1.803} {2.750} {3.718} {4.698} {5.686} 
\NEWDSCMACRO{8} {0.9633} {1.925} {2.925} {3.946} {4.978} {6.019} 
\NEWDSCMACRO{9} {1.035} {2.049} {3.101} {4.174} {5.261} {6.356} 
\NEWDSCMACRO{10} {1.109} {2.176} {3.284} {4.416} {5.566} {6.730} 
\NEWDSCMACRO{11} {1.188} {2.312} {3.484} {4.693} {5.938} {7.233} 
\NEWDSCMACRO{12} {1.274} {2.463} {3.720} {5.058} {6.563} {8.35} 
\NEWDSCMACRO{13} {1.371} {2.641} {4.037} {5.831} {6.683} {7.48} 
\NEWDSCMACRO{14} {1.484} {2.871} {4.678} {5.520} {6.347} {7.36} 
\NEWDSCMACRO{15} {1.625} {3.237} {4.536} {5.248} {6.261} {7.361} 
\NEWDSCMACRO{16} {1.816} {4.371} {4.196} {5.155} {6.251} {7.41} 
\NEWDSCMACRO{17} {2.131} {3.262} {4.063} {5.125} {6.282} {7.49} 
\NEWDSCMACRO{18} {4.508} {3.021} {3.998} {5.130} {6.340} {7.61} 
\NEWDSCMACRO{19} {2.152} {2.895} {3.969} {5.158} {6.424} {7.76} 
\NEWDSCMACRO{20} {1.864} {2.817} {3.962} {5.206} {6.533} {7.97} 
\NEWDSCMACRO{21} {1.699} {2.766} {3.972} {5.272} {6.678} {8.31} 
\NEWDSCMACRO{22} {1.585} {2.733} {3.995} {5.359} {6.880} {\redd{$\geq 10$}} 

\end{tabular}
\captionof{table}{\footnotesize{Effective number of accurate digits for each of the successive refined estimates for the massive macroscopic propagation function \bbd{q\ll n\uprm{\textsc{mmp}} \equiv q\ll n - q\ll n\uprm{eft}}, at the coupling \bbd{\t = {{25 i}\over{\pi}}}.  The first column represents the accuracy of the strict infinite-\bbd{n} limit at fixed double-scaling parameter \bbd{\ddsc},
\bbd{q\ll n\uprm{\textsc{mmp}} \simeq \MacroFLPLower 0[\ddsc] = F\urm{inst}\lp{\smgkt}[\ddsc]}.  The other columns represent the accuracy of the leading approximation corrected by adding successive terms \bbd{n\uu{-p}\cc \MacroFLPLower p[\ddsc]} for \bbd{p = 1,\cdots 5}.  The "effective number of accurate digits" is defined here as the log of the relative error of the estimate of \bbd{q\ll n\uprm{\textsc{mmp}}} , times \bbd{-{1\over{{\rm Log}[10]}}}. 
That is, the table entries are given by \bbd{- {1\over{{\rm Log}[10]}}\cc {\rm Log}\left | \cc {{q\ll n\uprm{\textsc{mmp}} - (q\ll n\uprm{\textsc{mmp}})\lrm{estimate}}\over{q\ll n\uprm{\textsc{mmp}}  }}    \cc \right | }. In the case \bbd{n=22} we can only give
a lower bound for the accuracy of the estimate carried to order \bbd{n\uu{-5},} as the agreement between the estimate and the exact value exceeds the \bbd{10} significant digits to which we have computed the estimates.}}
\label{TypesOfLiouvilleAmplitudeTransformationTableSpecificParameterChoiceOfInterestToUs}
\end{center}
\end{table}

\newpage
 \vskip-1in
 \begin{table}[H]
\begin{center}
\begin{tabular}{ |c||c|c|c|c|c|c|c| } 
 \hline\hline
 ~ & ${{{\rm accuracy~of}}\atop{\MacroFLPLower 0[\ddsc] 
 }}$  
 &${{\rm accuracy~incl.}\atop{n\uu{-1}\MacroFLPLower 1[\ddsc]}}$  &${{\rm accuracy~incl.}\atop{n\uu{-2}\MacroFLPLower 2[\ddsc]}}$ & ${{\rm accuracy~incl.}\atop{n\uu{-3}\MacroFLPLower 3[\ddsc]}}$& ${{\rm accuracy~incl.}\atop{n\uu{-4}\MacroFLPLower 4[\ddsc]}}$ & ${{\rm accuracy~incl.}\atop{n\uu{-5}\MacroFLPLower 5[\ddsc]}}$
    \\ \hline\hline

\NEWDSCMACRO{23} {1.498} {2.712} {4.030} {5.470} {7.220} {8.4} 
\NEWDSCMACRO{24} {1.429} {2.701} {4.078} {5.621} {8.5} {8.14} 
\NEWDSCMACRO{25} {1.371} {2.699} {4.140} {5.844} {7.231} {8.02} 
\NEWDSCMACRO{26} {1.322} {2.703} {4.219} {6.290} {6.976} {7.95} 
\NEWDSCMACRO{27} {1.280} {2.713} {4.322} {6.460} {6.839} {7.91} 
\NEWDSCMACRO{28} {1.242} {2.730} {4.461} {5.943} {6.754} {7.89} 
\NEWDSCMACRO{29} {1.209} {2.752} {4.672} {5.730} {6.699} {7.89} 
\NEWDSCMACRO{30} {1.179} {2.782} {5.098} {5.600} {6.662} {7.90} 
\NEWDSCMACRO{31} {1.152} {2.818} {5.282} {5.511} {6.640} {7.92} 
\NEWDSCMACRO{32} {1.127} {2.864} {4.740} {5.446} {6.629} {7.95} 
\NEWDSCMACRO{33} {1.104} {2.920} {4.511} {5.397} {6.627} {7.99} 
\NEWDSCMACRO{34} {1.083} {2.990} {4.365} {5.360} {6.633} {8.04} 
\NEWDSCMACRO{35} {1.064} {3.080} {4.260} {5.333} {6.647} {8.11} 
\NEWDSCMACRO{36} {1.046} {3.203} {4.179} {5.313} {6.668} {8.19} 
\NEWDSCMACRO{37} {1.029} {3.384} {4.113} {5.300} {6.696} {8.30} 
\NEWDSCMACRO{38} {1.013} {3.725} {4.059} {5.291} {6.732} {8.4} 
\NEWDSCMACRO{39} {0.9975} {4.349} {4.014} {5.288} {6.777} {8.7} 
\NEWDSCMACRO{40} {0.9832} {3.543} {3.976} {5.290} {6.832} {9.1} 
\NEWDSCMACRO{41} {0.9698} {3.271} {3.943} {5.296} {6.901} {9.2} 
\NEWDSCMACRO{42} {0.9570} {3.102} {3.915} {5.306} {6.987} {8.7} 
\NEWDSCMACRO{43} {0.9448} {2.977} {3.890} {5.320} {7.100} {8.5} 
\NEWDSCMACRO{44} {0.9332} {2.879} {3.869} {5.339} {7.258} {8.35} 

\end{tabular}
\captionof{table}{\footnotesize{Effective number of accurate digits for each of the successive refined estimates for the massive macroscopic propagation function \bbd{q\ll n\uprm{\textsc{mmp}} \equiv q\ll n - q\ll n\uprm{eft}}, at the coupling \bbd{\t = {{25 i}\over{\pi}}}.  The first column represents the accuracy of the strict infinite-\bbd{n} limit at fixed double-scaling parameter \bbd{\ddsc},
\bbd{q\ll n\uprm{\textsc{mmp}} \simeq \MacroFLPLower 0[\ddsc] = F\urm{inst}\lp{\smgkt}[\ddsc]}.  The other columns represent the accuracy of the leading approximation corrected by adding successive terms \bbd{n\uu{-p}\cc \MacroFLPLower p[\ddsc]} for \bbd{p = 1,\cdots 5}.  The "effective number of accurate digits" is defined here as the log of the relative error of the estimate of \bbd{q\ll n\uprm{\textsc{mmp}}} , times \bbd{-{1\over{{\rm Log}[10]}}}.
That is, the table entries are given by \bbd{- {1\over{{\rm Log}[10]}}\cc {\rm Log}\left | \cc {{q\ll n\uprm{\textsc{mmp}} - (q\ll n\uprm{\textsc{mmp}})\lrm{estimate}}\over{q\ll n\uprm{\textsc{mmp}}  }}    \cc \right | }.}}
\label{TypesOfLiouvilleAmplitudeTransformationTableSpecificParameterChoiceOfInterestToUs}
\end{center} 
\end{table}

\newpage
 \vskip-1in
 \begin{table}[H]
\begin{center}
\begin{tabular}{ |c||c|c|c|c|c|c|c| } 
 \hline\hline
 ~ & ${{{\rm accuracy~of}}\atop{\MacroFLPLower 0[\ddsc] 
 }}$  
 &${{\rm accuracy~incl.}\atop{n\uu{-1}\MacroFLPLower 1[\ddsc]}}$  &${{\rm accuracy~incl.}\atop{n\uu{-2}\MacroFLPLower 2[\ddsc]}}$ & ${{\rm accuracy~incl.}\atop{n\uu{-3}\MacroFLPLower 3[\ddsc]}}$& ${{\rm accuracy~incl.}\atop{n\uu{-4}\MacroFLPLower 4[\ddsc]}}$ & ${{\rm accuracy~incl.}\atop{n\uu{-5}\MacroFLPLower 5[\ddsc]}}$
    \\ \hline\hline 

\NEWDSCMACRO{45} {0.9222} {2.797} {3.852} {5.363} {7.52} {8.25} 
\NEWDSCMACRO{46} {0.9116} {2.727} {3.837} {5.392} {8.25} {8.18} 
\NEWDSCMACRO{47} {0.9015} {2.665} {3.824} {5.426} {7.71} {8.12} 
\NEWDSCMACRO{48} {0.8918} {2.611} {3.814} {5.468} {7.354} {8.07} 
\NEWDSCMACRO{49} {0.8824} {2.562} {3.805} {5.518} {7.162} {8.03} 
\NEWDSCMACRO{50} {0.8735} {2.517} {3.799} {5.579} {7.032} {8.00} 
\NEWDSCMACRO{51} {0.8648} {2.476} {3.795} {5.655} {6.933} {7.97} 
\NEWDSCMACRO{52} {0.8565} {2.438} {3.792} {5.752} {6.855} {7.95} 
\NEWDSCMACRO{53} {0.8485} {2.403} {3.792} {5.883} {6.791} {7.93} 
\NEWDSCMACRO{54} {0.8408} {2.370} {3.793} {6.082} {6.737} {7.92} 
\NEWDSCMACRO{55} {0.8333} {2.339} {3.795} {6.476} {6.691} {7.91} 
\NEWDSCMACRO{56} {0.8260} {2.310} {3.800} {6.761} {6.651} {7.91} 
\NEWDSCMACRO{57} {0.8190} {2.283} {3.806} {6.159} {6.616} {7.90} 
\NEWDSCMACRO{58} {0.8122} {2.257} {3.815} {5.913} {6.586} {7.90} 
\NEWDSCMACRO{59} {0.8056} {2.232} {3.825} {5.755} {6.560} {7.91} 
\NEWDSCMACRO{60} {0.7992} {2.209} {3.838} {5.638} {6.537} {7.91} 
\NEWDSCMACRO{61} {0.7930} {2.187} {3.852} {5.545} {6.516} {7.92} 
\NEWDSCMACRO{62} {0.7869} {2.165} {3.869} {5.468} {6.499} {7.93} 
\NEWDSCMACRO{63} {0.7810} {2.145} {3.889} {5.402} {6.484} {7.95} 
\NEWDSCMACRO{64} {0.7753} {2.125} {3.912} {5.345} {6.471} {7.97} 
\NEWDSCMACRO{65} {0.7697} {2.106} {3.938} {5.295} {6.460} {7.99} 

\end{tabular}
\captionof{table}{\footnotesize{Effective number of accurate digits for each of the successive refined estimates for the massive macroscopic propagation function \bbd{q\ll n\uprm{\textsc{mmp}} \equiv q\ll n - q\ll n\uprm{eft}}, at the coupling \bbd{\t = {{25 i}\over{\pi}}}.  The first column represents the accuracy of the strict infinite-\bbd{n} limit at fixed double-scaling parameter \bbd{\ddsc},
\bbd{q\ll n\uprm{\textsc{mmp}} \simeq \MacroFLPLower 0[\ddsc] = F\urm{inst}\lp{\smgkt}[\ddsc]}.  The other columns represent the accuracy of the leading approximation corrected by adding successive terms \bbd{n\uu{-p}\cc \MacroFLPLower p[\ddsc]} for \bbd{p = 1,\cdots 5}.  The "effective number of accurate digits" is defined here as the log of the relative error of the estimate of \bbd{q\ll n\uprm{\textsc{mmp}}} , times \bbd{-{1\over{{\rm Log}[10]}}}.
That is, the table entries are given by \bbd{- {1\over{{\rm Log}[10]}}\cc {\rm Log}\left | \cc {{q\ll n\uprm{\textsc{mmp}} - (q\ll n\uprm{\textsc{mmp}})\lrm{estimate}}\over{q\ll n\uprm{\textsc{mmp}}  }}    \cc \right | }.}}
\label{TypesOfLiouvilleAmplitudeTransformationTableSpecificParameterChoiceOfInterestToUs}
\end{center} 
\end{table}

\newpage
 \vskip-1in
 \begin{table}[H]
\begin{center}
\begin{tabular}{ |c||c|c|c|c|c|c|c| } 
 \hline\hline 
 ~ & ${{{\rm accuracy~of}}\atop{\MacroFLPLower 0[\ddsc] 
 }}$  
 &${{\rm accuracy~incl.}\atop{n\uu{-1}\MacroFLPLower 1[\ddsc]}}$  &${{\rm accuracy~incl.}\atop{n\uu{-2}\MacroFLPLower 2[\ddsc]}}$ & ${{\rm accuracy~incl.}\atop{n\uu{-3}\MacroFLPLower 3[\ddsc]}}$& ${{\rm accuracy~incl.}\atop{n\uu{-4}\MacroFLPLower 4[\ddsc]}}$ & ${{\rm accuracy~incl.}\atop{n\uu{-5}\MacroFLPLower 5[\ddsc]}}$
    \\ \hline\hline 

\NEWDSCMACRO{66} {0.7643} {2.088} {3.969} {5.250} {6.450} {8.01} 
\NEWDSCMACRO{67} {0.7590} {2.070} {4.004} {5.210} {6.443} {8.04} 
\NEWDSCMACRO{68} {0.7538} {2.053} {4.045} {5.173} {6.438} {8.08} 
\NEWDSCMACRO{69} {0.7488} {2.037} {4.094} {5.139} {6.434} {8.12} 
\NEWDSCMACRO{70} {0.7439} {2.021} {4.151} {5.108} {6.431} {8.16} 
\NEWDSCMACRO{71} {0.7390} {2.006} {4.222} {5.080} {6.431} {8.22} 
\NEWDSCMACRO{72} {0.7343} {1.991} {4.312} {5.054} {6.432} {8.28} 
\NEWDSCMACRO{73} {0.7297} {1.977} {4.430} {5.029} {6.434} {8.4} 
\NEWDSCMACRO{74} {0.7252} {1.963} {4.603} {5.007} {6.438} {8.5} 
\NEWDSCMACRO{75} {0.7209} {1.949} {4.914} {4.986} {6.444} {8.6} 
\NEWDSCMACRO{76} {0.7165} {1.936} {5.975} {4.966} {6.451} {8.9} 
\NEWDSCMACRO{77} {0.7123} {1.923} {4.829} {4.948} {6.460} {9.} 
\NEWDSCMACRO{78} {0.7082} {1.911} {4.536} {4.931} {6.471} {9.2} 
\NEWDSCMACRO{79} {0.7042} {1.899} {4.358} {4.915} {6.484} {8.8} 
\NEWDSCMACRO{80} {0.7002} {1.887} {4.228} {4.901} {6.499} {8.5} 
\NEWDSCMACRO{81} {0.6963} {1.876} {4.126} {4.887} {6.516} {8.4} 
\NEWDSCMACRO{82} {0.6925} {1.864} {4.041} {4.874} {6.536} {8.30} 
\NEWDSCMACRO{83} {0.6887} {1.853} {3.967} {4.862} {6.558} {8.21} 
\NEWDSCMACRO{84} {0.6851} {1.843} {3.903} {4.851} {6.584} {8.14} 
\NEWDSCMACRO{85} {0.6814} {1.832} {3.846} {4.841} {6.613} {8.08} 
\NEWDSCMACRO{86} {0.6779} {1.822} {3.794} {4.832} {6.646} {8.02} 
\NEWDSCMACRO{87} {0.6744} {1.812} {3.746} {4.823} {6.684} {7.97} 

\end{tabular}
\captionof{table}{\footnotesize{Effective number of accurate digits for each of the successive refined estimates for the massive macroscopic propagation function \bbd{q\ll n\uprm{\textsc{mmp}} \equiv q\ll n - q\ll n\uprm{eft}}, at the coupling \bbd{\t = {{25 i}\over{\pi}}}.  The first column represents the accuracy of the strict infinite-\bbd{n} limit at fixed double-scaling parameter \bbd{\ddsc},
\bbd{q\ll n\uprm{\textsc{mmp}} \simeq \MacroFLPLower 0[\ddsc] = F\urm{inst}\lp{\smgkt}[\ddsc]}.  The other columns represent the accuracy of the leading approximation corrected by adding successive terms \bbd{n\uu{-p}\cc \MacroFLPLower p[\ddsc]} for \bbd{p = 1,\cdots 5}.  The "effective number of accurate digits" is defined here as the log of the relative error of the estimate of \bbd{q\ll n\uprm{\textsc{mmp}}} , times \bbd{-{1\over{{\rm Log}[10]}}}.
That is, the table entries are given by \bbd{- {1\over{{\rm Log}[10]}}\cc {\rm Log}\left | \cc {{q\ll n\uprm{\textsc{mmp}} - (q\ll n\uprm{\textsc{mmp}})\lrm{estimate}}\over{q\ll n\uprm{\textsc{mmp}}  }}    \cc \right | }.}}
\label{TypesOfLiouvilleAmplitudeTransformationTableSpecificParameterChoiceOfInterestToUs}
\end{center} 
\end{table}

\newpage
 \vskip-1in
 \begin{table}[H]
\begin{center}
\begin{tabular}{ |c||c|c|c|c|c|c|c| } 
 \hline\hline 
 ~ & ${{{\rm accuracy~of}}\atop{\MacroFLPLower 0[\ddsc] 
 }}$  
 &${{\rm accuracy~incl.}\atop{n\uu{-1}\MacroFLPLower 1[\ddsc]}}$  &${{\rm accuracy~incl.}\atop{n\uu{-2}\MacroFLPLower 2[\ddsc]}}$ & ${{\rm accuracy~incl.}\atop{n\uu{-3}\MacroFLPLower 3[\ddsc]}}$& ${{\rm accuracy~incl.}\atop{n\uu{-4}\MacroFLPLower 4[\ddsc]}}$ & ${{\rm accuracy~incl.}\atop{n\uu{-5}\MacroFLPLower 5[\ddsc]}}$
    \\ \hline\hline 

\NEWDSCMACRO{88} {0.6710} {1.802} {3.703} {4.815} {6.729} {7.93} 
\NEWDSCMACRO{89} {0.6676} {1.792} {3.662} {4.808} {6.781} {7.89} 
\NEWDSCMACRO{90} {0.6643} {1.783} {3.624} {4.801} {6.843} {7.86} 
\NEWDSCMACRO{91} {0.6611} {1.774} {3.588} {4.796} {6.920} {7.83} 
\NEWDSCMACRO{92} {0.6579} {1.765} {3.555} {4.790} {7.016} {7.79} 
\NEWDSCMACRO{93} {0.6548} {1.756} {3.523} {4.786} {7.147} {7.77} 
\NEWDSCMACRO{94} {0.6517} {1.747} {3.493} {4.782} {7.342} {7.74} 
\NEWDSCMACRO{95} {0.6486} {1.739} {3.464} {4.779} {7.72} {7.72} 
\NEWDSCMACRO{96} {0.6456} {1.731} {3.436} {4.776} {8.07} {7.70} 
\NEWDSCMACRO{97} {0.6427} {1.722} {3.410} {4.775} {7.43} {7.68} 
\NEWDSCMACRO{98} {0.6398} {1.714} {3.385} {4.773} {7.182} {7.66} 
\NEWDSCMACRO{99} {0.6369} {1.706} {3.361} {4.773} {7.020} {7.64} 
\NEWDSCMACRO{100} {0.6341} {1.699} {3.337} {4.773} {6.900} {7.62} 
\NEWDSCMACRO{101} {0.6313} {1.691} {3.315} {4.774} {6.804} {7.61} 
\NEWDSCMACRO{102} {0.6286} {1.684} {3.293} {4.775} {6.724} {7.59} 
\NEWDSCMACRO{103} {0.6259} {1.676} {3.272} {4.777} {6.656} {7.58} 
\NEWDSCMACRO{104} {0.6232} {1.669} {3.252} {4.780} {6.596} {7.57} 
\NEWDSCMACRO{105} {0.6206} {1.662} {3.232} {4.784} {6.542} {7.56} 
\NEWDSCMACRO{106} {0.6180} {1.655} {3.213} {4.788} {6.494} {7.55} 
\NEWDSCMACRO{107} {0.6154} {1.648} {3.194} {4.793} {6.450} {7.54} 
\NEWDSCMACRO{108} {0.6129} {1.641} {3.176} {4.799} {6.409} {7.53} 
\NEWDSCMACRO{109} {0.6104} {1.634} {3.159} {4.806} {6.372} {7.52} 
\NEWDSCMACRO{110} {0.6080} {1.628} {3.141} {4.814} {6.337} {7.51} 
\end{tabular}
\captionof{table}{\footnotesize{Effective number of accurate digits for each of the successive refined estimates for the massive macroscopic propagation function \bbd{q\ll n\uprm{\textsc{mmp}} \equiv q\ll n - q\ll n\uprm{eft}}, at the coupling \bbd{\t = {{25 i}\over{\pi}}}.  The first column represents the accuracy of the strict infinite-\bbd{n} limit at fixed double-scaling parameter \bbd{\ddsc},
\bbd{q\ll n\uprm{\textsc{mmp}} \simeq \MacroFLPLower 0[\ddsc] = F\urm{inst}\lp{\smgkt}[\ddsc]}.  The other columns represent the accuracy of the leading approximation corrected by adding successive terms \bbd{n\uu{-p}\cc \MacroFLPLower p[\ddsc]} for \bbd{p = 1,\cdots 5}.  The "effective number of accurate digits" is defined here as the log of the relative error of the estimate of \bbd{q\ll n\uprm{\textsc{mmp}}} , times \bbd{-{1\over{{\rm Log}[10]}}}.
That is, the table entries are given by \bbd{- {1\over{{\rm Log}[10]}}\cc {\rm Log}\left | \cc {{q\ll n\uprm{\textsc{mmp}} - (q\ll n\uprm{\textsc{mmp}})\lrm{estimate}}\over{q\ll n\uprm{\textsc{mmp}}  }}    \cc \right | }.}}
\label{TypesOfLiouvilleAmplitudeTransformationTableSpecificParameterChoiceOfInterestToUs}
\end{center} 
\end{table}

\newpage
 \vskip-1in
 \begin{table}[H]
\begin{center}
\begin{tabular}{ |c||c|c|c|c|c|c|c| } 
 \hline\hline 
 ~ & ${{{\rm accuracy~of}}\atop{\MacroFLPLower 0[\ddsc] 
 }}$  
 &${{\rm accuracy~incl.}\atop{n\uu{-1}\MacroFLPLower 1[\ddsc]}}$  &${{\rm accuracy~incl.}\atop{n\uu{-2}\MacroFLPLower 2[\ddsc]}}$ & ${{\rm accuracy~incl.}\atop{n\uu{-3}\MacroFLPLower 3[\ddsc]}}$& ${{\rm accuracy~incl.}\atop{n\uu{-4}\MacroFLPLower 4[\ddsc]}}$ & ${{\rm accuracy~incl.}\atop{n\uu{-5}\MacroFLPLower 5[\ddsc]}}$
    \\ \hline\hline

\NEWDSCMACRO{111} {0.6056} {1.621} {3.125} {4.823} {6.304} {7.51} 
\NEWDSCMACRO{112} {0.6032} {1.615} {3.109} {4.832} {6.274} {7.50} 
\NEWDSCMACRO{113} {0.6008} {1.609} {3.093} {4.843} {6.245} {7.50} 
\NEWDSCMACRO{114} {0.5985} {1.603} {3.077} {4.855} {6.218} {7.49} 
\NEWDSCMACRO{115} {0.5962} {1.596} {3.062} {4.869} {6.192} {7.49} 
\NEWDSCMACRO{116} {0.5939} {1.590} {3.047} {4.884} {6.168} {7.49} 
\NEWDSCMACRO{117} {0.5917} {1.584} {3.033} {4.900} {6.145} {7.48} 
\NEWDSCMACRO{118} {0.5895} {1.579} {3.019} {4.919} {6.123} {7.48} 
\NEWDSCMACRO{119} {0.5873} {1.573} {3.005} {4.939} {6.102} {7.48} 
\NEWDSCMACRO{120} {0.5851} {1.567} {2.991} {4.961} {6.082} {7.48} 
\NEWDSCMACRO{121} {0.5830} {1.561} {2.978} {4.986} {6.063} {7.48} 
\NEWDSCMACRO{122} {0.5809} {1.556} {2.965} {5.014} {6.045} {7.48} 
\NEWDSCMACRO{123} {0.5788} {1.550} {2.952} {5.045} {6.027} {7.49} 
\NEWDSCMACRO{124} {0.5767} {1.545} {2.940} {5.081} {6.011} {7.49} 
\NEWDSCMACRO{125} {0.5747} {1.540} {2.928} {5.121} {5.994} {7.49} 
\NEWDSCMACRO{126} {0.5727} {1.534} {2.916} {5.168} {5.979} {7.50} 
\NEWDSCMACRO{127} {0.5707} {1.529} {2.904} {5.223} {5.964} {7.50} 
\NEWDSCMACRO{128} {0.5687} {1.524} {2.892} {5.288} {5.950} {7.51} 
\NEWDSCMACRO{129} {0.5668} {1.519} {2.881} {5.368} {5.936} {7.51} 
\NEWDSCMACRO{130} {0.5648} {1.514} {2.869} {5.471} {5.923} {7.52} 

\end{tabular}
\captionof{table}{\footnotesize{Effective number of accurate digits for each of the successive refined estimates for the massive macroscopic propagation function \bbd{q\ll n\uprm{\textsc{mmp}} \equiv q\ll n - q\ll n\uprm{eft}}, at the coupling \bbd{\t = {{25 i}\over{\pi}}}.  The first column represents the accuracy of the strict infinite-\bbd{n} limit at fixed double-scaling parameter \bbd{\ddsc},
\bbd{q\ll n\uprm{\textsc{mmp}} \simeq \MacroFLPLower 0[\ddsc] = F\urm{inst}\lp{\smgkt}[\ddsc]}.  The other columns represent the accuracy of the leading approximation corrected by adding successive terms \bbd{n\uu{-p}\cc \MacroFLPLower p[\ddsc]} for \bbd{p = 1,\cdots 5}.  The "effective number of accurate digits" is defined here as the log of the relative error of the estimate of \bbd{q\ll n\uprm{\textsc{mmp}}} , times \bbd{-{1\over{{\rm Log}[10]}}}.
That is, the table entries are given by \bbd{- {1\over{{\rm Log}[10]}}\cc {\rm Log}\left | \cc {{q\ll n\uprm{\textsc{mmp}} - (q\ll n\uprm{\textsc{mmp}})\lrm{estimate}}\over{q\ll n\uprm{\textsc{mmp}}  }}    \cc \right | }.}}
\label{TypesOfLiouvilleAmplitudeTransformationTableSpecificParameterChoiceOfInterestToUs}
\end{center} 
\end{table}

\newpage
 \vskip-1in
 \begin{table}[H]
\begin{center}
\begin{tabular}{ |c||c|c|c|c|c|c|c| } 
 \hline\hline 
 ~ & ${{{\rm accuracy~of}}\atop{\MacroFLPLower 0[\ddsc] 
 }}$  
 &${{\rm accuracy~incl.}\atop{n\uu{-1}\MacroFLPLower 1[\ddsc]}}$  &${{\rm accuracy~incl.}\atop{n\uu{-2}\MacroFLPLower 2[\ddsc]}}$ & ${{\rm accuracy~incl.}\atop{n\uu{-3}\MacroFLPLower 3[\ddsc]}}$& ${{\rm accuracy~incl.}\atop{n\uu{-4}\MacroFLPLower 4[\ddsc]}}$ & ${{\rm accuracy~incl.}\atop{n\uu{-5}\MacroFLPLower 5[\ddsc]}}$
    \\ \hline\hline 

\NEWDSCMACRO{131} {0.5629} {1.509} {2.858} {5.612} {5.910} {7.53} 
\NEWDSCMACRO{132} {0.5610} {1.504} {2.848} {5.831} {5.898} {7.54} 
\NEWDSCMACRO{133} {0.5592} {1.499} {2.837} {6.324} {5.886} {7.55} 
\NEWDSCMACRO{134} {0.5573} {1.494} {2.826} {6.251} {5.875} {7.56} 
\NEWDSCMACRO{135} {0.5555} {1.490} {2.816} {5.788} {5.864} {7.58} 
\NEWDSCMACRO{136} {0.5537} {1.485} {2.806} {5.563} {5.854} {7.59} 
\NEWDSCMACRO{137} {0.5519} {1.480} {2.796} {5.412} {5.844} {7.61} 
\NEWDSCMACRO{138} {0.5501} {1.476} {2.786} {5.297} {5.834} {7.63} 
\NEWDSCMACRO{139} {0.5483} {1.471} {2.776} {5.204} {5.825} {7.65} 
\NEWDSCMACRO{140} {0.5466} {1.467} {2.767} {5.126} {5.816} {7.67} 
\NEWDSCMACRO{141} {0.5449} {1.462} {2.757} {5.057} {5.808} {7.69} 
\NEWDSCMACRO{142} {0.5432} {1.458} {2.748} {4.997} {5.800} {7.72} 
\NEWDSCMACRO{143} {0.5415} {1.454} {2.739} {4.942} {5.792} {7.75} 
\NEWDSCMACRO{144} {0.5398} {1.449} {2.730} {4.893} {5.784} {7.78} 
\NEWDSCMACRO{145} {0.5382} {1.445} {2.721} {4.847} {5.777} {7.82} 
\NEWDSCMACRO{146} {0.5365} {1.441} {2.712} {4.805} {5.770} {7.87} 
\NEWDSCMACRO{147} {0.5349} {1.437} {2.703} {4.766} {5.764} {7.92} 
\NEWDSCMACRO{148} {0.5333} {1.433} {2.695} {4.729} {5.758} {7.98} 
\NEWDSCMACRO{149} {0.5317} {1.429} {2.686} {4.694} {5.752} {8.05} 
\NEWDSCMACRO{150} {0.5301} {1.425} {2.678} {4.661} {5.746} {8.14} 

\end{tabular}
\captionof{table}{\footnotesize{Effective number of accurate digits for each of the successive refined estimates for the massive macroscopic propagation function \bbd{q\ll n\uprm{\textsc{mmp}} \equiv q\ll n - q\ll n\uprm{eft}}, at the coupling \bbd{\t = {{25 i}\over{\pi}}}.  The first column represents the accuracy of the strict infinite-\bbd{n} limit at fixed double-scaling parameter \bbd{\ddsc},
\bbd{q\ll n\uprm{\textsc{mmp}} \simeq \MacroFLPLower 0[\ddsc] = F\urm{inst}\lp{\smgkt}[\ddsc]}.  The other columns represent the accuracy of the leading approximation corrected by adding successive terms \bbd{n\uu{-p}\cc \MacroFLPLower p[\ddsc]} for \bbd{p = 1,\cdots 5}.  The "effective number of accurate digits" is defined here as the log of the relative error of the estimate of \bbd{q\ll n\uprm{\textsc{mmp}}} , times \bbd{-{1\over{{\rm Log}[10]}}}.
That is, the table entries are given by \bbd{- {1\over{{\rm Log}[10]}}\cc {\rm Log}\left | \cc {{q\ll n\uprm{\textsc{mmp}} - (q\ll n\uprm{\textsc{mmp}})\lrm{estimate}}\over{q\ll n\uprm{\textsc{mmp}}  }}    \cc \right | }.}}
\label{TypesOfLiouvilleAmplitudeTransformationTableSpecificParameterChoiceOfInterestToUs}
\end{center} 
\end{table}

\section{Discussion and conclusions}\label{ConclusionsSection}
\heading{Summary}

In this paper we have done the following:

\bii
 
  \item{We considered the large-R-charge, fixed \bbd{\t} expansion of the Coulomb branch correlators of \bbd{{\cal N} = 2} superconformal SQCD,
decomposing the log of the correlation function into a known \textsc{eft}
term \bbd{q\ll n\uprm{\textsc{eft}}} and a massive macroscopic propagation
term \bbd{q\ll n\uprm{\textsc{mmp}}}, focusing our attention on
the latter.}
\item{We derived an asymptotic expansion of \bbd{q\ll n\uprm{\textsc{mmp}}} by:
\bii
\item{expanding \bbd{- {\rm Log}[q\ll n\uprm{\textsc{mmp}}]} as \bbd{\sqrt{{{8\pi n}\over {{\rm Im}[\s]}}}
+ \g[\s]\cc {\rm log}[n] + w\ll 0[\s] + \sum\ll {p\geq 1}\cc n\uu{-{p\over 2}}\cc w\ll p[s]};}
\item{using the recursion relations at large \bbd{n} and fixed \bbd{\t} to get a first-order ODE for each
coefficient function \bbd{w\ll p[s]}; and}
\item{solving each ODE in for \bbd{w\ll p[\s]} in closed form up to an integration constant \bbd{c\ll p} muliplying \bbd{s\uu{+{p\over 2}}}; and}
\item{fixing the value of \bbd{c\ll p} by taking the double-scaling limit and comparing with the order \bbd{\ddsc\uu{{p\over 2}}} term in the lage-\bbd{\ddsc} expansion
of the negative of the logarithm of ref. \cite{Grassi:2019txd}'s function \bbd{F\lp{\smgkt}\urm{inst}[\ddsc]}.}
\ei
}
\item{We gave explicit expressions for \bbd{w\ll p[\s]} up to and including \bbd{p=5} using this algorithm;}
\item{We evaluated at \bbd{\t = {{25}\over \pi}\times i} and calculated the estimate of \bbd{q\ll n\uprm{\textsc{mmp}}} up to and including (next-to)${}\uu 6$leading order, by which we mean absolute order \bbd{n\uu{-{5\over 2}}} in
the exponent of the estimate for \bbd{q\ll n\uprm{\textsc{mmp}}}.}

  \item{We used numerically-computed localization results \cite{LocData} at \bbd{\t = {{25}\over{\pi}}\times i} and \bbd{1 \leq n \leq  150} as
a basis to judge the accuracy of our estimates.}

 \item{We showed that there are no "additional" instanton corrections to the expressions \bbd{w\ll p[\s]} in addition to the instanton "corrections" already
present in the relationship between \bbd{s} and \bbd{\t}.  In
other words, we showed that the fixed-coupling
large-R-charge expansion of the MMP function is independent
of the \emm{infrared} \bbd{\th-}angle \bbd{\th\lrm{IR} \equiv 2\pi {\rm Re}[\s]}. At \bbd{\t = {{25}\over \pi}\times i,} the instanton corrections are in any case
too small to be visible and have not been included in the numerical
computations of correlators by localization and the method of \cite{Gerchkovitz:2016gxx}; it would be valuable to work at a somewhat
stronger gauge coupling and compare our asymptotic estimates
with fully instanton-corrected correlation functions.  We hope this
can be attempted in the future.}
 \item{We found remarkably good agreement between the asymptotic expansion to (next-to-)${}\uu 6$leading order on the one hand, and the actual value of the MMP correction as computed \cite{LocData} numerically
via the algorithm of  \cite{Gerchkovitz:2016gxx}
on the other hand.  At the highest values \bbd{n\sim 150} reached by the data available to us \cite{LocData}, the agreement is within between one part in one hundred million and one part in a billion, of the size of
MMP correction itself, which is already exponentially small compared to the log of the
full correlator. }
\item{We have also computed subleading large-\bbd{n} corrections at fixed \bbd{\ddsc,} up to and including order \bbd{n\uu{-5}}.  These corrections have a more complicated functional form, but can 
nonetheless be solved for explicitly}
\item{We compared the double-scaled large-charge expansion of
the MMP function to localization results at \bbd{\ddsc = {n\over{100}}} and for values of at least \bbd{n\gtrsim 10} we have
found agreement to a few parts in a hundred million or less, of the size of the
exponentially small correction itself, at the largest values of \bbd{n} where we can compare with localization results.  (This is excluding "accidental accuracies" in which the estimate transitions between 
slightly overestimating and slightly underestimating the exact result as \bbd{n} is varied, generating an atypically precise estimate in certain small ranges of \bbd{n} at a given order in the expansion.  These
accidental accuracies show up as logarithmic spikes on the accuracy charts displaying the effective number of accurate digits of the various estimates.)}
\ei

  \heading{Conclusions}

In the present
paper, we have rather mechanically executed an algorithm to
generate the hyperasymptotic MMP corrections to the exact \textsc{eft}
result for the logarithm of the correlation function of the $n\uu{\rm{\underline{th}}}$ power of the chiral ring generator.  We have
done this both for the fixed-\bbd{\t} large-charge expansion
and for its double-scaled ({\it i.e. \rm} fixed-\bbd{\ddsc}) counterpart.
In future work there are many interesting conceptual issues to be explored
relating to the convergence of both types of asymptotic series, particularly
in light of recent progress in the application \cite{Dondi:2021buw} of resurgence theory to the large-charge expansion.  But
here we have set aside larger and deeper questions and
restricted ourselves to executing the steps to generate higher
order terms as the    algorithm  has instructed us.  Following this recipe
produces asymptotic  estimates of eight or more digits of accuracy relative to the exponentially small correction itself.

\section*{Acknowledgments}%

We thank Domenico Orlando and Susanne Reffert for collaboration 
on closely related work.  We particularly thank Domenico Orlando for indispensable comments and advice, and especially for providing numerical results of exact calculation of correlators by localization that
allowed us to check our asymptotic expansion to high precision.  In addition we thank Zohar Komargodski for discussions on the
scheme-dependence of the sphere partition function in \bbd{{\cal N} = 2} superconformal SQCD.  The work of S.H. is supported by the World Premier International Research
Center Initiative (\textsc{wpi} Initiative), \textsc{mext}, Japan; by the \textsc{jsps} Program for Advancing Strategic International Networks to Accelerate the Circulation of Talented Researchers;
and also supported in part by \textsc{jsps kakenhi} Grant Numbers \textsc{jp22740153, jp26400242}.
We are also grateful to the Simons Center for Geometry and Physics for hospitality during the program, ``Quantum Mechanical Systems at Large Quantum Number,'' during which this work was initiated.

\end{document}